\documentclass[11pt]{article}
\usepackage[letterpaper,margin=1in]{geometry}
\usepackage[T1]{fontenc}
\usepackage{lmodern,microtype}
\usepackage{amsmath,amssymb,amsthm,mathtools}
\usepackage{aliascnt,booktabs,tabularx,array,enumitem}
\usepackage{algorithm,algpseudocode,xspace,needspace}
\usepackage[dvipsnames]{xcolor}
\usepackage[colorlinks=true,linkcolor=blue!45!black,citecolor=blue!45!black,urlcolor=blue!45!black]{hyperref}
\usepackage[capitalize,noabbrev,nameinlink]{cleveref}
\usepackage{graphicx}
\usepackage{tikz}
\usetikzlibrary{arrows.meta,decorations.pathreplacing,calc}
\usepackage{placeins}
\setlist{topsep=5pt,itemsep=3pt,parsep=0pt}
\allowdisplaybreaks[1]
\newtheorem{theorem}{Theorem}[section]
\newaliascnt{lemma}{theorem}
\newtheorem{lemma}[lemma]{Lemma}
\aliascntresetthe{lemma}
\crefname{lemma}{Lemma}{Lemmas}
\newaliascnt{corollary}{theorem}
\newtheorem{corollary}[corollary]{Corollary}
\aliascntresetthe{corollary}
\crefname{corollary}{Corollary}{Corollaries}
\newaliascnt{proposition}{theorem}
\newtheorem{proposition}[proposition]{Proposition}
\aliascntresetthe{proposition}
\crefname{proposition}{Proposition}{Propositions}
\theoremstyle{definition}
\newaliascnt{definition}{theorem}
\newtheorem{definition}[definition]{Definition}
\aliascntresetthe{definition}
\crefname{definition}{Definition}{Definitions}
\theoremstyle{remark}
\newtheorem*{remark}{Remark}
\newcommand{\E}{\mathbb E}
\newcommand{\calM}{\mathcal M}
\newcommand{\calI}{\mathcal I}
\newcommand{\OPT}{\operatorname{Greedy}}
\newcommand{\ALG}{\operatorname{ALG}}

\newcommand{\rank}{\operatorname{rank}}
\newcommand{\cl}{\operatorname{cl}}
\newcommand{\ind}{\mathbf 1}
\newcommand{\ee}{\mathrm e}
\newcommand{\Unif}{\operatorname{Unif}}
\newcommand{\Bin}{\operatorname{Bin}}
\newcommand{\Law}{\mathcal D}
\newcommand{\IGNORE}[1]{}
\newcommand{\Surv}{\mathcal A}
\newcommand{\FK}{\mathsf F}
\newcommand{\RK}{\mathsf R}
\newcommand{\Dist}{\mathsf D}
\title{As Easy as Rank One: Optimal Algorithms for \\ Matroid Secretary and Single-Sample Matroid Prophet%
\footnote{
This paper merges the preprints of Singla~\cite{Singla-arXiv26};
Abdi, Banihashem, Hajiaghayi, and Mittal~\cite{ABHM-arXiv26}; and
Huang~\cite{Huang-arXiv26}. Beyond unifying their presentation, it
simplifies several proofs and adds new results.}}
\author{%
Hamed Abdi\thanks{University of Maryland, College Park. Email: \texttt{hamedabdi37@gmail.com}.}
\and Kiarash Banihashem\thanks{University of Maryland, College Park. Email: \texttt{kiarash@umd.edu}.}
\and MohammadTaghi Hajiaghayi\thanks{University of Maryland, College Park. Email: \texttt{hajiaghayi@gmail.com}.}
\and Zhiyi Huang\thanks{The University of Hong Kong. Email: \texttt{zhiyi@cs.hku.hk}.}
\and Danny Mittal\thanks{University of Maryland, College Park. Email: \texttt{dannymittal@gmail.com}.}
\and Sahil Singla\thanks{School of Computer Science, Georgia Institute of Technology. Email: \texttt{ssingla@gatech.edu}. Supported in part by NSF awards CCF-2327010 and CCF-2440113.}}

\usepackage{color-edits}
\addauthor{kb}{cyan}
\addauthor{dm}{ForestGreen}
\addauthor{ss}{red}
\addauthor{zh}{blue}

\date{November 2026}

\begin{document}
\maketitle

\begin{abstract}
We resolve the matroid secretary conjecture in its strong form
and the single-sample matroid prophet inequality problem, achieving
competitive ratios $\ee$ and $2$, respectively.  Both ratios are
optimal, even for rank-one matroids.

\smallskip

For matroid secretary, our ordinal algorithm accepts the elements
of the canonical optimum independently, each with probability
$1/\ee$. It needs only the number of elements in advance and uses
an expected polynomial number of independence queries on arrived
elements. For single-sample matroid prophet inequalities, our
algorithm uses one sample per independent value distribution and
$O(n^2)$ independence queries on a known $n$-element matroid. It works for every
arrival order fixed independently of the samples, values, and
algorithmic randomness.
The selected elements and their values have exactly the joint
distribution obtained by independently retaining each
element--value pair in the canonical optimum of a fresh product
draw with probability $1/2$.

\smallskip

Our main algorithms maintain a reference state whose greedy
solution contains every accepted element. 
Its updates, made even after rejections, preserve exact
distributional invariants.
A short reversible-update argument gives a $4$-probability-competitive
secretary algorithm, already resolving the original conjecture; a
related invariant gives the optimal single-sample prophet inequality.
For secretary, we characterize the distributions achievable by
ordinal algorithms through a finite polytope. 
A continuous-time formulation then lets us design an accepted-set
distribution achieving the optimal ratio $\ee$ and realize it
efficiently.
We also obtain guarantees for matroid intersections.
\end{abstract}

\clearpage
 
\section{Introduction}

The classic secretary problem~\cite{Dynkin1963,Lindley1961} asks
us to select the largest of $n$ distinct values revealed in uniformly
random order, with immediate and irrevocable decisions. Rejecting
roughly the first $n/\ee$ values and then selecting the next value larger than all earlier ones
gives the asymptotically optimal success probability $1/\ee$.
Motivated by applications in auction design, Hajiaghayi, Kleinberg,
and Parkes~\cite{HKP-EC04} and Kleinberg~\cite{Kleinberg-SODA05}
studied the generalization in which up to $r$ values may be selected.
Babaioff, Immorlica, and Kleinberg~\cite{BIK-SODA07}, with Kempe
in the journal version~\cite{BIKK-JACM18}, introduced the
\emph{matroid secretary problem}: fixed nonnegative values arrive
in uniformly random order, and the selected elements must form an
independent set of a matroid. Does such a constraint make online selection any harder than choosing a single element?

Matroids are natural constraints for this question. Offline,
decreasing-value greedy finds a maximum-value independent set;
indeed, among downward-closed systems, matroids are exactly those
for which greedy is optimal for every nonnegative value
assignment~\cite{Edmonds1971}. 
An algorithm is \emph{$c$-competitive}, for $c\geq1$, if its expected
value is at least $1/c$ times the offline optimum. 

The \emph{matroid secretary conjecture}~\cite{BIK-SODA07} asks
whether a universal constant suffices for every matroid; its
strong form asks for the optimal constant $\ee$, matching rank one.
Before the results presented here, the best general guarantee
was $O(\log\log r)$, where $r$ is the matroid
rank~\cite{Lachish-FOCS14,FSZ-SODA15}. This bound had stood since
2014, despite constant guarantees for
graphic~\cite{BIK-SODA07}, laminar~\cite{IW-SODA11}, and
transversal~\cite{DP-ICALP08} matroids, among many other classes
(see \Cref{tab:msp-special-cases-compact}). Natural greedy and
partition-based approaches also face provable
barriers~\cite{BBSW-WINE21,AKKG-ITCS23}.

A related question concerns the information needed for online
selection. In a \emph{matroid prophet inequality}, the values are
independent draws from known distributions, and the arrival order
may be any fixed order. Kleinberg and Weinberg~\cite{KW-STOC12}
gave a $2$-competitive algorithm for every matroid, matching
the classical single-element prophet inequality~\cite{KS1978}.
In the \emph{single-sample} version~\cite{AKW-SODA14}, the distributions are unknown
and the algorithm receives only one independent sample from each.
For rank-one matroids, one sample  per distribution
already suffices for the optimal ratio~$2$~\cite{RWW-ITCS20}.
For general matroids, however, constant ratios were known only with polylogarithmically many samples per distribution~\cite{FLTWWZ-EC24,FSZ-SODA26}.
Does one sample suffice for a constant competitive ratio,
or even for the optimal ratio $2$? A positive answer would also resolve the matroid secretary
conjecture for known matroids, by Li's reduction~\cite{FLTWWZ-EC24}
(\Cref{sec:reduction}).

We resolve both problems with optimal ratios $\ee$ and $2$,
matching the rank-one guarantees.
Since natural greedy rules that compare arrivals with a fixed sample
can fail (\Cref{sec:lookingBack}), our algorithms maintain a
reference state whose greedy solution contains every accepted element,
updating it even after rejections to preserve an exact
distributional invariant.
Our secretary algorithms use only \emph{ordinal} information:
the relative order of observed values.
Moreover, our optimal algorithm accepts each element of the canonical
optimum\footnote{The canonical optimum is obtained by
decreasing-value greedy, with ties broken by a fixed order of
element labels.} independently with probability $1/\ee$.
Such a per-element guarantee of $1/c$, called
$c$-\emph{probability-competitiveness}, implies ordinary
$c$-competitiveness.

\subsection{Results}

\begin{theorem}[Optimal matroid secretary; informal
\Cref{thm:survival-efficient}] \label{thm:intro-survival-efficient}
For every $n$-element matroid revealed online, there is an ordinal
secretary algorithm that accepts the elements of the canonical
optimum independently, each with probability $1/\ee$.
Knowing only $n$ in advance, it uses an expected polynomial
number of independence queries on subsets of arrived elements.
\end{theorem}

Our second result considers the standard single-sample prophet model with an oblivious arrival order: the matroid and one independent sample per element are given before arrivals.

\begin{theorem}[Optimal single-sample matroid prophet inequality; informal
\Cref{thm:ssmpi-tight}] \label{thm:intro-prophet}
For every known matroid and independent nonnegative value
distributions, one independent sample from each distribution
suffices for a $2$-competitive algorithm using comparisons and
$O(n^2)$ independence queries. The arrival order may be any fixed
order independent of the samples, values, and algorithmic
randomness; the algorithm need not know that order.
\end{theorem}

More strongly, its accepted elements and values are distributed
exactly as an independent half-sample of the canonical optimum of
fresh independent values, so it obtains exactly half the expected
offline optimum.

A short combinatorial  argument  proves the original matroid secretary conjecture (\Cref{sec:snapshot}).

\begin{theorem}[Combinatorial matroid secretary; informal
\Cref{thm:four-secretary}] \label{thm:intro-combinatorial}
For every $n$-element matroid revealed online, there is an ordinal
secretary algorithm that accepts each element of the canonical
optimum with probability at least $1/4$.
Knowing only $n$ in advance, it uses $O(n^2)$ independence
queries on subsets of arrived elements.
\end{theorem}

\noindent\textbf{Characterizing online acceptance distributions.}
For a known matroid, we also show that a finite polytope exactly
characterizes the conditional accepted-set distributions of ordinal secretary algorithms
(\Cref{thm:polytope-realization}). 
Its continuous-time version turns algorithm design into distribution
design: every family satisfying its constraints is realized online.
Constructing one such family already resolves the strong matroid
secretary conjecture; we then give an efficient implementation
(\Cref{thm:tight-secretary,thm:survival-efficient}).
The same characterization also gives exact LPs for two benchmarks
among ordinal algorithms (\Cref{sec:lp-objective}): 
the optimal per-element
selection probability (\emph{probability competitiveness}) and the
best expected-value fraction (\emph{ordinal competitiveness}).
These differ even on uniform matroids: for rank $k$ and $n\geq2k$,
the first is $1-\Theta(\sqrt{\log k/k})$ as $k\to\infty$, whereas
ordinal algorithms achieve expected-value fraction $1-O(k^{-1/2})$
(\Cref{cor:uniform-separation}). The matching lower bound resolves
a question of Soto, Turkieltaub, and Verdugo~\cite{SotoTV-SODA18}.

\medskip\noindent\textbf{Finite-size guarantees and extensions.}
For linear matroids known in advance, one can beat $1/\ee$ for
every finite $n\geq2$: a separate construction selects each
optimum element with the classical rank-one success probability
for $n$ elements (\Cref{thm:linear}).
Computations
on small matroids suggest the same finite-size comparison for
ordinal competitiveness more generally (\Cref{sec:lp-computations}).
Beyond a single matroid, combining our reference-set method with
the preprocessing of Feldman, Svensson, and
Zenklusen~\cite{FSZ-SICOMP22} gives ratio $8k^2(k+1)$ for
intersections of $k$ matroids (\Cref{thm:intersection}), avoiding the exponential dependence
on $k$ in their general combination bound.

\subsection{Technical overview}

Inserting an optimum element into a greedy solution displaces at
most one element, its \emph{exchange partner}. If this partner were
fixed independently of a uniformly random processing order, then
with probability $1/2$ the optimum element would precede it and could
be accepted safely. But the partner depends on the algorithm's state,
which depends on the order. Instead, our algorithms maintain a
reference state whose greedy solution contains every accepted
element, and update it even after rejections so that the state seen
by any fixed element, its \emph{snapshot}, has an explicit distribution.

\smallskip\noindent\textbf{Reversible updates (\Cref{sec:snapshot}).}
The combinatorial algorithm of \Cref{thm:intro-combinatorial}  first rejects a random half-sample,
which becomes the reference set. It then tries to insert each real
arrival and, interleaved with these, to delete each sample element,
updating only when greedy membership of every previously processed
element is preserved. This test is symmetric, so every update can
be undone: for any fixed processing order, the updates merely
permute the equally likely initial samples. Hence each snapshot is
uniform and independent of the processing order, and conditioning
on it fixes the exchange partner without biasing their relative
order. An optimum element is a real arrival with probability $1/2$
and is then accepted with probability at least $1/2$, giving $1/4$.

\smallskip\noindent\textbf{Exchangeable values (\Cref{sec:ssmpi-tight}).}
The prophet algorithm of \Cref{thm:intro-prophet} stores one value per element, initially its
sample, and each arrival proposes replacing its sample by its real
value. If the update would remove a processed element from the
greedy basis, that element's acceptance blocks it and its rejection
forces it; otherwise a fair coin decides. The arrival is accepted
if the update occurs and it belongs to the new basis. By exchangeability of sample and real values, the stored values
remain a fresh product draw, and the accepted set a half-sample of
its processed greedy elements.

\smallskip\noindent\textbf{Distributions as algorithms (\Cref{sec:lp}).}
For the optimal ratio $\ee$, we reverse the approach: design the
accepted-set distributions first, and then realize them online.
In a continuous-time formulation, two linear conditions
characterize which distributions are realizable: \emph{capacity},
which prevents moving more probability mass than a state contains,
and \emph{consistency}, which ensures that the transitions produce
the prescribed distributions. Every feasible family is realized by an explicit rule (\Cref{thm:continuous-realization}), so it suffices to construct one.

\smallskip\noindent\textbf{Survivor distributions (\Cref{sec:survival}).}
We construct this family from an offline experiment with independent auxiliary arrival times:
 sample each element independently with
probability $1/\ee$, and then thin the sample's greedy solution to obtain the \emph{survivors}.
The thinning probabilities are motivated by the rank-one analysis
and preserve every sampled global optimum element. The survivors
therefore form an independent set and include each optimum element
independently with probability $1/\ee$.
The main obstacle is showing that this family is feasible. Consistency follows from a
direct calculation as time advances. The crux is capacity: for every
possible preceding accepted set, there must be enough probability
mass to support the next acceptance. We prove it by induction after removing the largest element, comparing the survivor experiment with the same experiment in the matroid contracted by the arriving element. 
The rule realizing this family (\Cref{alg:survival-rule}) already  resolves
the strong matroid secretary conjecture
(\Cref{thm:tight-secretary}).

\smallskip\noindent\textbf{From proof to efficient algorithm (\Cref{sec:implementation}).}
This rule needs probabilities of entire survivor sets, which we do
not know how to compute efficiently. We therefore turn the capacity proof into a sampler: inductive comparisons
become recursive calls, and unused mass becomes rejection. The
sampler needs the auxiliary randomness conditioned on the accepted
set, so we maintain a sample of it, one coordinate per element.
Forward updates make new acceptances survive; reverse updates
prevent old rejections from surviving. 
A continuous snapshot argument shows these updates keep the sample
correctly distributed, giving an expected-polynomial-time algorithm (\Cref{thm:intro-survival-efficient}).

\begin{table}[htbp]
\centering
\small
\setlength{\tabcolsep}{4pt}
\renewcommand{\arraystretch}{1.05}
\begin{tabularx}{\linewidth}{@{}>{\raggedright\arraybackslash}p{0.29\linewidth}
>{\raggedright\arraybackslash}X@{}}
\toprule
\textbf{Matroid class} & \textbf{Guarantees and references}\\
\midrule
General matroids
& $O(\log\log r)$\,\cite{Lachish-FOCS14,FSZ-SODA15}
  $\longrightarrow$ $\ee^{\mathrm P}$
  (this paper, \Cref{thm:survival-efficient}). \\
\addlinespace[3pt]
Rank one / partition & $\ee$\,\cite{Lindley1961,Dynkin1963,BIKK-APPROX07}. \\
\addlinespace[3pt]
Uniform (rank $r$) & $O(1)$\,\cite{HKP-EC04}; $1+O(r^{-1/2})$\,\cite{Kleinberg-SODA05}; $[1+O(\sqrt{\log r/r})]^{\mathrm P}$\,\cite{SotoTV-SODA18}; finite-rank refinements\,\cite{BJS-MOR14,CCJ-SODA15,AL-TCS21}. \\
\addlinespace[3pt]
Truncated partition & $O(1)$\,\cite{BIK-SODA07,BIKK-JACM18}; also covered by laminar results. \\
\addlinespace[3pt]
Transversal & $16$\,\cite{DP-ICALP08}\ $\to8$\,\cite{KP-ICALP09}\ $\to \ee$\,\cite{KRTV-ESA13}; $\ee^{\mathrm P}$\,\cite{SotoTV-SODA18}. \\
\addlinespace[3pt]
Laminar & $O(1)$\,\cite{IW-SODA11}; $1/0.053$\,\cite{HP-Laminar13}; $3\sqrt3\ee$\,\cite{JSZ-IPCO13}; $9.6$\,\cite{MTW-TOCS16}; $(3\sqrt3)^{\mathrm P}$\,\cite{SotoTV-SODA18}; $4.75$\,\cite{HPZ-ESA24}; $[1/(1-\ln2)]^{\mathrm P}$\,\cite{BLSV-IPCO25}. \\
\addlinespace[3pt]
Graphic & $16$\,\cite{BIK-SODA07}; $3\ee$\,\cite{BDGIT-SODA09}; $2\ee$\,\cite{KP-ICALP09}; $4^{\mathrm P}$\,\cite{SotoTV-SODA18}; $(1/0.2504)^{\mathrm P}$\,\cite{BLSV-IPCO25}; $3.95$\,\cite{BHKKMO-ESA25}. \\
\addlinespace[3pt]
Simple / high-girth graphic & Simple: $3.77$\,\cite{BHKKMO-ESA25}, $(1/0.2693)^{\mathrm P}$\,\cite{BLSV-IPCO25}; girth $g$: $\ee+o_g(1)$\,\cite{BHKKMO-ESA25}. \\
\addlinespace[3pt]
Cographic; regular / MFMC & $3\ee$\,\cite{Soto-SICOMP13}; $9\ee$ for regular and max-flow min-cut\,\cite{DK-SICOMP14}. \\
\addlinespace[3pt]
Hypergraphic; matching & $4^{\mathrm P}$ for both\,\cite{SotoTV-SODA18}; matching matroids have a vertex ground set. \\
\addlinespace[3pt]
$k$-column-sparse; $k$-framed & $k\ee$ for column-sparse\,\cite{Soto-SICOMP13}; $[k^{k/(k-1)}]^{\mathrm P}$ for both ($k\ge2$)\,\cite{SotoTV-SODA18}. \\
\addlinespace[3pt]
Gammoids
& $[\mu^{\mu/(\mu-1)}]^{\mathrm P}$ for exchangeability $\mu\ge2$;
  semiplanar: $(4^{4/3})^{\mathrm P}$\,\cite{SotoTV-SODA18}. \\
\addlinespace[3pt]
Rank two & $(1/0.3462)^{\mathrm P}$\,\cite{BLSV-IPCO25}; all rank-two matroids. \\
\addlinespace[3pt]
Density / cocircuits
& $\gamma(\calM)$; $c$ if each nonloop lies in a cocircuit
  of size $\leq c$\,\cite{Soto-SICOMP13}. \\
\addlinespace[3pt]
Paving; almost all matroids & $1+O(r^{-1/2})$ for paving; $2+o(1)$ for almost all labelled matroids\,\cite{HN-SIDMA20}. \\
\addlinespace[3pt]
Proper minor-closed $\mathcal C$
& $O_{\mathcal C}(1)$ over $\mathbb F_p$ (fixed prime $p$),
  \emph{under a structural hypothesis}\,\cite{HN-SIDMA20}. \\
\addlinespace[3pt]
$k$-fold unions &  $1+O(\sqrt{\log n/k})$ for $k\gg\log n$\,\cite{GHKQ-kFold25}. \\
\addlinespace[3pt]
Graphic, arrived-elements-only  & $36$ \,\cite{DPLPP-Graphic26}. \\
\addlinespace[3pt]
\bottomrule
\end{tabularx}
\caption{Earlier guarantees and our optimal general-matroid ratio.
$\mathrm P$ denotes a per-element probability guarantee.
Some results require advance knowledge of the matroid or its
representation. Parameters $\mu$ and $\gamma(\calM)$ are defined
in the cited works.}
\label{tab:msp-special-cases-compact}
\end{table}

\subsection{Related work and roadmap}

\noindent\textbf{Matroid secretary and related models.}
For general matroids, the guarantee improved from
$O(\log r)$~\cite{BIK-SODA07} to $O(\sqrt{\log r})$~\cite{CL-SODA12}
and then to $O(\log\log r)$~\cite{Lachish-FOCS14,FSZ-SODA15};
\Cref{tab:msp-special-cases-compact} lists progress on special classes. Constant guarantees were known in models that weaken the adversary: free-order secretary~\cite{JSZ-IPCO13,AKW-SODA14}, where the algorithm chooses the order; randomly assigned values~\cite{Soto-SICOMP13,GV-Journal13}; and prophet secretary~\cite{EHKS-SICOMP24},
with known value distributions.

\smallskip\noindent\textbf{Sample-based prophet inequalities.}
Beyond rank one~\cite{RWW-ITCS20}, single-sample guarantees were  known
for several matroid classes and matchings~\cite{CDFFLLP-SODA22}.
For general matroids, \cite{FLTWWZ-EC24} obtained ratio
$4+\varepsilon$ using $O_\varepsilon(\log^4 n)$ samples per
distribution. Feldman, Svensson, and Zenklusen~\cite{FSZ-SODA26}
reduced this to $O_\varepsilon(\log n+\log r\,\log^2\log r)$
samples with the same ratio. 
Both handle an \emph{almighty} adversary, which sees the samples,
values, and algorithmic randomness before choosing the order.
We assume an oblivious order.
Complementing Li's reduction, Azar, Kleinberg, and Weinberg~\cite{AKW-SODA14} convert
\emph{order-oblivious} secretary algorithms, whose guarantee survives
an arbitrary order after a random initial sample, into single-sample prophet algorithms with the same ratio.

\smallskip\noindent\textbf{LP and distributional approaches.}
Buchbinder, Jain, and Singh~\cite{BJS-MOR14} give finite LP
characterizations for classical and multiple-choice secretary
problems; Chan, Chen, and Jiang~\cite{CCJ-SODA15} develop
continuous LPs for these variants. These LPs track only the number of accepted elements, which suffices under cardinality constraints; for general matroids, feasibility depends on the accepted set itself, so our polytope tracks distributions over accepted sets.
Random-order contention
resolution~\cite{AW-FOCS18,LS-ESA18} provides related per-element
guarantees, and Dughmi~\cite{Dughmi-SICOMP25} establishes an
equivalence between matroid secretary and correlated random-order
contention resolution. Aminian, Niazadeh, and Nuti~\cite{ANN-arXiv26}
study stationary online contention resolution schemes,
whose output distribution is independent of arrival order.
  Our characterization instead conditions on the observed elements
  and their relative value order, averaging over the arrival order.

\smallskip\noindent\textbf{Concurrent and subsequent work.}
The linear-matroid guarantee, including the finite-size
probabilities in \Cref{thm:linear}, was obtained independently
in~\cite{BDLSV-arXiv26}. 
Building on Singla's preprint~\cite{Singla-arXiv26},
Chan, Lin, and Wang~\cite{CLW-arXiv26} improved the
general-matroid competitive ratio from $4$ to 
$\approx 3.1462$.
OpenAI~\cite{OpenAI-AlmightyProphet26} subsequently obtained a
constant-competitive single-sample matroid prophet inequality
against the stronger almighty adversary.

\smallskip\noindent\textbf{Roadmap.}
  \Cref{sec:snapshot,sec:ssmpi-tight} prove the $4$-competitive
  secretary and optimal single-sample prophet guarantees.
  \Cref{sec:lp,sec:survival,sec:implementation} develop the
  polyhedral characterization, the optimal survivor distributions,
  and their efficient implementation, respectively.
     \Cref{sec:extensions} gives finite-size results and extensions.
   AI models played a material role in developing the main algorithmic    guarantees; the AI methodology statement describes this role.
   The appendices contain examples, deferred proofs, and computations.

\section{Matroid Secretary via Reversible Updates}
\label{sec:snapshot}

We begin with a combinatorial $4$-probability-competitive algorithm
for matroid secretary. Its proof illustrates two ideas that will recur:
protect irrevocable decisions through a changing greedy solution,
and make the updates to a reference state reversible, so that they preserve a useful distribution.

\subsection{Preliminaries}

Fix a matroid $\calM=(V,\calI)$ with nonnegative values $v_e$.
Ties are broken by a fixed order of element labels, independent of
arrival order; ``larger'' and ``smaller'' refer to this tie-broken
value order. Write $T+e=T\cup\{e\}$, $T-e=T\setminus\{e\}$, and
$v(A)=\sum_{e\in A}v_e$. For $T\subseteq V$, let $\OPT(T)$ be the
output of \emph{greedy}, which scans $T$ in decreasing value order
and adds each element that keeps the current set independent.
\emph{Greedy includes zero-valued elements whenever possible}, so $\OPT(T)$
is a basis of the restriction to $T$; this does not change the
optimum value.

The \emph{rank} of $T$ is the largest size of an independent subset
of $T$. An element is \emph{spanned} by $T$ if adding it does not
increase the rank; the \emph{closure} $\cl(T)$ is the set of all
such elements, and an element spanned by $\emptyset$ is a \emph{loop}.
The elements greedy selects from a prefix $P$ of its scan span $P$,
so an element is selected precisely when the elements preceding it
do not span it, and greedy selects exactly $\rank(P)$ elements
from $P$. We use the following standard
consequences~\cite{Edmonds1971,Brualdi1969,Schrijver2003}.

\begin{lemma}[Greedy monotonicity and one-element sensitivity]
\label{lem:greedy}
With fixed tie-breaking, the following properties hold.
\begin{enumerate}[label=(\roman*)]
\item If $e\in T\subseteq V$ and $e\in\OPT(V)$, then
      $e\in\OPT(T)$. More generally, reducing other values while
      keeping a selected element's value fixed does not remove it
      from greedy.
\item For $e\notin T$, the set $\OPT(T+e)$ is either $\OPT(T)$,
      $\OPT(T)+e$, or $(\OPT(T)-f)+e$ for one $f\in\OPT(T)$.
\item Changing one value on a fixed ground set either leaves the
      greedy basis unchanged or exchanges that element with one
      other element. An increase can only bring the changed element
      into the basis; a decrease can only remove it.
\end{enumerate}
\end{lemma}
\begin{proof}
For (i), restricting the ground set or reducing other values can
only shrink the selected element's predecessor set. A subset cannot
span an element that the larger set did not span.

For (ii), inserting $e$ into the scan can only add a predecessor
to each old element. Thus no old element can enter greedy, and the
only possible new member is $e$. Greedy returns bases of $T$ and
$T+e$, whose ranks differ by at most one, giving the three possibilities.

For (iii), increasing the value of $e$ shrinks its predecessor
set and can only add $e$ to the others', so only $e$ can enter and
only other elements can leave; since both bases have the same size,
any change is one exchange. A decrease is symmetric.
\end{proof}

\subsection{The algorithm}

\paragraph{Model.}
The $n=|V|$ elements arrive in a uniformly random order. On each
arrival, the algorithm observes the element and its value and must
irrevocably accept or reject it before the next arrival. The accepted
set must be independent. The number $n$ is known; values and the
matroid are otherwise unknown, and independence queries may involve
only observed elements. A $c$-probability-competitive algorithm
accepts each $e\in\OPT(V)$ with probability at least $1/c$.
An algorithm is \emph{ordinal} if it uses values only through their
tie-broken order.

\medskip\noindent\textbf{Algorithm.} 
The algorithm first rejects and stores a random initial sample $S$,
with $|S|\sim\Bin(n,1/2)$;
call the remaining arrivals \emph{real}. It then runs $n$ processing
steps, each either receives the next real arrival or
internally processes the next stored sample element.
Its reference state is a \emph{configuration} $C\subseteq V$, initially $S$; it also maintains the set $F$ of processed elements, initially empty.
A sample element enters $F$ only when internally processed and
remains rejected.
Write $C\oplus e$ for adding $e$ to $C$ if absent and removing it
if present, so $(C\oplus e)\oplus e=C$. Processing $e$ proposes $C'=C\oplus e$, a deletion for
a sample element and an insertion for a real arrival. The update
occurs exactly when it passes the \emph{preservation test}:
the greedy membership of every element of $F$ remains unchanged.

\begin{algorithm}[htbp]
\caption{Matroid Secretary via Reversible Updates}\label{alg:matroidSec}
\begin{itemize}[leftmargin=*,itemsep=2pt,parsep=0pt,topsep=3pt]
\item Draw $X$ uniformly from $\{0,1\}^n$, and let $K=\sum_i X_i$ denote the number of ones.
\item Reject the first $K$ arrivals. Store this sample $S$
      in arrival order; call the remaining elements real.
\item Initialize $\ALG=\emptyset$, configuration $C=S$, and processed set $F=\emptyset$.
\item For $i=1,\ldots,n$:   \Comment{Process $n$ elements one by one}
\begin{enumerate}[itemsep=1pt,parsep=0pt,topsep=3pt]
\item If $X_i=1$, let $e$ be the next stored sample element;
      otherwise, receive the next real arrival $e$. Set $C'=C\oplus e$.
\item If $\OPT(C')\cap F=\OPT(C)\cap F$:
\label{alg:direct-test}\label{alg:secretary-test}
\Comment{Preservation test}
\begin{enumerate}[itemsep=1pt,parsep=0pt,topsep=3pt]
\item If $e\in\OPT(C')$, set $\ALG\gets\ALG+e$. \Comment{Accept $e$}
\item Set $C\gets C'$. \Comment{Update even without acceptance}
\end{enumerate}
\item Set $F\gets F+e$. \Comment{Record every processed element}
\end{enumerate}
\item Return $\ALG$.
\end{itemize}
\end{algorithm}

Only the step processing $e$ can change its membership in $C$,
so immediately before that step $e\in C$ if and only if $e\in S$.
At a sample processing step, therefore, $e\notin C'$ and cannot be accepted in step 2(a).
All queried sets consist of observed elements.

The preservation test on all of $F$ protects earlier acceptances.
Its symmetry also makes each configuration update \emph{reversible}:
with $e$ and $F$ fixed, it swaps $C$ and $C'$ when the test passes
and leaves both unchanged otherwise.
Thus, for a fixed processing order, the updates permute the
possible configurations.  As a result, the configuration
before any fixed element is processed is a uniformly random subset
of $V$, independent of the processing order
(\Cref{cor:snapshot-independent}).
\Cref{sec:lookingBack} illustrates this mechanism and explains
why natural greedy alternatives fail.

\begin{theorem}\label{thm:four-secretary}
\Cref{alg:matroidSec} always returns an independent set and accepts
each $e\in\OPT(V)$ with probability at least $1/4$.
It is ordinal and uses $O(n^2)$ independence queries.
\end{theorem}

\begin{lemma}[Feasibility]\label{lem:secretary-feasibility}
After every processing step,
$\ALG\subseteq\OPT(C)\cap F$.
\end{lemma}
\begin{proof}
The inclusion holds initially, since $\ALG=F=\emptyset$. If a step's
preservation test fails, $C$ and $\ALG$ are unchanged and $F$ only grows. If it
passes, every previous acceptance lies in $F$, so it stays in the
greedy solution, and a new acceptance lies in $\OPT(C')$ and is then
added to $F$.
\end{proof}

\subsection{The snapshot lemma}

Let $\Pi$ be the processing order. For each $e\in V$, the
\emph{snapshot} $C_e$ is the configuration immediately before
processing $e$.

\begin{lemma}[Snapshot lemma]\label{lem:snapshot}
Fix an element $e$ and a processing order $\pi$.
For every $T\subseteq V$, exactly one initial sample set $S\subseteq V$ gives $C_e=T$.
\end{lemma}
\begin{proof}
Fix the processing order and consider a step before $e$.
Let $f$ be the element processed at this step and $F$ the set
processed before it. Both $f$ and $F$ are determined by the fixed
order.

At this step, the algorithm either leaves the configuration $C$
unchanged or replaces it by $C\oplus f$. The test
$\OPT(C)\cap F=\OPT(C\oplus f)\cap F$
has the same answer at $C$ and $C\oplus f$: the two sides are
simply interchanged. If it passes, starting from $C$ gives
$C\oplus f$, and starting from $C\oplus f$ gives $C$.
If it fails, each of these two configurations stays unchanged. Thus each step maps configurations bijectively onto themselves and is its own inverse; this is the reversibility of the updates.

Consequently, given the configuration $D$ immediately after this
step, we can recover the preceding configuration explicitly.
Compare $\OPT(D)\cap F$ and $\OPT(D\oplus f)\cap F$.
If they are equal, the preceding configuration was $D\oplus f$;
otherwise it was $D$. These are the only two possible predecessors,
since this step can change only the membership of $f$, and the test
selects exactly one of them.

Now start from the prescribed snapshot $T$ immediately before $e$ and
undo the preceding steps from last to first by the above rule. This
determines a unique configuration $S$ before the first step, which
must be the initial sample. Conversely, running the algorithm forward
from $S$ reverses each reconstruction step and reaches $T$, so $S$ indeed gives $C_e=T$.

The reconstruction does not require the accepted set: every test
uses only the current configuration, the current element, and the
processed set determined by the fixed order.
\end{proof} 

\begin{corollary}[Snapshot independence]
\label{cor:snapshot-independent}
For each fixed $e\in V$, the snapshot $C_e$ is a uniformly random
subset of $V$, independent of the uniformly random processing order $\Pi$.
\end{corollary}
\begin{proof}
Fix $T\subseteq V$ and a permutation $\pi$. By
\Cref{lem:snapshot}, there is exactly one initial sample $S_0$
giving $C_e=T$ under order $\pi$.
Exactly one bit vector and arrival order produce $(S_0,\pi)$:
$X_i=\ind\{\pi(i)\in S_0\}$, and the arrival order lists $S_0$
first and then $V\setminus S_0$, each in its order within $\pi$.
This choice has probability $2^{-n}/n!$. Hence
$\Pr[C_e=T,\Pi=\pi]=1/(2^n n!)$, which proves the claim.
\end{proof}

\subsection{The competitive guarantee}

\begin{proof}[Proof of \Cref{thm:four-secretary}]
By \Cref{lem:secretary-feasibility}, $\ALG$ lies in a greedy basis and is independent.
Fix $e\in\OPT(V)$ and condition on $C_e=T$ with $e\notin T$.
Since only $e$'s own step changes its membership in the configuration, $e\notin S$, so $e$ is processed at its real arrival. By \Cref{lem:greedy}(i),
$e\in\OPT(T+e)$, so it is accepted whenever the preservation test passes.

If inserting $e$ removes no element, the preservation test passes.
Otherwise, \Cref{lem:greedy}(ii) gives a unique displaced $f$ with
$\OPT(T+e)=(\OPT(T)-f)+e.$
Since $e\notin F$, the test passes precisely when $f\notin F$,
or equivalently when $e$ precedes $f$ in $\Pi$.
The partner $f$ is determined by $T$, while conditional on
$C_e=T$, the order $\Pi$ remains uniform by
\Cref{cor:snapshot-independent}. Hence $e$ precedes $f$ with
probability $1/2$. In either case,
$\Pr[e\in\ALG\mid C_e=T]\geq1/2$.
Finally, $\Pr[e\notin C_e]=\Pr[e\notin S]=1/2$. Averaging over these snapshots gives
$\Pr[e\in\ALG]\geq1/4$.

The algorithm uses only comparisons and independence queries on
observed sets, with at most two greedy scans of at most $n$ queries
at each of the $n$ steps.
\end{proof}

\section{Single-Sample Matroid Prophet Inequalities}
\label{sec:ssmpi-tight}

We next obtain the optimal $2$-competitive single-sample matroid
prophet inequality. 
As in \Cref{sec:snapshot}, we prove feasibility,
a snapshot lemma, and snapshot independence; exchangeability of
sample and real values now replaces reversibility. The invariant
has two parts: the configuration keeps its initial product
distribution, and, conditional on it, each processed greedy element
is accepted independently with probability $1/2$.

\subsection{The algorithm}

\paragraph{Model.} 
Fix a matroid $\calM=(V,\calI)$ with $n=|V|$ elements. Each element
$e$ has a nonnegative \emph{real value} $R_e$ drawn independently from an
unknown distribution $\Dist_e$. Before the arrivals, the algorithm
receives one sample $S_e\sim\Dist_e$ for each $e\in V$.
These samples are mutually independent and independent of all real
values. The matroid is given by an independence oracle on all subsets
of $V$. The real values arrive in an arbitrary fixed order $\pi$,
independent of the samples, real values, and algorithmic randomness.
The algorithm need not know $\pi$ in advance. It must immediately
accept or reject each arrival, and its accepted set must be independent.

\smallskip\noindent\textbf{Algorithm.}
The reference state is again a \emph{configuration}: $C$ assigns each $f\in V$ a value
$C(f)\in\{S_f,R_f\}$, initially $C(f)=S_f$. Once $R_e$ is revealed, write
$C\oplus e$ for switching coordinate $e$ between its sample and real
values, leaving the other coordinates unchanged. Thus
$(C\oplus e)\oplus e=C$. If the values are equal, $C\oplus e$ changes nothing numerically, but the analysis distinguishes the copies.
Only processing $e$ changes its
coordinate, so when $e$ arrives, $C(e)=S_e$ and $C\oplus e$
installs $R_e$. 
Write $\OPT(C)$ for the greedy basis of $V$ with values
$C(f)$ and the fixed tie-breaking from \Cref{sec:snapshot}.

Maintain the configuration $C$, the accepted set $\ALG$, and the processed set $F$.
The default is to update with probability $1/2$. If the update
would remove a processed element from the greedy basis, its
earlier decision determines the outcome: an acceptance blocks
the update, and a rejection forces it. Accept the arriving element
exactly when the update occurs and it belongs to the new basis.

\begin{algorithm}[htbp]
\caption{Single-Sample Matroid Prophet}\label{alg:ssmpi-tight}
\begin{itemize}[leftmargin=*,itemsep=2pt,parsep=0pt,topsep=3pt]
\item Initialize $\ALG=F=\emptyset$, $C(f)=S_f$ for every $f\in V$,
and $B=\OPT(C)$.
\item For each arriving element $e$, when $R_e$ is revealed:
\begin{enumerate}[itemsep=1pt,parsep=0pt,topsep=3pt]
\item Set $C'=C\oplus e$ and compute $B'=\OPT(C')$.
\item Draw a fresh independent fair bit $Z$. If
$(B\setminus B')\cap F=\{f\}$, set
$Z\gets\ind\{f\notin\ALG\}$.
\item If $Z=1$, set $C\gets C'$ and $B\gets B'$.
\Comment{Update even without acceptance}
\item If $Z=1$ and $e\in B'$,  set
$\ALG\gets\ALG+e$.      \Comment{Accept $e$}
\item Set $F\gets F+e$.
\Comment{Record every processed element}
\end{enumerate}
\item Return $\ALG$.
\end{itemize}
\end{algorithm}

By \Cref{lem:greedy}(iii), $(B\setminus B')\cap F$ has at most one
element. The override in step~2 applies only to a processed element
\emph{leaving} the basis, not one reentering it.

\begin{theorem}\label{thm:ssmpi-tight}
\Cref{alg:ssmpi-tight} always returns an independent set and satisfies
\[
\E\Big[\sum_{e\in\ALG}R_e\Big]
=\frac12\,\E\Big[\max_{I\in\calI}\sum_{e\in I}R_e\Big].
\]
It uses $O(n^2)$ independence queries and no distributional
knowledge beyond the given samples. More strongly, the selected
elements and their values have exactly the distribution obtained by drawing
$Y\sim\prod_{e\in V}\Dist_e$ and retaining each
$(e,Y_e)$ with $e\in\OPT(Y)$ independently with probability $1/2$.
\end{theorem}

The ratio $2$ is optimal even for rank-one matroids with known
distributions, by the classical prophet inequality~\cite{KS1978}. 

\begin{lemma}[Feasibility]\label{lem:ssmpi-feasibility}
After every arrival,
\[
\ALG\subseteq\OPT(C)\cap F,
\qquad C(e)=R_e\quad\text{for every }e\in\ALG.
\]
\end{lemma}
\begin{proof}
The statements hold initially. Any earlier acceptance removed
by an update would belong to $(B\setminus B')\cap F$, and
the override blocks exactly such an update. Thus every update
preserves earlier acceptances. 
A newly accepted element belongs to $B'$ and has its real value
in $C$. Later steps leave its coordinate unchanged.
Every arrival is added to $F$.
\end{proof}

\subsection{The snapshot lemma}

As in \Cref{sec:snapshot}, the \emph{snapshot} $C_e$ is the configuration that $e$ faces, immediately before it is processed. Two independent draws from the
same distribution can equivalently be generated by drawing their
unordered pair and then assigning its entries to sample and real
by a fair coin. 
For the next lemma, fix the arrival order and, for each $e$,
the unordered pair $0\leq\ell_e\leq u_e$.
The assignments of each pair's entries to $S_e$ and $R_e$
remain independent fair coins; the algorithm's fresh coins
also remain random and are independent of these assignments.
The pairs are fixed only for analysis; the algorithm does not
know them before the corresponding arrivals.

Let the bit $\xi_e$ record which \emph{endpoint} currently supplies
$C(e)$: zero means $\ell_e$ and one means $u_e$.
Distinguish the endpoint copies even when their numerical values
are equal. Initially, $\xi$ is uniform on $\{0,1\}^V$.
A performed update at $e$ in step~3 flips $\xi_e$, installing its real value.
For fixed value pairs, $\xi$ determines the configuration $C$, and
in particular each snapshot $C_e$.
An \emph{independent half-sample} of a set is obtained by retaining
each element independently with probability $1/2$.

\begin{lemma}[Snapshot lemma]\label{lem:ssmpi-snapshot}
Fix the value pairs and arrival order. After any fixed prefix,
let $C$, $F$, and $\xi$ denote the current configuration,
processed set, and endpoint vector. Then $\xi$ is uniform on
$\{0,1\}^V$, and conditional on $\xi$, the accepted set is an
independent half-sample of $\OPT(C)\cap F$.
\end{lemma}
\begin{proof}
Induct on the processed prefix. Initially, $\xi$ is uniform and
$F=\ALG=\emptyset$.
Let $e$ be the next element. Condition only on the current
coordinates $\xi_{-e}=(\xi_f)_{f\ne e}$, averaging over the
preceding execution rather than fixing its decisions.
Let $B_0$ and $B_1$ be the greedy bases of the configuration given
by $\xi_{-e}$ together with $C(e)=\ell_e$ or $C(e)=u_e$,
respectively. Both are determined by $\xi_{-e}$, and the algorithm's
current and proposed bases are $B=B_{\xi_e}$ and $B'=B_{1-\xi_e}$.
The induction hypothesis makes $\xi_e$ fair and specifies the
conditional accepted-set distribution in each orientation.
Write $\xi'_e$ for this bit after processing $e$. We split the
analysis into two cases.

\smallskip
\noindent\emph{Case 1: $B_0\cap F=B_1\cap F$.}
Write $P=B_0\cap F=B_1\cap F$. 
The old accepted set is an independent half-sample of $P$,
independent of $\xi_e$. No processed element would leave the basis, so step~2 does not
override the fresh fair bit $Z$. Step~3 flips $\xi_e$ when
$Z=1$ and leaves it unchanged when $Z=0$. Thus
\[
\xi'_e=(\xi_e+Z)\bmod 2.
\]
The bits $\xi'_e$ and $Z$ are independent and fair, and both
are independent of the old accepted set. Earlier acceptances
are unchanged, and $e$ is accepted exactly when $Z=1$ and
$e\in B_{\xi'_e}$. Thus, conditional on the new endpoint vector,
the accepted set is an independent half-sample of
$B_{\xi'_e}\cap(F+e)$.

\smallskip
\noindent\emph{Case 2: $B_0\cap F\ne B_1\cap F$.}
By \Cref{lem:greedy}(iii), increasing $e$'s value from $\ell_e$
to $u_e$ exchanges a unique processed element $f\in F$ for $e$:
$B_1=B_0-f+e$.
Put $P=(B_0\cap F)-f=B_1\cap F$, the processed elements
common to both bases.
 By induction, the accepted subset of $P$ is
an independent half-sample of $P$ in both orientations. Condition
on this subset; then $\xi_e$ remains fair, and only $e$ and $f$
remain to be tracked.

If $\xi_e=0$, the current basis is $B_0\ni f$, and by induction
$f$ was accepted with probability $1/2$, even under this conditioning. The update would install $u_e$
and remove the processed $f$, so step~2 overrides $Z$: acceptance
of $f$ blocks the update, while rejection of $f$ forces it, and then
$e\in B_1$ is accepted. If $\xi_e=1$, the current basis is
$B_1\not\ni f$, so $f$ was rejected by feasibility. The update would
install $\ell_e$ and remove only the unprocessed $e$, so the fresh
bit $Z$ decides. Either way $e$ is rejected: with $Z=0$ nothing
changes, and with $Z=1$ the new basis $B_0$ excludes $e$.
Hence each row below has probability $1/4$, conditional on the
accepted subset of $P$.
\begin{center}
\begin{tabular}{@{}lccc@{}}
\toprule
$\xi_e$ & Deciding event & New basis &
$\ALG_{\mathrm{new}}\cap\{e,f\}$\\
\midrule
$0$ & $f$ accepted & $B_0$ & $\{f\}$\\
$1$ & $Z=1$        & $B_0$ & $\emptyset$\\
\addlinespace[3pt]
$0$ & $f$ rejected & $B_1$ & $\{e\}$\\
$1$ & $Z=0$        & $B_1$ & $\emptyset$\\
\bottomrule
\end{tabular}
\end{center}
Each new basis appears in two rows, so $\xi'_e$ is fair.
Given the new basis, its unique member of $\{e,f\}$ is accepted
in exactly one of those rows, hence with probability $1/2$.
These probabilities hold for every accepted subset of $P$.
Thus conditioning on the new basis leaves that subset an
independent half-sample of $P$, independent of this additional
acceptance. This gives the required distribution on
$B_{\xi'_e}\cap(F+e)$.

In both cases, $\xi'_e$ is fair conditional on the unchanged
coordinates $\xi_{-e}$. Hence the full endpoint vector remains
uniform, with the required conditional accepted-set distribution.
\end{proof}

\begin{corollary}[Snapshot independence]
\label{cor:ssmpi-snapshot-independent}\label{lem:ssmpi-thinning}
In the original sampling model, after every fixed prefix,
$C\sim\prod_{e\in V}\Dist_e$.
Conditional on $C=w$, the accepted set is an independent
half-sample of $\OPT(w)\cap F$. In particular, every snapshot
$C_e$ has this product distribution for every fixed arrival order.
\end{corollary}
\begin{proof}
Generate the sample and real values as described before \Cref{lem:ssmpi-snapshot}; conditional on the value pairs, the lemma applies. Averaging over the pairs gives
$C\sim\prod_{e\in V}\Dist_e$.
The conditional accepted-set distribution from the lemma depends on the
pairs and orientations only through the numerical configuration
$C$: it is an independent half-sample of $\OPT(C)\cap F$.
It therefore persists after averaging over the pairs, even when
some endpoint values are equal.
\end{proof}

\subsection{The competitive guarantee}

\begin{proof}[Proof of \Cref{thm:ssmpi-tight}]
At the final prefix, $F=V$. By
\Cref{cor:ssmpi-snapshot-independent}, $(C,\ALG)$ has the
same joint distribution as a fresh product draw $Y$ together with an
independent half-sample $T$ of $\OPT(Y)$.
By \Cref{lem:ssmpi-feasibility}, every accepted element has its
real value stored in $C$. Consequently,
\[
\{(e,R_e):e\in\ALG\}
\overset{\mathrm d}{=}
\{(e,Y_e):e\in T\}.
\]
Taking expected total values, which are nonnegative, and using greedy optimality gives the claimed expected-value identity.
The distributional identity uses a fresh
draw $Y$; the output need not be a half-sample of the optimum
of the realized values $(R_e)_{e\in V}$.

Every decision uses only the samples, the current real value,
earlier observations, and fair coins, so no distributional
knowledge is needed. Since the algorithm stores $B$, it computes
one new greedy basis per arrival: at most $n+1$ greedy scans,
hence $n(n+1)$ independence queries.
\end{proof}

\section{Polyhedral Characterization of Ordinal Secretary Algorithms}
\label{sec:lp}

The preceding sections construct algorithms by preserving suitable
distributions. We now characterize exactly the accepted-set distributions achievable by ordinal secretary algorithms on a fixed known matroid: they form a finite
polytope, and every feasible point yields an online algorithm.
Constraints on the \emph{terminal distribution}, the accepted-set
distribution after all arrivals, give exact LPs over ordinal
algorithms, which we use to study small matroids
(\Cref{sec:lp-examples,sec:lp-computations}). Analytically, the
continuous-time version reduces the design of our optimal algorithm
to constructing feasible distributions (\Cref{sec:survival,sec:implementation}).

Fix a matroid $\calM=(V,\calI)$ with $n$ elements and positive rank.
Values are fixed and nonnegative, arrivals form a uniformly random
permutation, and ties are broken by a fixed order of labels. An
algorithm is \emph{ordinal} if it uses values only through the
tie-broken order among observed elements; it may otherwise use its
entire history and internal randomness. 
The characterization allows advance knowledge of the matroid;
the family in \Cref{sec:survival} uses only its observed restriction.

\subsection{Distributions and feasible transitions}
\label{sec:lp-laws}

We describe an ordinal algorithm by two families of numbers: the
distribution of its accepted set given the elements observed so far,
and the probability mass it moves at each arrival. Since an ordinal
algorithm sees only the relative order of values, both are indexed
by ordered sets. An \emph{ordered set} $O$ is a subset of $V$
equipped with its decreasing-value order, not its arrival order. We
use the same symbol for its underlying set, and $O-e$ inherits the
restricted order. Write $\omega$ for a full value order, identified
with the ordered set $V$, and $\OPT(O)$ for the greedy basis under
the order of $O$.

Write $\ALG_m$ for the set accepted after the first $m$ arrivals.
For an ordered set $O$ of size $m$, the \emph{accepted-set
distribution} $\mu_O$ is the intended conditional distribution of
$\ALG_m$, with probabilities
\begin{equation}\label{eq:lp-mu}
\mu_O(A)=\Pr[\ALG_m=A\mid
                 \text{the first $m$ arrivals form }O].
\end{equation}
It averages over the uniform arrival order on $O$ and the
algorithm's randomness, and is the same for every full value order
extending $O$, since an ordinal algorithm cannot see unobserved
comparisons.

For $e\in O$ and $A\subseteq O-e$, the \emph{accepting mass}
$y_{O,e}(A)$ is the probability that the previous accepted set is $A$
and $e$ is accepted, conditional on the first $|O|$ arrivals forming
$O$ and the last arrival being $e$. Thus $y_{O,e}(A)$ is the
probability mass moved from $A$ to $A+e$. Using these joint
probabilities, rather than acceptance probabilities, keeps the
constraints linear; the corresponding acceptance rule from accepted
set $A$ is
\begin{equation}\label{eq:lp-rule}
q_{O,e}(A)=\frac{y_{O,e}(A)}{\mu_{O-e}(A)},
\qquad\text{with $0/0=0$.}
\end{equation}
All entries outside their stated supports are zero. Two linear
conditions constrain these numbers.

\medskip\noindent\textbf{Capacity.}
Only available mass can be moved, and an accepted extension must
remain independent. Hence, for $e\in O$ and $A\subseteq O-e$,
\begin{equation}\label{eq:lp-cap}
0\leq y_{O,e}(A)\leq\mu_{O-e}(A),\qquad
y_{O,e}(A)=0\quad\text{if }A+e\notin\calI.
\end{equation}

\medskip\noindent\textbf{Consistency.}
Conditional on the first $|O|$ arrivals forming $O$, the last element
is uniform in $O$. To finish with a particular set $A$, an element
$e\notin A$ must be rejected from $A$, whereas $e\in A$ must be
accepted from $A-e$. Therefore, for every nonempty $O$ and
$A\subseteq O$,
\begin{equation}\label{eq:lp-flow}
\mu_O(A)=\frac{1}{|O|}\Big[
\sum_{e\in O\setminus A}
\bigl(\mu_{O-e}(A)-y_{O,e}(A)\bigr)
+\sum_{e\in A}y_{O,e}(A-e)
\Big].
\end{equation}
This equation enforces irreversibility. For example, in rank one,
the family that always places all mass on the best observed element
cannot satisfy it: accepting a first element leaves no way to replace
it when a larger element arrives.

Let $\mathcal P(\calM)$ be the polytope of families $(\mu,y)$ with
$\mu_O$ supported on independent subsets of $O$ and satisfying
\begin{equation}\label{eq:polytope}
\begin{aligned}
&\mu_\emptyset(\emptyset)=1,\qquad
  \mu_O(A)\geq0,\qquad \sum_{A\subseteq O}\mu_O(A)=1
                              &&\text{for every }O,\\
&\text{capacity \eqref{eq:lp-cap}},\\
&\text{consistency \eqref{eq:lp-flow}}.
\end{aligned}
\end{equation}

\begin{theorem}[Exact characterization]
\label{thm:polytope-realization}\label{lem:lp-consistency}
The points of $\mathcal P(\calM)$ are exactly the accepted-set distributions and accepting masses of ordinal algorithms.
Every point $(\mu,y)$ is realized by the rule \eqref{eq:lp-rule},
which uses only the observed ordered set, the current arrival, and
the previously accepted set.
\end{theorem}
\begin{proof}
For necessity, fix an ordinal algorithm and a full value order.
Conditional on the preceding observed set being $O-e$, learning that
the next arrival is $e$ does not change the uniform order of those
preceding arrivals or the distribution of past internal randomness.
The preceding accepted set therefore has distribution $\mu_{O-e}$. Its
accepting masses satisfy capacity, and averaging the two possible
transitions to each $A$ over the uniform last arrival gives
consistency. Normalization and independent support are immediate.

Conversely, fix $(\mu,y)\in\mathcal P(\calM)$ and apply
\eqref{eq:lp-rule} at each arrival. Capacity makes this a feasible
probability and preserves independence. Induct on $m$ to prove that
the conditional distribution after $m$ arrivals is $\mu_O$. The claim is
immediate for $m=0$. Conditional on a nonempty observed set $O$ and
last arrival $e$, the induction hypothesis gives preceding distribution
$\mu_{O-e}$. The rule moves exactly $y_{O,e}(B)$ from each $B$
to $B+e$, including when $\mu_{O-e}(B)=0$, since then capacity
forces $y_{O,e}(B)=0$. Averaging over $e$ gives $\mu_O(A)$ by
\eqref{eq:lp-flow}. This also proves that every accepting mass is
the prescribed $y_{O,e}(B)$.
\end{proof}

Thus an ordinal algorithm gains nothing from remembering the arrival
order of earlier elements: the rule \eqref{eq:lp-rule} realizes its
accepted-set distributions using only $O$, $e$, and $A$. This justifies
indexing the variables by ordered sets rather than arrival sequences.
For example, \Cref{alg:matroidSec} induces a feasible point
whose terminal distribution includes every optimum element with
probability at least $1/4$.

\begin{remark}[Scope of the realization theorem]
The proof of \Cref{thm:polytope-realization} applies to any known
downward-closed feasibility family, with ``independent'' replaced
by ``feasible.'' The matroid-specific work lies in constructing
distributions that satisfy capacity and consistency, and in
implementing them efficiently
(\Cref{lem:survival-comparison,lem:kernel}). The prefix objective
in \Cref{cor:lp-ordinal} uses greedy optimality and is stated only
for matroids.
\end{remark}

\subsection{Optimal competitive guarantees}
\label{sec:lp-objective}

Both competitive guarantees below can be expressed as linear
constraints on the terminal distributions $\mu_\omega$, the
distributions of the final accepted set $\ALG_n$ for each full value
order $\omega$. Thus \Cref{thm:polytope-realization} turns the search for optimal ordinal algorithms into linear programs.

Let $\rho(\calM)$ be the infimum probability-competitive ratio among
ordinal algorithms for the known matroid $\calM$. Equivalently,
$1/\rho(\calM)$ is the largest fraction $\gamma$ for which one such
algorithm selects every element of $\OPT(V)$ with probability at
least $\gamma$, for every value assignment. 
The per-element guarantee therefore gives the LP
\begin{equation}\label{eq:lp}
\begin{aligned}
\text{maximize}\quad &\gamma\\
\text{subject to}\quad
 &\gamma\geq0,\qquad (\mu,y)\in\mathcal P(\calM),\\
 &\sum_{\substack{A\subseteq V\\e\in A}}\mu_\omega(A)
       \geq\gamma
                 &&\text{for every }\omega,\ e\in\OPT(\omega).
\end{aligned}
\end{equation}

\begin{corollary}[Probability competitiveness]\label{thm:lp-exact}
The value of \eqref{eq:lp} is $1/\rho(\calM)$, and an optimum is
attained. A feasible point of value $\gamma>0$ yields a
$(1/\gamma)$-probability-competitive algorithm by \eqref{eq:lp-rule}.
\end{corollary}
\begin{proof}
Both directions follow from \Cref{thm:polytope-realization} and the
terminal inclusion constraints. The feasible region is nonempty
and compact: all entries and $\gamma$ lie in $[0,1]$. Its optimum
is positive: accepting the element at a uniform arrival position,
if it is a nonloop, gives $1/n$.
\end{proof}

 The expected-value objective leads to a similar LP. Let
 $\rho_{\mathrm{ord}}(\calM)$ be the infimum expected-value
 competitive ratio among ordinal algorithms, the \emph{ordinal
 competitive ratio}. Following~\cite{SotoTV-SODA18}, consider each
 prefix $P$ of
the decreasing-value order $\omega$, consisting of its largest
elements. The prefix constraints below require the algorithm to
select, in expectation, a $\gamma$ fraction of the maximum
feasible number of elements from every such prefix.

\begin{corollary}[Ordinal competitiveness]\label{cor:lp-ordinal}
The largest expected-value fraction $\gamma=1/\rho_{\mathrm{ord}}(\calM)$
 achievable by an ordinal algorithm is
 the value of \eqref{eq:lp} with its terminal constraints
replaced by
\begin{equation}\label{eq:lp-prefix}
\sum_{A\subseteq V}|A\cap P|\,\mu_\omega(A)
\geq\gamma\rank(P)
\qquad\text{for every full order $\omega$ and prefix $P$ of $\omega$}.
\end{equation}
An optimum is attained and is realized by \eqref{eq:lp-rule}.
\end{corollary}
\begin{proof}
For a fixed order, nonnegative values are nonnegative combinations
of prefix indicators, and $|\OPT(\omega)\cap P|=\rank(P)$ by greedy.
Thus the prefix inequalities imply the expected-value guarantee.
Conversely, approximate a prefix indicator by strictly ordered values
with the same order; the algorithm's distribution stays fixed, so the guarantee
implies the prefix inequality in the limit. This is the prefix
characterization of Soto, Turkieltaub, and
Verdugo~\cite{SotoTV-SODA18}. Apply
\Cref{thm:polytope-realization}; compactness is as above.
\end{proof}

The probability constraints imply the prefix constraints by summing over $\OPT(\omega)\cap P$, so $\rho_{\mathrm{ord}}(\calM)\leq\rho(\calM)$.  \Cref{sec:lp-examples} applies these
programs to special classes. For our optimal construction, we now pass to continuous time, where the distributions also vary with time.

\subsection{Continuous-time acceptance distributions}
\label{sec:continuous-lp}

For the construction in \Cref{sec:survival,sec:implementation},
it is convenient to express the characterization of
\Cref{thm:polytope-realization} in continuous time.
In the \emph{continuous-time model}, each element $e$ arrives at
an independent uniform time $T_e\in[0,1]$. The algorithm observes
this time when $e$ arrives and immediately accepts or rejects $e$,
using only observations so far and independent private randomness.
When $n$ is known, any algorithm in this model can be simulated
in the ordinary secretary model: independently sample and sort
$n$ uniform times, and reveal the next time to the algorithm
at each arrival. Conversely, an ordinary secretary algorithm
can ignore the times. The same capacity and consistency principles
therefore apply, now averaging over the last arrival's time
as well as its identity.

For $0<t<1$, the \emph{accepted-set distribution} $\Law_{O,t}$ is the
intended distribution of the accepted set, conditional on precisely
the ordered set $O$ having arrived by $t$. The \emph{accepting mass}
$z_{O,e,t}(A)$ is the joint probability that the previous accepted
set is $A$ and $e$ is accepted, conditional on $e$ arriving at time
$t$ with preceding observed set $O-e$.\footnote{Accepting masses
are equivalent if they differ only on a probability-zero set of times.}
These are the continuous-time counterparts of $\mu_O$ and
$y_{O,e}$; as before, they are shared by all full value orders
extending $O$.
Each $\Law_{O,t}$ is a probability distribution on independent subsets of $O$,
with $\Law_{\emptyset,t}(\emptyset)=1$.
We define $\Law_{O,1}$ by continuity. In particular,
$\Law_{V,1}$ is the terminal distribution, since all elements
have arrived by time $1$.

\medskip\noindent\textbf{Capacity.}
For $e\in O$ and $A\subseteq O-e$, the measurable accepting
masses must satisfy
\begin{equation}\label{eq:clock-capacity}
0\leq z_{O,e,t}(A)\leq\Law_{O-e,t}(A),\qquad
z_{O,e,t}(A)=0\quad\text{if }A+e\notin\calI.
\end{equation}

\medskip\noindent\textbf{Consistency.}
Averaging over the last arrival gives, for $m=|O|>0$,
$A\subseteq O$, and $0<t\leq1$,
\begin{equation}\label{eq:clock-balance}
\begin{aligned}
t^m\Law_{O,t}(A)
=\int_0^t u^{m-1}\Bigl[
\sum_{e\in O\setminus A}
\bigl(\Law_{O-e,u}(A)-z_{O,e,u}(A)\bigr) 
+\sum_{e\in A}z_{O,e,u}(A-e)\Bigr]\,du.
\end{aligned}
\end{equation}

\begin{theorem}[Continuous-time realization]
\label{thm:continuous-realization}
The accepted-set distributions $\Law_{O,t}$ and accepting masses $z_{O,e,t}(A)$
are realizable by an ordinal algorithm in the continuous-time model
if and only if
they satisfy capacity \eqref{eq:clock-capacity} and consistency
\eqref{eq:clock-balance}.
Call such a family \emph{feasible}; every feasible family is realized by accepting the current arrival
$e$ at time $t$, from accepted set $A$, with probability
\begin{equation}\label{eq:continuous-rule}
q_{O,e,t}(A)=\frac{z_{O,e,t}(A)}{\Law_{O-e,t}(A)},
\qquad\text{with $0/0=0$.}
\end{equation}
\end{theorem}
\begin{proof}
Necessity of capacity is as in the discrete formulation.

For consistency, fix a nonempty $O$, let $m=|O|$, and condition
on precisely $O$ having arrived by $t$. Under this conditioning,
its arrival times are independent uniforms on $[0,t]$.
For a specified element $e\in O$ to be last at time $u$,
its own arrival time contributes density $1/t$, while all other
$m-1$ arrival times must be at most $u$, an event of probability
$(u/t)^{m-1}$. Thus the joint density of the last identity $e$
and its time $u$ is $(1/t)(u/t)^{m-1}=u^{m-1}/t^m$.
Conditional on this last arrival, the preceding arrival times are
independent uniforms on $[0,u]$. The observations before $u$
therefore have the same distribution as when precisely $O-e$
has arrived by $u$. Since decisions cannot depend on future
arrivals, the accepted set immediately before $e$ arrives
has distribution $\Law_{O-e,u}$.

Fix $A\subseteq O$. If $e\notin A$, ending in $A$ requires
the preceding accepted set to be $A$ and $e$ to be rejected.
This has probability $\Law_{O-e,u}(A)-z_{O,e,u}(A)$.
If $e\in A$, the preceding set must be $A-e$ and $e$ must
be accepted, with probability $z_{O,e,u}(A-e)$.
Summing over the last element $e$, integrating over its time
$u\in[0,t]$, and multiplying by $t^m$ gives
\eqref{eq:clock-balance}.

Conversely, use \eqref{eq:continuous-rule}, which is feasible
by capacity. Induct on $|O|$, starting from the empty set.
Once the preceding distributions are realized, the joint masses are
$\Law_{O-e,u}(A)q_{O,e,u}(A)=z_{O,e,u}(A)$.
The same last-arrival calculation gives \eqref{eq:clock-balance},
so the resulting distribution is $\Law_{O,t}$.
Continuity gives the terminal distribution at $t=1$.
\end{proof}

If the distributions and masses depend only on $n$, the current time,
the observed order, and the arrived matroid restriction, so does
the realizing rule. The next two sections construct such a family
and implement it efficiently.

\section{A Feasible Family with Guarantee \texorpdfstring{$1/\ee$}{1/e}}
\label{sec:survival}
 
We now construct a feasible family for \Cref{sec:continuous-lp} in
two steps. First, an offline experiment on an ordered set $O$ at horizon
$t$ defines the \emph{survivor distribution} $\Law_{O,t}$ over independent sets, our target accepted-set distribution. The experiment is not an online
selection rule: its auxiliary arrival times are independent of
actual arrivals, and its survival decisions are not online
acceptances.
Second, we prescribe accepting masses $z_{O,e,t}(A)$ and prove capacity
and consistency. \Cref{thm:continuous-realization} then gives an 
online algorithm whose accepted set, conditional on precisely $O$
having arrived by $t$, has distribution $\Law_{O,t}$. 
The resulting guarantee already resolves the strong
matroid secretary conjecture. \Cref{sec:implementation} gives
an efficient implementation without evaluating probabilities
of entire survivor sets.
 
\subsection{The survivor distributions}
\label{sec:survival-target}\label{sec:survival-family}
 
\paragraph{Motivation from rank one.}
Fix $p\in[1/\ee,1)$ throughout. The classical single-item secretary rule in the
continuous-time model rejects
arrivals before $p$, then selects the first subsequent element
larger than all earlier arrivals. Suppose $e$ arrives at $t>p$,
is larger than every predecessor, and has at least one predecessor.
It is selected exactly when the largest predecessor arrived before
$p$. Conditional on the predecessor set and $t$, the largest
predecessor's arrival time is uniform on $[0,t]$. Hence $e$
is selected with probability $p/t$.\footnote{If $e$ is the first
arrival overall and arrives after $p$, the classical rule
will accept it; our construction instead uses $p/t$.}
This averages over earlier arrival times; it is not conditional
on having made no selection.

For a matroid, let $O$ contain the elements observed up to and
including the current arrival $e$, ordered by value. Call $e$
\emph{eligible} if $e\in\OPT(O)$, or equivalently if the larger
observed elements do not span it. This generalizes being larger
than all predecessors in rank one. We target acceptance
probability $p/t$ for each eligible arrival at $t>p$, conditional
on $O$, $e$, and $t$ but averaged over earlier arrival times and decisions.
By greedy monotonicity (\Cref{lem:greedy}(i)), every global
optimum element is eligible whenever it arrives, giving the
target selection probability
\begin{equation}\label{eq:rankOneMotiv}
\int_p^1\frac pt\,dt ~=~ p\ln(1/p).
\end{equation}
This is maximized at $p=1/\ee$. The challenge is to achieve these
probabilities while keeping the accepted set independent.
We first construct a survivor set, always independent in $\calM$,
with these inclusion probabilities, then realize its distribution
online as an accepted-set distribution.

\begin{definition}[Survivor distribution $\Law_{O,t}$]
\label{def:survivor-distribution}
For an ordered set $O$ and \emph{horizon} $t\in[p,1]$, perform the
following offline \emph{survivor experiment}:
\begin{enumerate}[leftmargin=*,itemsep=3pt,parsep=0pt]
\item Draw independent \emph{auxiliary times} $T'_e\sim\Unif[0,t]$
for $e\in O$. Set $E=\{e\in O:T'_e\leq p\}$. Call these elements
\emph{early} and the others \emph{late}.
\item For each $e\in O$, let its \emph{capped spanning time} $\sigma_e$ be the first time
$u\in[p,t]$ when the larger elements already present in the
survivor experiment span $e$. These are the elements
$\{f\in O:f\succ e,\ T'_f\leq u\}$, where $f\succ e$ means
that $f$ is larger in the tie-broken value order.
Set $\sigma_e=t$ if they never span $e$ by $t$.
  \item Conditional on all auxiliary times, each \emph{early element}
  $e$ survives independently with probability $\ln(\sigma_e/p)$, its
  \emph{conditional survival probability}. No late element survives.
\end{enumerate}
The surviving elements form the \emph{survivor set}, with distribution $\Law_{O,t}$ and output probabilities $\Law_{O,t}(A)$. For $0<t\leq p$
or $O=\emptyset$, $\Law_{O,t}$ is the point mass at $\emptyset$.
\end{definition}

We choose the logarithm in step~3 to reproduce the rank-one
integral \eqref{eq:rankOneMotiv}, stopped when larger elements
span $e$: an eligible arrival at time $u>p$ should be accepted with
probability $p/u$, and $e$ remains eligible until $\sigma_e$.
At horizon $1$, the larger elements' auxiliary times fix $\sigma_e$
but not $T'_e$, so $e$ is still early with probability $p$. Its
survival probability is therefore
$p\ln(\sigma_e/p)=\int_p^{\sigma_e}(p/u)\,du$:
early sampling supplies the factor $p$, and step~3 the logarithm.

In step~2, all larger elements present contribute to spanning, whether
or not they survive. Since $p\leq\sigma_e\leq1$ and
$p\geq1/\ee$, the conditional survival probabilities lie in $[0,1]$.
An early element outside $\OPT(E)$ has $\sigma_e=p$ and cannot
survive. Thus the survivor set is a subset of $\OPT(E)$ and is
independent.

\smallskip
\noindent\textbf{The accepting masses $z_{O,e,t}$.}
Recall from \Cref{sec:continuous-lp} that $z_{O,e,t}(A)$ specifies
the joint probability of the preceding accepted set being $A$
and accepting $e$, conditional on the observed ordered set $O$
and the arrival of $e$ at time $t$.
For a nonloop $e$, the contraction $\calM/e$ has independent sets
$A\subseteq V-e$ with $A+e\in\calI$. A set $S$ spans $f\ne e$
in $\calM/e$ exactly when $S+e$ spans $f$ in $\calM$.
For any ordered set $T\subseteq V-e$, let $\Law_{T,t,e}$ denote
the survivor distribution on $T$ in $\calM/e$, with the restricted
value order. To determine whether $f$ is spanned, this experiment
adds $e$ to the larger elements present, regardless of how $e$ compares with $f$.

Every output $A$ can be extended by $e$. We therefore use the
contracted survivor distribution to allocate the target
probability $p/t$ among feasible predecessor sets:
\begin{equation}\label{eq:survival-z}
z_{O,e,t}(A)=
\begin{cases}
\displaystyle\frac pt\Law_{O-e,t,e}(A),
    &t>p\text{ and }e\in\OPT(O),\\[3pt]
0,  &\text{otherwise}.
\end{cases}
\end{equation}
For an eligible arrival at $t>p$, summing over predecessor sets
gives $\sum_A z_{O,e,t}(A)=p/t$, matching the rank-one target.
This choice will also give consistency in \Cref{sec:survival-consistency}: with auxiliary times fixed,
an early $e\in\OPT(O)$ has conditional survival probability
$\ln(t/p)$, with derivative $1/t$. Averaging over auxiliary times
gives the $z$-terms in \eqref{eq:clock-balance}.
The realizing rule divides $z_{O,e,t}(A)$ by
$\Law_{O-e,t}(A)$ to obtain the acceptance probability from
set $A$, as in \eqref{eq:continuous-rule}; capacity bounds it by $1$.

\begin{theorem}[Feasible survivor distributions]
\label{thm:survival-lp}\label{thm:tight-secretary}
With $z$ given by \eqref{eq:survival-z}, the survivor distributions
satisfy capacity \eqref{eq:clock-capacity} and consistency
\eqref{eq:clock-balance}. Hence \Cref{alg:survival-rule} is ordinal
and accepts the elements of $\OPT(V)$ independently, each with
probability $p\ln(1/p)$, which equals $1/\ee$ for $p=1/\ee$.
\end{theorem}

\begin{algorithm}[htbp]
\caption{Survival secretary rule}
\label{alg:survival-rule}
\begin{itemize}[leftmargin=*,itemsep=3pt,parsep=0pt,topsep=3pt]
\item Initialize $\ALG=\emptyset$, and generate arrival times in the continuous-time model (\Cref{sec:continuous-lp}).
\item When $e$ arrives at time $t$, let $O$ be the arrived set,
including $e$. If $t>p$ and $e\in\OPT(O)$, accept $e$ with probability
\[
\frac pt\cdot\frac{\Law_{O-e,t,e}(\ALG)}{\Law_{O-e,t}(\ALG)}
\qquad(\text{with }0/0=0),
\]
and reject it otherwise.
On acceptance, set $\ALG\gets\ALG+e$.
\item Return $\ALG$.
\end{itemize}
\end{algorithm}
 
\noindent\textbf{Why this resolves the strong conjecture.}
For $t\in[p,1]$ and every $e\in\OPT(O)$, the larger elements
never span $e$, so $\sigma_e=t$ regardless of the auxiliary
times. Conditional on being early, $e$ survives with
the fixed probability $\ln(t/p)$.
The early indicators are independent, each with probability
$p/t$, and the survival coins are independent. Thus the
optimum elements survive independently, each with probability
$(p/t)\ln(t/p)$.

At $t=1$ and $p=1/\ee$, 
an optimum element survives surely 
when early, giving independent inclusion with probability $1/\ee$.
Once capacity and consistency are proved in
\Cref{sec:survival-comparison-proof,sec:survival-consistency},
\Cref{thm:continuous-realization} transfers this guarantee to
an ordinal online algorithm. Observe that the algorithm may also accept
elements outside $\OPT(V)$.

\subsection{Continuous-time capacity}
\label{sec:survival-comparison-proof}
 
For an ordered set $O$ and a nonloop $e\notin O$, we need to show that
\begin{align} \label{eq:targetContCap}
\Law_{O,t}(A)\geq\frac pt\Law_{O,t,e}(A)
\qquad(A\subseteq O,\ t\in[p,1]).
\end{align}
Applied to the predecessor set, this says that the mass prescribed
for accepting $e$ is available in each state $A$.

To prove this, we would like to induct on $|O|$, removing its largest element $f$ after
conditioning on its auxiliary arrival time. We handle $f$'s
survival separately and apply induction to $O-f$. However, simply
deleting $f$ would change the remaining elements' spanning times:
once present, $f$ contributes to spanning smaller elements whether
or not it survives. We therefore strengthen the induction hypothesis to allow
\emph{background elements} with fixed auxiliary arrival times.

\medskip\noindent\textbf{Background elements.}
Choose some elements of $V\setminus O$ and assign them fixed
auxiliary arrival times in $[0,t]$. For $u\in[p,t]$, let $H_u$
be those whose auxiliary times are at most $u$, and write
$H=(H_u)_{u\in[p,t]}$ for this \emph{background schedule}.
Background elements contribute to spanning once present,
regardless of their values, but are not candidates for survival.

For each $f\in O$, its spanning time is now the first
$u\in[p,t]$ at which
\[
H_u\cup\{f'\in O:f'\succ f,\ T'_{f'}\leq u\}
\]
spans $f$, or $t$ if this never occurs. All other parts of the
survivor experiment are unchanged. Denote its distribution by
$\Law_{O,t,H}$: the background is fixed, while the auxiliary
times of elements in $O$ remain independent uniforms on $[0,t]$,
and their survival  coins remain random.
Empty background is omitted from the notation. A constant
background $H_u=\{e\}$ is denoted by $e$; for a nonloop $e$,
this gives the contracted experiment in $\calM/e$.

Let $\sigma_{a,H}$ be the first $u\in[p,t]$ at which $H_u$
spans $a$, or $t$ if this never occurs. For $a\notin O$,
write $H^{a,u}$ for adding $a$ to the background with auxiliary
time $u\in[p,t]$, and abbreviate $H+a=H^{a,p}$. If $a$ is already in the background, keep the
earlier of its old and new auxiliary times.

\begin{lemma}[Capacity comparison]
\label{lem:survival-comparison}%
For every background schedule $H$ disjoint from $O$, every
$e\notin O$ (possibly already in the background, in which case
$H+e$ lowers its auxiliary time to $p$), and every $A\subseteq O$,
\begin{equation}\label{eq:general-survival-comparison}
\Law_{O,t,H}(A)\geq
\frac p{\sigma_{e,H}}\Law_{O,t,H+e}(A).
\end{equation}
\end{lemma}

With empty background and a nonloop $e$, $\sigma_{e,H}=t$,
giving the desired capacity bound with factor $p/t$ in \eqref{eq:targetContCap}.
 
\begin{proof}
We induct on $|O|$, assuming the comparison for every smaller
candidate set, every fixed background, and every element
outside that set. The empty case is immediate: both experiments
output $\emptyset$ and $p/\sigma_{e,H}\leq1$.
 
\medskip\noindent\textbf{Condition on the largest element.}
Let $f$ be the largest element of $O$ and fix $A\subseteq O-f$.
To shorten the formulas, write $P_J(A)=\Law_{O-f,t,J}(A)$ for the smaller experiment
with background $J$. The induction hypothesis applies to these
distributions on $O-f$. The induction step must show
\begin{align}   \label{eq:comparisonCases}
\Law_{O,t,H}(B)\geq
\frac p{\sigma_{e,H}}\Law_{O,t,H+e}(B)
\qquad\text{for }B=A\text{ and }B=A+f.
\end{align}
We first express these probabilities by conditioning on $f$.
Let
\[
s_0=\sigma_{f,H},\qquad s_1=\sigma_{f,H+e}\leq s_0.
\]
Since $f$ is largest in $O$, its spanning time depends only on
the background. If it is early, it survives with probability
$\ln(s_0/p)$, while the remaining experiment has background
$H+f$ whether or not it survives: every early $f$ is present
throughout $[p,t]$. If its auxiliary arrival time is $u>p$,
it cannot survive, and the remaining background is $H^{f,u}$.
Hence
\begin{align}
\Law_{O,t,H}(A+f)
 &=\frac pt\,\ln(s_0/p)\,P_{H+f}(A),
 \label{eq:survive-recursion}\\
\Law_{O,t,H}(A)
 &=\frac pt\bigl(1-\ln(s_0/p)\bigr)P_{H+f}(A)
   +\frac1t\int_p^t P_{H^{f,u}}(A)\,du.
   \label{eq:reject-recursion}
\end{align}
\noindent Here $p/t$ is the probability that $f$ is early, and $du/t$ that its auxiliary time lies in $du$. The same recurrences hold under $H+e$, with $s_1$ in place
of $s_0$.
To prove \eqref{eq:comparisonCases}, we compare each expression with $p/\sigma_{e,H}$ 
times its counterpart under $H+e$. Induction compares the
$P_J(A)$ terms, but its factors depend on spanning times in the
modified backgrounds, which we compute next.
 
\medskip\noindent\textbf{How the spanning times change.}
For any set $S\subseteq V$, closure exchange gives
\[
f\in\cl(S+e)\setminus\cl(S)
\quad\Longleftrightarrow\quad
e\in\cl(S+f)\setminus\cl(S).
\]
Both sides say that $S+e$, $S+f$, and $S+e+f$ have rank
$\rank(S)+1$. Applied to $S=H_u$ for $u<t$, this says that
adding $e$ newly spans $f$ on exactly the same interval as
adding $f$ newly spans $e$. The former interval is $[s_1,s_0)$.

 If $s_0=s_1$, this interval is empty. Adding $f$ to the background
with any auxiliary time therefore leaves the capped spanning
time of $e$ unchanged. 
Apply induction on $O-f$, adding $e$ to the backgrounds
$H+f$ and $H^{f,u}$. In both cases the comparison factor is
$p/\sigma_{e,H}$. Since $s_0=s_1$, the coefficients in the
two recurrences agree and are nonnegative, proving \eqref{eq:comparisonCases}.
 
From now on, suppose $s_1<s_0$. Equality of the two nonempty intervals gives
\begin{equation}\label{eq:span-pair}
\sigma_{e,H}=s_0,\qquad \sigma_{e,H+f}=s_1.
\end{equation}
In $H^{f,u}$, the fixed auxiliary time of $f$ is $u$, so it can
help span $e$ only from $\max\{u,s_1\}$ onward; without that
help the spanning time is $s_0$. In $H^{f,u}+e$, the background
spans $f$ at $s_1$ or when $f$ itself becomes present at $u$,
whichever comes first. Consequently,
\begin{equation}\label{eq:span-release}
\sigma_{e,H^{f,u}}=\min\{s_0,\max\{s_1,u\}\},\qquad
\sigma_{f,H^{f,u}+e}=\min\{s_1,u\}.
\end{equation}
Thus the required factor for the full experiment is $p/s_0$,
while the identities above determine the factors available
from induction on $O-f$. We first prove the comparison for
output $A+f$, and then for output $A$.
 
\medskip\noindent\textbf{Outputs $A+f$ containing $f$.}
Induction under $H+f$ and \eqref{eq:span-pair} give
$P_{H+f}(A)\geq(p/s_1)P_{H+e+f}(A)$. Therefore
\[
\Law_{O,t,H}(A+f)
~\geq~\frac{p^2}{ts_1}\ln(s_0/p)P_{H+e+f}(A)
~\geq~\frac{p^2}{ts_0}\ln(s_1/p)P_{H+e+f}(A)
 ~=~\frac p{s_0}\Law_{O,t,H+e}(A+f).
\]
The second inequality uses $s_1\leq s_0$ and nonnegativity of
the logarithms. This proves the comparison in \eqref{eq:comparisonCases} for output $A+f$.
 
\medskip\noindent\textbf{Outputs $A$ without $f$: the early deficit.}
It remains to prove the comparison for output $A$.
Subtracting $p/s_0$ times \eqref{eq:reject-recursion} under $H+e$
from the recurrence under $H$, and multiplying by $t$, gives
\begin{equation}\label{eq:capacity-split}
t\Bigl[\Law_{O,t,H}(A)-\frac p{s_0}\Law_{O,t,H+e}(A)\Bigr]
~=~\mathcal E+\int_p^t\Bigl[P_{H^{f,u}}(A)
 -\frac p{s_0}P_{H^{f,u}+e}(A)\Bigr]du,
\end{equation}
where
$\mathcal E=p\bigl(1-\ln(s_0/p)\bigr)P_{H+f}(A)
-(p^2/s_0)\bigl(1-\ln(s_1/p)\bigr)P_{H+e+f}(A)$
is the contribution from early auxiliary arrivals of $f$;
the integral is the contribution from late arrivals.
We must show that their sum is nonnegative.
The early term may be negative: adding $e$ makes an early $f$
more likely to be absent from the output. We first bound this
possible loss, then show that the late contribution compensates.
 
Since $s_0\leq1$ and $p\geq1/\ee$, we have $1-\ln(s_0/p)\geq0$, so
the induction bound $P_{H+f}(A)\geq(p/s_1)P_{H+e+f}(A)$ gives
\begin{equation}\label{eq:early-deficit}
\begin{aligned}
\mathcal E
\,\geq\,\frac{p^2P_{H+e+f}(A)}{s_0s_1}
 \Bigl[s_0\bigl(1-\ln(s_0/p)\bigr)
       -s_1\bigl(1-\ln(s_1/p)\bigr)\Bigr]
       \,=\, -\frac{p^2P_{H+e+f}(A)}{s_0s_1}
       \int_{s_1}^{s_0}\ln(v/p)\,dv,
\end{aligned}
\end{equation}
using $\frac{d}{dv}[v(1-\ln(v/p))]=-\ln(v/p)$. It therefore suffices to show that the integral in
\eqref{eq:capacity-split} satisfies %
\[
\int_p^t\Bigl[P_{H^{f,u}}(A)
 -\frac p{s_0}P_{H^{f,u}+e}(A)\Bigr]du 
 ~\geq~ 
\frac{p^2P_{H+e+f}(A)}{s_0s_1}
\int_{s_1}^{s_0}\ln(v/p)\,dv.
\]
 
\medskip\noindent\textbf{The late surplus pays for the deficit.}
The required bound involves $P_{H+e+f}(A)$, so we compare
the late probabilities to this same quantity.
For $p\leq u\leq s_0$, apply the induction hypothesis on $O-f$
twice: first add $e$ to the background with auxiliary time $p$,
then change $f$'s fixed auxiliary time from $u$ to $p$, using
the case where the added element already lies in the background.
By \eqref{eq:span-release},
  \[
  P_{H^{f,u}}(A)\geq
   \frac p{\max\{s_1,u\}}P_{H^{f,u}+e}(A) \qquad \text{and} \qquad 
  P_{H^{f,u}+e}(A)\geq
   \frac p{\min\{s_1,u\}}P_{H+e+f}(A).
  \]
Since $\max\{s_1,u\}\leq s_0$, the first bound leaves a
nonnegative coefficient, so we may substitute the second:
\begin{equation}\label{eq:late-surplus}
\begin{aligned}
P_{H^{f,u}}(A)-\frac p{s_0}P_{H^{f,u}+e}(A)
&~\geq~\left(\frac p{\max\{s_1,u\}}-\frac p{s_0}\right)
       \frac p{\min\{s_1,u\}}P_{H+e+f}(A)\\
&=~\frac{p^2P_{H+e+f}(A)}{s_0s_1}
       \frac{s_0-\max\{s_1,u\}}u.
\end{aligned}
\end{equation}
Here $\max\{s_1,u\}\min\{s_1,u\}=s_1u$. For $u\geq s_0$,
\eqref{eq:span-release} instead gives $\sigma_{e,H^{f,u}}=s_0$,
so the induction hypothesis alone makes the same difference nonnegative.
It remains to integrate this lower bound. Reversing the order
of integration gives exactly the scalar integral required above:
\begin{equation}\label{eq:scalar-slack}
\int_p^{s_0}\frac{s_0-\max\{s_1,u\}}u\,du
~=~\int_p^{s_0}\int_{\max\{s_1,u\}}^{s_0}\frac1u\,dv\,du
~=~\int_{s_1}^{s_0}\int_p^v\frac1u\,du\,dv
~=~\int_{s_1}^{s_0}\ln(v/p)\,dv.
\end{equation}

Both double integrals range over the region $\{(u,v):p\leq u\leq v,\ s_1\leq v\leq s_0\}$.
Integrating the lower bound in \eqref{eq:late-surplus} over
$u\in[p,s_0]$ exactly cancels the negative lower bound in
\eqref{eq:early-deficit}. The remaining late contribution,
from $u\geq s_0$, is nonnegative. Hence the right side of
\eqref{eq:capacity-split} is nonnegative, proving the comparison in \eqref{eq:comparisonCases}
for outputs without $f$ and completing the induction.
\end{proof}

For an eligible arrival $e\in\OPT(O)$ at time $t>p$, apply
\Cref{lem:survival-comparison} to $O-e$ with empty background.
Since $e$ is a nonloop, $\sigma_{e,\emptyset}=t$, giving
$z_{O,e,t}(A)=\frac pt\Law_{O-e,t,e}(A)
\leq\Law_{O-e,t}(A).$
All other accepting masses are zero, and the contracted experiment
outputs only sets extendible by $e$. Thus
\eqref{eq:clock-capacity} holds.

\begin{remark}[A constructive proof]
The proof of \Cref{lem:survival-comparison} is constructive:
\Cref{sec:kernel-construction} turns each use of the induction
hypothesis into a recursive call of the efficient implementation,
and each lowered coefficient into a random rejection.
\end{remark}
 
\subsection{Continuous-time consistency}
\label{sec:survival-consistency}
 
We verify consistency by increasing $t$ while keeping $O$ and
$p$ fixed. Multiplying $\Law_{O,t}(A)$ by $t^{|O|}$ removes
the uniform-density factor, leaving an integral over $[0,t]^O$.
For $t>p$, increasing $t$ has two effects. The expanding
integration domain contributes configurations with an auxiliary
arrival at time $t$; that element is late and cannot survive.
For fixed auxiliary times inside the domain, each early
$e\in\OPT(O)$ has conditional survival probability $\ln(t/p)$, which
increases at rate $1/t$. We compute both contributions and
show that their sum gives the differential balance identity,
then integrate.
 
\begin{lemma}[Consistency]\label{lem:survival-balance}
The survivor distributions and accepting masses satisfy
\eqref{eq:clock-balance}.
\end{lemma}
\begin{proof}
Fix $O$, let $m=|O|>0$, and fix $A\subseteq O$.
First take $p<t<1$. For a vector of auxiliary arrival times
$x=(x_f)_{f\in O}\in[0,t]^O$, let $\psi_A(x,t)$ be the
probability of survivor set $A$, averaging only over the
survival  coins. The vector has density $t^{-m}$, so
\[
t^m\Law_{O,t}(A)=\int_{[0,t]^O}\psi_A(x,t)\,dx.
\]

 \noindent\textbf{Expanding the integration domain.}
As $t$ increases, new auxiliary-time vectors enter $[0,t]^O$.
The contribution with $x_e=t$ places $e$ at the horizon.
This late element cannot survive and changes no other conditional survival
probability: any spanning time it reduces to $t$ was already
capped at $t$. The remaining experiment is therefore the one
on $O-e$. Integrating over the other $m-1$ coordinates gives
zero if $e\in A$, and $t^{m-1}\Law_{O-e,t}(A)$ otherwise.
Summing over $e\in O$ gives
$t^{m-1}\sum_{e\in O\setminus A}\Law_{O-e,t}(A)$.

\medskip\noindent\textbf{Changing the survival probabilities.}
Fix the auxiliary times $x$, all strictly below $t$.
An early element outside $\OPT(O)$ is already spanned before
$t$, so its conditional survival probability is unchanged.
For an early $e\in\OPT(O)$, however, $\sigma_e=t$, so that
probability is $\ln(t/p)$ and increases at rate $1/t$.
Late elements never survive.

When $e\in\OPT(O)$ is early, the other survivors have the same
conditional distribution as the experiment in $\calM/e$
with the same auxiliary times. Indeed, for $f$ smaller than $e$, the larger elements determining its
spanning time already include $e$ from $p$. For $f$ larger
than $e$, adding $e$ cannot
make $f$ spanned sooner: if the larger elements $S$ present do not
span $f$ but $S+e$ does, closure exchange implies that $S+f$
spans $e$. All elements of $S+f$ are larger than $e$, contrary to
$e\in\OPT(O)$. Thus the other spanning times and conditional survival
probabilities agree with the contracted experiment, irrespective
of the precise value of $x_e\in[0,p]$.
 
Conditional on $x$, $\psi_A(x,t)$ is a product of Bernoulli
probabilities. The factor for early $e\in\OPT(O)$ is $\ln(t/p)$
if $e\in A$ and $1-\ln(t/p)$ otherwise. Differentiating it gives
$+1/t$ or $-1/t$, multiplied by the probability that the other
survivors are $A-e$ or $A$, respectively.
 
For $e\in A\cap\OPT(O)$, integrating this contribution gives
\[
p\cdot\frac1t\cdot t^{m-1}\Law_{O-e,t,e}(A-e)
~=~ t^{m-1}z_{O,e,t}(A-e).
\]
The three factors come from integrating the early coordinate
$x_e\in[0,p]$, differentiating its conditional survival probability, and
integrating the other coordinates. For $e\in\OPT(O)\setminus A$,
the contribution is instead $-t^{m-1}z_{O,e,t}(A)$.
Elements outside $\OPT(O)$ contribute zero to this interior
term and have $z_{O,e,t}=0$.
For $e\notin A$, the boundary contribution and the negative contribution above combine to give
$t^{m-1}(\Law_{O-e,t}(A)-z_{O,e,t}(A))$, the rejected-arrival
term. For $e\in A$, the positive contribution is the
accepted-arrival term. Together,
\[
\begin{aligned}
\frac{d}{dt}\bigl(t^m\Law_{O,t}(A)\bigr)
~=~ t^{m-1}\Bigl[&\sum_{e\in O\setminus A}
\bigl(\Law_{O-e,t}(A)-z_{O,e,t}(A)\bigr)  + \sum_{e\in A}z_{O,e,t}(A-e)\Bigr],
\end{aligned}
\]
which is the differential form of \eqref{eq:clock-balance}.
For $t<p$, the survivor set is empty and all accepting masses
vanish, so both sides equal $mt^{m-1}\ind\{A=\emptyset\}$.
 
\medskip\noindent\textbf{Integrating the balance equation.}
For auxiliary times in $[0,s]^O$, increasing the horizon from
$s$ to $t$ changes each conditional survival probability by at most
$(t-s)/p$. A product of Bernoulli probabilities changes by at
most the sum of the parameter changes. Comparing the integral over $[0,s]^O$ with the added domain
gives, for $p\leq s\leq t\leq1$,
\[
\bigl|t^m\Law_{O,t}(A)-s^m\Law_{O,s}(A)\bigr|
~\leq~ \frac mp(t-s)s^m+(t^m-s^m).
\]
These bounds justify the differentiation above and show that
increments of $t^m\Law_{O,t}(A)$ on $[p,1]$ equal integrals
of its derivative.
On $[0,p]$, it equals $t^m\ind\{A=\emptyset\}$, which is zero
at $0$ and agrees with the survivor experiment at $p$.
Integrating the displayed differential identity from $0$ to $t$
therefore gives \eqref{eq:clock-balance} for every $0<t\leq1$.
\end{proof}

 \begin{proof}[Proof of \Cref{thm:survival-lp}]
The experiments give measurable distributions on independent sets.
\Cref{lem:survival-comparison,lem:survival-balance} verify
capacity and consistency; \Cref{thm:continuous-realization}
realizes them via \eqref{eq:continuous-rule}, which is exactly \Cref{alg:survival-rule}.
The rule uses only the observed value order, the current time, and
the arrived matroid restriction; $n$ suffices to simulate the arrival times.
The terminal guarantee was established after the theorem.
\end{proof}

 \section{Efficient Survival Simulation}
\label{sec:implementation}

\paragraph{Toward an efficient implementation.}
Fix $p=1/\ee$. 
We want to accept an eligible element $e$ arriving at time $t>p$
with probability 
\[\frac{p}{t}\cdot \frac{\Law_{O,t,e}(A)}{\Law_{O,t}(A)},
\]
where $O$ and $A$ are the sets of observed and accepted elements
before $e$'s arrival. We do not know how to compute this probability
efficiently. Instead, we sample a binary acceptance decision with
this probability without computing it.

The inductive proof of \Cref{lem:survival-comparison} is almost a
recursive algorithm for sampling such a decision. Its inductive step
\begin{enumerate}[leftmargin=*,itemsep=1pt,parsep=0pt,topsep=3pt]
\item conditions on the auxiliary time $T'_f$ (denoted by $u$ in the
proof) of the largest element $f\in O$, expressing both distributions
as mixtures of survivor-set probabilities on $O-f$ with modified
backgrounds;
\item decreases selected nonnegative coefficients in the expression
for $\Law_{O,t}(A)$; and
\item applies the induction hypothesis once or twice to each term in
the expression for $\Law_{O,t}(A)$, matching the terms for
$\Law_{O,t,e}(A)$ up to the factor $p/t$.
\end{enumerate}
The coefficient reductions in step~2 correspond to rejecting $e$ with
an appropriate probability. 
The uses of induction in step~3 become recursive calls on smaller
candidate sets, with modified backgrounds and comparison factor
$p/\sigma_{e,H}$.
The main
obstacle is step~1: it needs a sample of $T'_f$ given $O$, $t$, and
the survivor set $A$, and an independent uniform sample does not
suffice.

This motivates maintaining a configuration that encodes the
experiment's auxiliary times and survival coins, and whose survivor
set equals the accepted set. If, conditional on the observed set at
horizon $t$, these encoded values are independent and uniform as in
the survivor experiment, then conditioning additionally on $A$ gives
exactly the sample of $T'_f$ needed above.

The natural source of randomness is the arrival times. However, using
them directly as auxiliary times has two problems. First, an eligible
element $e$ accepted at time $t>p$ cannot keep $T'_e=T_e$, or it
would be late and could not survive. We must make it early, adjusting
the other coordinates without changing their survivors. Second,
increasing the horizon can increase an early element's survival
probability. If a previously rejected element $f$ would become a
survivor, we cannot accept it retroactively; we must instead make it
late, again adjusting the other coordinates. Forward and reverse
procedures perform these two updates. Their probability transfers balance, preserving uniformity
conditional on the observed set at each fixed horizon.

\medskip\noindent\textbf{Computational model.}
We work in the randomized real RAM: exact arithmetic operations,
comparisons, and independent uniform draws from $[0,1]$ cost one
unit each, as do evaluations of $\ln$ and $\exp$, which go beyond
the standard model. Independence queries are counted separately.
All running-time bounds below are expected bounds in this model,
not bit-complexity bounds.

\begin{theorem}[Efficient realization]\label{thm:survival-efficient}
\Cref{alg:survival-efficient} realizes the same survivor distributions
and accepting masses as \Cref{alg:survival-rule} with $p=1/\ee$. It is
ordinal, uses only $n$ and comparison and independence queries on
arrived elements, and has expected polynomial arithmetic and
independence-oracle complexity.
Its terminal set has distribution $\Law_{V,1}$; in particular, the
elements of $\OPT(V)$ are accepted independently, each with
probability $1/\ee$.
\end{theorem}

\subsection{Auxiliary configurations and update procedures}
\label{sec:kernel-interface}

    The \emph{auxiliary configuration} is a vector with one coordinate
$X_f\in[0,t]$ per element $f$: 
if $X_f>p$, it specifies a late auxiliary arrival at $X_f$; 
if $X_f\leq p$, it specifies an early element and $X_f/p$ is its survival coin. 
Lowercase $x_f$ denotes a particular value of this coordinate. The exact auxiliary time of an early element is irrelevant,
since it is present throughout $[p,t]$. A late element never
survives, so its survival coin is irrelevant. Thus one coordinate
suffices. For independent uniform coordinates on $[0,t]$, the
early set and late auxiliary times have the same distribution
as in \Cref{def:survivor-distribution}. Conditional on them,
the variables $X_f/p$ for early elements are independent
uniforms on $[0,1]$, providing the required survival randomness.

    Given $x\in[0,t]^O$ and a background schedule $H$, let $\sigma_f(x)$ be the first time
$u\in[p,t]$ at which
\[
H_u\cup\{f'\in O:f'\succ f,\ x_{f'}\leq u\}
\]
    spans $f$, or $t$ if this never occurs.
    As in \Cref{def:survivor-distribution}, $\sigma_f(x)$ is the capped spanning time of $f$.
    Note that $\sigma_f(x)$ does not depend on $x_f$, and we suppress its dependence on $O$, $t$, and $H$.
Further, write
\[
g(u)=p\ln \frac{u}{p},\qquad g'(u)=\frac pu,\qquad
      g^{-1}(x)=p\exp(x/p).
\]
For $p\leq u\leq1$, we have $0\leq g(u)\leq g(1)=p$.
    By the definition of the auxiliary configuration, $f$ is early and the coin $\frac{x_f}{p}$ is at most the survival probability $\ln (\sigma_f(x)/p)$ if and only if $x_f \le g(\sigma_f(x))$.
Hence, the survivor set determined by configuration $x$ is
\begin{equation}\label{eq:latent-survivors}
\Surv_{O,t,H}(x)=\{f\in O:x_f\leq g(\sigma_f(x))\}.
\end{equation}
We omit $H$ when it is empty, and write $e$ for constant background
$\{e\}$, as in \Cref{sec:survival}.
For a uniform configuration on $[0,t]^O$, this set has
distribution $\Law_{O,t,H}$.

The forward procedure $\FK_{O,t,H,e}$ takes a configuration $x\in[0,t]^O$ and either returns failure, denoted $\perp$, or a configuration $y$ for the experiment with background $H+e$. 
    The reverse procedure $\RK_{O,t,H,e}$ takes $y$ and always returns an $x$.
We write $\FK(x,dy)$ for the probability of a successful forward output in $dy$, and $\RK(y,dx)$ for the reverse output distribution, suppressing parameters when unambiguous. 
    The forward measure can have total mass less than one, in which case the missing mass corresponds to failure.
    The notation $dx$ denotes ordinary volume measure on the coordinate domain $[0,t]^O$.
The next lemma summarizes the properties we need for these two procedures.

\begin{lemma}[Forward and reverse procedures]%
\label{lem:kernel}
      For every ordered set $O$, horizon $t \in (p,1]$, background $H$
disjoint from $O$, and $e\notin O$, there are forward and reverse
procedures with the following properties.
\begin{enumerate}[label=(\roman*),leftmargin=*,itemsep=4pt,parsep=0pt]
\item \textbf{Probability-mass balance.} Reverse calls always return
an output, and the measures on successful input--output pairs satisfy
\begin{equation}\label{eq:adjoint}
dx\,\FK_{O,t,H,e}(x,dy)
=\frac p{\sigma_{e,H}}\,dy\,\RK_{O,t,H,e}(y,dx).
\end{equation}
In particular, a forward call on uniform input succeeds with
probability $p/\sigma_{e,H}$ and, conditional on success, has
uniform output.
        \item \textbf{Survivors are preserved.} Every pair from a successful forward call or a reverse call satisfies
\[
\Surv_{O,t,H}(x)=\Surv_{O,t,H+e}(y)=A.
\]
\item \textbf{Coordinates move only when allowed.} For such a pair,
$x_f=y_f$ for $f\in A$, and $x_f\geq y_f$ for $f\notin A$.
Thus survivors' coordinates never change; other coordinates only
decrease forward and increase in reverse.
        \item \textbf{Expected cost.} On uniform input, either procedure makes $O(|O|)$ recursive calls in expectation.
              Excluding recursion, the number of independence queries and the expected number of arithmetic operations per call are polynomial in the number of available elements.
\end{enumerate}
\end{lemma}

We construct the procedures and prove \Cref{lem:kernel} in \Cref{sec:kernel-construction}.
Identity \eqref{eq:adjoint} is stronger than the capacity comparison:
it matches individual successful input--output pairs, not just the
probabilities of survivor sets. For an eligible arrival $e$ at $t>p$,
empty background gives $\sigma_{e,\emptyset}=t$. Integrating the
identity over pairs with survivor set $A$ yields, for uniform $X_O$,
\[
\Pr[\Surv_{O,t}(X_O)=A,\ \FK_{O,t,\emptyset,e}
       \text{ succeeds}]
=\frac pt\Law_{O,t,e}(A)=z_{O+e,e,t}(A).
\]
    If $\Law_{O,t}(A)>0$, dividing by $\Law_{O,t}(A) = \Pr[\Surv_{O,t}(X_O)=A]$ gives the acceptance probability from the opening paragraph.
    This probability averages over the coordinates conditional on survivor set $A$; it need not be the success probability for each individual vector.
    If $\Law_{O,t}(A)=0$, use the convention $0/0=0$ from \Cref{alg:survival-rule}.

\subsection{The efficient online algorithm}
\label{sec:latent-algorithm}

We use the continuous-time model of \Cref{sec:continuous-lp},
simulating arrival times internally in the ordinary secretary model.
The algorithm maintains the observed ordered set $O$, the accepted
set $\ALG$, and coordinates $X_O$. Its internal updates via reverse procedures are called
\emph{reverse events}.

\begin{algorithm}[htbp]
\caption{Efficient survival simulation}
\label{alg:survival-efficient}
\begin{itemize}[leftmargin=*,itemsep=3pt,parsep=0pt,topsep=3pt]
        \item Let $p = 1/\ee$ and $g(t)=p\ln(t/p)$.
Initialize $O=\ALG=\emptyset$ and an empty vector $X_O$.
Use the forward and reverse procedures of \Cref{lem:kernel}.
\item Process arrivals and reverse events in chronological order.
\begin{enumerate}[leftmargin=*,label=\arabic*.,itemsep=4pt,parsep=0pt]
\item \textbf{Arrival $e$ at $t\leq p$.}
Reject $e$, set $X_e=t$, and add $e$ to $O$.
\item \textbf{Arrival $e$ at $t>p$.}
If $e\in\OPT(O+e)$, run $\FK_{O,t,\emptyset,e}(X_O)$.
On success, replace $X_O$ by the returned vector, set $X_e=g(t)$,
                      and accept $e$ by adding it to $\ALG$.
If $e$ is ineligible or the call fails, reject it, leave $X_O$
unchanged, and set $X_e=t$. In either case add $e$ to $O$.
\item \textbf{Reverse event.}
For each $f\in\OPT(O)\setminus\ALG$ with $X_f\leq p$, the
candidate event time is $g^{-1}(X_f)=p\exp(X_f/p)$.
At such an event at time $t$, set
\[
X_{O-f}\gets\RK_{O-f,t,\emptyset,f}(X_{O-f}),\qquad X_f\gets t.
\]
Leave $\ALG$ unchanged.
\end{enumerate}
\item Recompute candidate reverse times after every event.
Return $\ALG$ at time $1$.
\end{itemize}
\end{algorithm}

A rejected element can still have an early coordinate in the configuration, either because it arrived before $p$ or because a later forward call lowered its coordinate.
Reverse events prevent exactly these elements from becoming survivors.
They never reverse an online decision.
Throughout this section, the configuration at an event time refers to the one immediately after the update; we omit the probability-zero cases of coincident events.

\begin{lemma}[Feasibility invariant]
\label{lem:latent-invariant}
Almost surely, at every $t\in[p,1]$ the algorithm maintains
\begin{equation}\label{eq:latent-feasible}
\ALG=\Surv_{O,t}(X_O).
\end{equation}
Accepted coordinates never change, and every recomputed reverse
event lies strictly in the future.
\end{lemma}
\begin{proof}
At time $p$, the accepted and survivor sets are empty, since
$g(p)=0$ and the coordinates are positive almost surely.

\emph{Rejected arrival.} A rejected arrival at $t>p$ gets
$X_e=t>p$ and cannot survive. It enters the survivor experiment
only at the current horizon, so it changes no other capped
  spanning time $\sigma_f$ for $f\ne e$.

\emph{Accepted arrival.} Let $x,y$ be the input and output of the
forward call. Then $\Surv_{O,t}(x)=\Surv_{O,t,e}(y)$ (\Cref{lem:kernel}).
Setting $X_e=g(t)\leq p$ makes $e$ early. As shown in the proof of
\Cref{lem:survival-balance}, an early $e\in\OPT(O+e)$ affects the
other spanning times exactly as a constant background $e$.
Thus the old survivors are unchanged. 
  The new element $e$ has capped spanning time $\sigma_e=t$, so $X_e=g(t)$ makes it a survivor as well.

\emph{Between events.} With $O$ and the coordinates fixed, the
threshold $g(\sigma_f)$ increases only for $f\in\OPT(O)$, whose
  capped spanning time is $\sigma_f=t$.
  No survivor disappears.
A rejected element can become a survivor only if it lies in $\OPT(O)$
and $g(t)$ reaches $X_f\leq p$, which is precisely its scheduled
reverse event.

  \emph{Reverse event.} Just before the event, the triggering $f\in\OPT(O)$ is an early nonsurvivor and acts on the other elements as constant background $f$.
  The reverse call removes this background while preserving the survivor set on $O-f$ (\Cref{lem:kernel}).
  Appending $X_f=t>p$ to the returned configuration changes no capped spanning time $\sigma_{f'}$ for $f'\ne f$ and keeps $f$ a nonsurvivor.
  Hence the accepted set remains unchanged.

The procedures fix every survivor's coordinate, so an accepted $e$
keeps $X_e=g(T_e)$ thereafter. Finally, just after an update at time
$t$, a reverse-event candidate $f$ has $\sigma_f=t$ but is not a
survivor. Therefore $X_f>g(t)$, and its candidate time
$g^{-1}(X_f)$ is strictly greater than $t$.
\end{proof}

\paragraph{Number of events.}
At a reverse event, the triggering element's configuration coordinate becomes $t>p$.
Subsequent reverse calls only increase nonsurvivors' coordinates, so that same element cannot trigger again before the next arrival.
There are consequently at most $n$ forward calls and $n(n+1)$
reverse events.
In particular, the event schedule is finite. 

\subsection{The continuous snapshot lemma}
\label{sec:latent-snapshot}

\begin{lemma}[Continuous snapshot lemma]\label{lem:latent-uniform}
At every fixed $t\in[p,1]$, conditional on the observed ordered set
$O$, the coordinates $X_O$ are jointly uniform on $[0,t]^O$.
\end{lemma}

This is the continuous counterpart of \Cref{lem:snapshot}.
There, reversible updates permute equally likely configurations.
Here, the updates are randomized, and \eqref{eq:adjoint} replaces
the bijection by an equality of probability masses. Concretely, an
accepted arrival of $e$ at time $u$ moves its coordinate from $u$ to
$g(u)$, and a reverse event of $e$ at time $u$ moves it back from
$g(u)$ to $u$. Under uniform coordinates, the horizon reaches an
arrival time in $[u,u+du]$ with probability proportional to $du$,
while the threshold reaches a rejected early coordinate with
probability proportional to $g'(u)\,du=(p/u)\,du$. Since
$p/u=p/\sigma_{e,\emptyset}$ is exactly the factor in
\eqref{eq:adjoint}, these two flows of probability cancel. We make
this precise by reconstructing an execution backward.

\begin{proof}
  For analysis only, append the unarrived coordinates, setting $X_e=T_e$ until $e$ arrives.
  The full vector $X\in[0,1]^V$ is uniform at time $p$.
  At a horizon $v$ away from event times, it determines the arrived set $O=\{e:X_e\leq v\}$ and, by \eqref{eq:latent-feasible}, the accepted set.
  Between events the vector is constant.
  We show that its uniform distribution is preserved at every fixed later time.

\medskip\noindent\textbf{The event table.}
At time $u>p$, an arrival has $X_e=u$. A reverse event has
$X_e=g(u)$ with $e$ still rejected immediately before the update.
Call these the \emph{late} and \emph{early boundaries}, respectively.
For this table, let $O$ include the element $e$ involved in the
event, and let $x,y$ be vectors on $O-e$. Abbreviate
$\FK=\FK_{O-e,u,\emptyset,e}$ and
$\RK=\RK_{O-e,u,\emptyset,e}$, and let
$r(x)=\Pr[\FK(x)=\perp]$ for eligible $e$.
Each pair in the middle columns lists $(X_e,X_{O-e})$.
\begin{center}
\small
\renewcommand{\arraystretch}{1.2}
\begin{tabular}{@{}llll@{}}
\toprule
Event & Before & After & Update\\
\midrule
Ineligible arrival & $(u,x)$ & $(u,x)$ & Identity\\
Rejected eligible arrival & $(u,x)$ & $(u,x)$ & Probability $r(x)$\\
Accepted arrival & $(u,x)$ & $(g(u),y)$ & $\FK(x,dy)$\\
Reverse event & $(g(u),y)$ & $(u,x)$ & $\RK(y,dx)$\\
\bottomrule
\end{tabular}
\end{center}
  The first three rows describe three cases of $e$'s arrival; in the last row, $e$ has already arrived and remains rejected.

\medskip\noindent\textbf{Backward reconstruction.}
  We compare executions of the online algorithm from uniform vectors at time $p$ with the following backward reconstructions from uniform vectors at time $t$, and show that their measures agree after reversing time.
Start from a generic vector $Y\in[0,1]^V$ at the target time $t$.
Set $X=Y$, $O=\{e:X_e\leq t\}$, and
$\ALG=\Surv_{O,t}(X_O)$, and run the horizon backward.
Keep coordinates fixed until reaching one of the following boundaries.
\begin{itemize}[leftmargin=*,itemsep=3pt,parsep=0pt,topsep=3pt]
    \item At a late boundary $X_e=u>p$ of an arrived element, $e$ is a nonsurvivor.
          If $e\notin\OPT(O)$, remove it from $O$.
Otherwise apply $\FK_{O-e,u,\emptyset,e}$ to $X_{O-e}$.
          On failure, remove $e$ from $O$. On success, replace $X_{O-e}$ by the output,
set $X_e=g(u)$, and keep $e$ arrived but rejected.
\item At an early boundary $X_e=g(u)$ of an accepted element,
$e\in\OPT(O)$: otherwise $\sigma_e<u$, and survival would give
$X_e\leq g(\sigma_e)<g(u)$. Apply $\RK_{O-e,u,\emptyset,e}$ to $X_{O-e}$,
set $X_e=u$, and remove $e$ from both $O$ and $\ALG$.
\end{itemize}
  Once an element is removed from $O$ at time $u$, its coordinate remains fixed at $u$, which serves as its arrival time in the reconstructed execution.
  In the first rule, a removal reconstructs a rejected arrival and a successful call reconstructs a reverse event; the second rule reconstructs an accepted arrival.
  The reconstructed state corresponds to the moment just before the event at time $u$.

  \medskip\noindent\textbf{Validity of the
    reconstruction.}
  We check that the reconstructed state is \emph{valid} at every horizon $v$ between events: $O=\{e:X_e\leq v\}$ and $\ALG=\Surv_{O,v}(X_O)$, as in \eqref{eq:latent-feasible}.
  The first condition holds by construction: the procedures return coordinates in $[0,u]$, and the horizon reaches a coordinate in $O$ only at a late boundary, which either removes the element or lowers its coordinate to $g(u)\leq p$.
  For the second, an element kept rejected at $X_e=g(u)$ is a nonsurvivor just before $u$, since $g(v)<X_e$ for $v<u$.
The survivor-preserving property of \Cref{lem:kernel}(ii), together with the argument of \Cref{lem:latent-invariant}, shows that these steps leave the survivor status of every element other than $e$ unchanged and give valid states between events.

  \medskip\noindent\textbf{Termination of the
    reconstruction.}
  Between two removals, every update is a successful forward call, which only decreases nonsurvivors' coordinates, followed by setting the triggering coordinate to $g(u)\leq p$. Each element therefore triggers at most once in between.
  There are at most $n$ removals and $n(n+1)$ successful forward calls.
  Hence, the reconstruction reaches time $p$. There the survivor set is empty since $g(p)=0$, so $\ALG=\emptyset$ and the reconstruction ends in a valid initial state.

\medskip\noindent\textbf{Matching individual events.}
  We show that each event of the online algorithm and the backward step that reconstructs it carry the same mass when their input states are weighted by the corresponding boundary weights.
Under uniform volume, a late boundary has weight $du\,dx$, because
$X_e=u$. An early boundary has weight $g'(u)\,du\,dy$, because
$X_e=g(u)$. The coordinates of unarrived elements contribute the
same additional volume factor and are temporarily held fixed.
Since $g'(u)=p/u$ and $\sigma_{e,\emptyset}=u$,
\eqref{eq:adjoint} gives
  \begin{equation}\label{eq:event-matching}
du\,dx\,\FK_{O-e,u,\emptyset,e}(x,dy)
=g'(u)\,du\,dy\,\RK_{O-e,u,\emptyset,e}(y,dx).
  \end{equation}
  This matches each accepted arrival with the second rule, and each
  reverse event with a successful call in the first rule. A rejected eligible arrival
leaves the vector unchanged and has probability $r(x)$ in both
directions. An ineligible arrival is the identity in both directions.
Thus all four rows have matching weighted transition measures.

\medskip\noindent\textbf{From events to complete executions.}
  The boundary weights are ordinary volume on the full vector, with the event time $u$ in place of the coordinate $X_e$.
  Partition states at time $p$, or just after an event, according to the next boundary (late or early) and its element, with a separate part for reaching $t$ without another event.
  On each part, the vector remains unchanged until that boundary or time $t$.
  Hence the volume at the beginning of the interval matches the boundary weight if a boundary is reached, and the volume at time $t$ otherwise.
  The same holds backward, partitioning according to the previous boundary or reaching $p$.
  The two partitions pair up, since validity on both sides (\Cref{lem:latent-invariant} for the online algorithm) rules out early boundaries inside an interval: an accepted element would be a nonsurvivor before its boundary, and a rejected element of $\OPT(O)$ a survivor after it.

  Fix a sequence of event types and elements.
  Alternating these unchanged-vector intervals with the matched random updates gives equal measures to executions of the online algorithm and to the backward reconstructions that visit the same states in reverse order.
  Integrate over the
  intermediate states and sum over event sequences. The online algorithm
  and the backward reconstruction both have a uniform finite bound on
  the number of events; all terms are
  nonnegative.
  Coincident boundaries form a null set for the corresponding
boundary weights. By \eqref{eq:event-matching}, the weighted
transitions assign zero mass to these exceptional outgoing
configurations in either direction. Since executions have
finitely many events, coincidences occur with probability zero.
  We obtain
\begin{equation}\label{eq:forward-backward-history-duality}
dX\,P_{p,t}(X,dY)
=dY\,\overleftarrow P_{t,p}(Y,dX),
\end{equation}
where $P_{p,t}(X,dY)$ and $\overleftarrow P_{t,p}(Y,dX)$ are
  the output distributions of the online algorithm from time $p$ to $t$ and of the backward reconstruction from $t$ to $p$, respectively.
  This argument uses measures on pairs;
it does not assume that the randomized updates have densities.
  Since the backward reconstruction terminates with probability one,
integrating out $X$ gives
\[
\int dX\,P_{p,t}(X,dY)
=dY\int\overleftarrow P_{t,p}(Y,dX)=dY.
\]
Thus the full vector is uniform at time $t$. Conditioning on
$O=\{e:Y_e\leq t\}$ proves the lemma.
\end{proof}

Uniformity at a fixed time does not by itself describe inputs at
random internal events. We need those inputs both to identify the
accepting masses and to bound the total work. The same backward
argument supplies them. The following notation records expected
numbers of events in a specified time-and-coordinate region; it
counts every occurrence of a reverse event, not just the first.

\begin{corollary}[Inputs at algorithm events]
\label{cor:latent-event-inputs}
For a measurable region $B$ of pairs $(u,x)$, let
$\nu^{\mathrm{arr}}_{O,e}(B)$ be the expected number of arrivals
of $e$ with preceding observed set $O$ and $(u,X_O)\in B$
immediately before the arrival. For $p<u<1$,
\begin{equation}\label{eq:arrival-input}
\nu^{\mathrm{arr}}_{O,e}(du,dx)
=(1-u)^{n-|O|-1}\,du\,dx,
\qquad e\notin O,\quad x\in[0,u]^O.
\end{equation}
Similarly, let $\nu^{\mathrm{rev}}_{O,f}(B)$ be the expected number
  of reverse events with observed set $O$, triggering element $f$, and
$(u,X_{O-f})\in B$ just before the event. Then
\begin{equation}\label{eq:reverse-input}
\nu^{\mathrm{rev}}_{O,f}(du,dy)
=(1-u)^{n-|O|}\frac pu\,du\,dy,
\qquad f\in\OPT(O),\quad y\in[0,u]^{O-f}.
\end{equation}
For $f\notin\OPT(O)$ this measure is zero. Thus, for fixed event
time, observed set, and triggering element, the relevant inputs
are uniform under these expected-event measures.
\end{corollary}
\begin{proof}
  Stop an execution of the online algorithm immediately before a marked
  event and reconstruct that prefix backward, without undoing the marked
  event itself.
  The input state has the arriving element still unarrived, or
  the triggering element still rejected. The same interval and
  update matchings show that the expected-event measure of these input
  states is their boundary weight multiplied by the probability that the
  backward reconstruction terminates.
That probability is one by the validity and termination argument
above. The weights are therefore $du\,dx$ at arrivals and
  $g'(u)\,du\,dy$ at reverse events, times volume on the unarrived
coordinates. There are $n-|O|-1$ such coordinates in the first case
and $n-|O|$ in the second, each ranging over $(u,1)$. Integrating
them out gives the stated factors. Every marked occurrence gives
one prefix, so this also counts repeated reverse events correctly.
\end{proof}

\begin{proof}[Proof of \Cref{thm:survival-efficient}]
Before $p$ the accepted set is empty. Thereafter,
\Cref{lem:latent-uniform} and \eqref{eq:latent-feasible} give the conditional distribution
$\Law_{O,t}$ at each fixed horizon. At an eligible arrival,
\Cref{cor:latent-event-inputs} supplies the uniform input needed
for the calculation after \Cref{lem:kernel}; hence the accepting
mass is exactly $z_{O+e,e,t}(A)$. Before $p$ and at ineligible
arrivals it is zero. The algorithm therefore realizes the
prescribed distributions and accepting masses. The terminal
independence guarantee follows from \Cref{thm:survival-lp}.

For complexity, compare reverse events with accepted arrivals.
Fix an observed set $O$ and $f\in\OPT(O)$. An arrival of $f$ into
$O-f$ has input measure $(1-u)^{n-|O|}\,du\,dx$ by
\eqref{eq:arrival-input}. On successful forward calls,
\eqref{eq:adjoint}, integrated over $x$, gives output measure
\[
(1-u)^{n-|O|}\frac pu\,du\,dy,
\]
because reverse calls return an output with probability one.
This is precisely \eqref{eq:reverse-input}. Summing over $O$ and
$f$ and integrating gives
\[
\E[\text{number of reverse events}]=\E|\ALG|\leq n.
\]
This is an equality of expected counts, not a pairing within a single
execution. There are also at most $n$ forward calls. Integrating
the uniform-input cost bounds of \Cref{lem:kernel} against the
event measures gives $O(n^2)$ expected calls in total, including
recursion. Each call uses polynomial local work. Updating the
greedy solution and recomputing candidate event times also cost
polynomial work, using the pathwise $O(n^2)$ event bound.

Every comparison and independence query involves arrived elements:
recursive calls only move elements of their input into the fixed
background. All decisions use only observed value comparisons,
arrival times, and independent randomness. This proves ordinality,
the information restriction, and the asserted expected complexity.
\end{proof}

\subsection{Proof of \Cref{lem:kernel}}
\label{sec:kernel-construction}

We now construct the procedures of \Cref{lem:kernel}.
As in the proof of \Cref{lem:survival-comparison}, we remove the largest element $f$, remember
when it contributes to spanning, and recurse on $O-f$.
Here its configuration coordinate also specifies whether it survives. We give the scalar
updates explicitly and identify the smaller background comparisons
they require. Every recursive call below has candidate set $O-f$
and the same horizon $t$.
A failed forward subcall fails the enclosing call. %

\medskip\noindent\textbf{Base cases.}
If $\sigma_{e,H}=p$, then $e$ is already spanned by $H_p$, so adding
it changes no spanning time; both procedures directly return their input.
If $O=\emptyset$, the forward procedure returns the empty vector
with probability $p/\sigma_{e,H}$ and otherwise fails; the reverse
procedure always returns the empty vector.

\medskip\noindent\textbf{Unchanged spanning times.}
Beyond the base cases above, let $f$ be the largest element of $O$.
If $f$'s spanning time stays the same with $e$ added to the background, i.e., $\sigma_{f,H}=\sigma_{f,H+e}$, keep its coordinate and recurse with background $H^{f,u}$, where $u=\max\{p,x_f\}$ forward and
$u=\max\{p,y_f\}$ in reverse.

\begin{figure}[htbp]
  \centering
  \includegraphics[width=\linewidth]{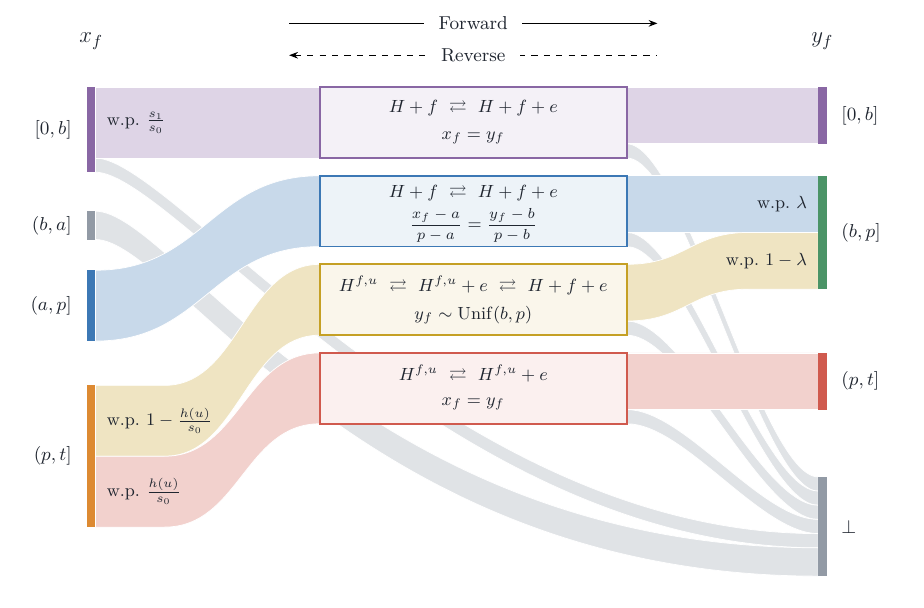}
  \caption{Forward and reverse coordinate updates in the main case $s_0>s_1$.
    Read the diagram from left to right for the forward procedure and from right to left for the reverse procedure.
    The middle rectangles specify recursive background changes on $O-f$; gray ribbons lead to forward failure $\perp$.
    Ribbon widths are schematic.}
  \label{fig:forward-reverse-transport}
\end{figure}

\medskip\noindent\textbf{The main case.}
Suppose
\[
  s_0=\sigma_{f,H}>s_1=\sigma_{f,H+e},\qquad
  a=g(s_0),\quad b=g(s_1).
\]
Then $\sigma_{e,H}=s_0$ by \eqref{eq:span-pair}.
Under $H$, the element $f$ survives exactly when $x_f\leq a$;
under $H+e$, exactly when $y_f\leq b$.
Let
\[
  h(u)=\min\{s_0,\max\{s_1,u\}\},\qquad
  \lambda=\frac{s_0(p-a)}{s_1(p-b)}.
\]
Here $h(u)=\sigma_{e,H^{f,u}}$ by \eqref{eq:span-release}, and
$0\leq\lambda\leq1$, as verified below. \Cref{fig:forward-reverse-transport}
specifies four branches, numbered from top to bottom, and the
failure outcomes. An arrow $J\rightleftarrows J+c$ means a
forward call $\FK_{O-f,t,J,c}$ on the remaining coordinates;
in reverse it means $\RK_{O-f,t,J,c}$. Two arrows are performed in
the displayed order forward and the opposite order in reverse.
The labels ``w.p.'' give the probabilities of choosing a branch,
before making its recursive calls; unlabeled choices have probability one.

\medskip\noindent\textbf{Reading the forward procedure.}
If $x_f\leq b$, use the first branch with probability $s_1/s_0$, keeping
$y_f=x_f$, and otherwise fail. If $b<x_f\leq a$, fail immediately. If
$a<x_f\leq p$, use the second branch and set
\[
  y_f=b+\frac{p-b}{p-a}(x_f-a).
\]
For $x_f=u>p$, choose branch~3 with probability $1-h(u)/s_0$
and branch~4 with probability $h(u)/s_0$. In the third branch, $y_f$ is an independent
uniform draw from $(b,p)$; in the fourth, $y_f=x_f$.

The first branch corresponds to outputs containing $f$ in the capacity
proof: it uses only part of the available surviving mass. The second
branch maps early nonsurviving coordinates from $(a,p]$ to $(b,p]$.
It covers only a fraction $\lambda$ of the required mass there;
the missing $1-\lambda$ fraction is exactly the early deficit in
\eqref{eq:early-deficit}.
The third branch uses the late surplus to cover the early deficit; it is
precisely the two successive comparisons in \eqref{eq:late-surplus}, not
an additional rejection rule. The fourth branch leaves $f$ late.

\medskip\noindent\textbf{Reading the reverse procedure.}
If $y_f\leq b$, use the first branch, keeping $x_f=y_f$. If
$b<y_f\leq p$, choose the second or third branch with
probabilities $\lambda$ and $1-\lambda$. In the second branch, set
\[
  x_f=a+\frac{p-a}{p-b}(y_f-b).
\]
In the third branch, draw $x_f=u\in(p,s_0)$ with density proportional to
\begin{equation}\label{eq:transport-residual}
  \frac{s_0-\max\{s_1,u\}}u,
\end{equation}
and perform the two smaller reverse calls in reverse order. If $y_f>p$, use the fourth branch, keeping $x_f=y_f$.
All reverse subcalls return an output.
If $a=p$, then $\lambda=0$, so the
second branch is unused and no division by $p-a$ is performed.
Endpoint cases of zero probability do not affect the identities.

To sample \eqref{eq:transport-residual} efficiently, draw $v$ uniformly from
$(s_1,s_0)$ and, independently, draw $u=p\exp(U\ln(s_0/p))$ with
$U\sim\Unif[0,1]$. Repeat until $u<v$. The successful pair has
density proportional to $1/u$ on
$\{(u,v):p<u<v,\ s_1<v<s_0\}$, so its $u$-marginal has the
required density. This is the integration region in
\eqref{eq:scalar-slack}. A trial succeeds with probability
\[
  \frac{\E[\ln(v/p)]}{\ln(s_0/p)}\geq\frac12:
\]
concavity puts $\ln(v/p)$ above its chord on $[p,s_0]$, and
$\E v=(s_1+s_0)/2\geq(p+s_0)/2$. Thus this scalar sampler uses
at most two trials in expectation.

\begin{proof}[Proof of \Cref{lem:kernel}]
  We prove the claims by induction on $|O|$, first checking balance,
  then the coordinate properties and costs.

  \medskip\noindent\textbf{Probability-mass balance.}
  The base cases satisfy \eqref{eq:adjoint} directly. When the largest
  element's spanning time is unchanged, its coordinate is copied, and
  the smaller call has the required factor
  $p/\sigma_{e,H^{f,u}}=p/\sigma_{e,H}$: by closure exchange, if $f$
  newly spanned $e$ at some time before $t$, then $e$ would newly span $f$ at that
  time, contradicting $\sigma_{f,H}=\sigma_{f,H+e}$. Induction applies.

  We now consider the main case.
  By \eqref{eq:span-pair} and \eqref{eq:span-release}, the recursive calls in the first, second, and fourth branches have probability-mass balance factors $p/s_1$, $p/s_1$, and
  $p/h(u)$. Multiplying by the scalar change-of-variable factors and branch probabilities gives, respectively,
  \[
    \frac{s_1}{s_0}\frac p{s_1} = \frac{p}{s_0},\qquad
    \frac{p-a}{p-b}\frac p{s_1} = \lambda \frac{p}{s_0},\qquad
    \frac{h(u)}{s_0}\frac p{h(u)} = \frac{p}{s_0}.
  \]
  These are the required factor times the respective reverse-branch
  probabilities.

  The third branch has two recursive factors, $p/h(u)$ and
  $p/\min\{s_1,u\}$, and scalar output density $1/(p-b)$.
  Its joint contribution, before normalizing the reverse choice of
  $u$, is
  \[
    \Bigl(1-\frac{h(u)}{s_0}\Bigr)\,
    \frac{p}{h(u)}\, \frac{p}{\min\{s_1,u\}}\, \frac{1}{p-b} \,du
    ~=~ \frac{p^2}{s_0s_1(p-b)}
    \frac{s_0-\max\{s_1,u\}}u\,du,
    \qquad p<u<s_0.
  \]
  By \eqref{eq:scalar-slack}, the antiderivative in
  \eqref{eq:early-deficit}, and the definitions of $a,b$, its integral is
  \[
    \frac{p^2}{s_0s_1(p-b)}
    \int_{s_1}^{s_0}\ln(v/p)\,dv
    = \frac p{s_0} \left(1-\frac{s_0(p-a)}{s_1(p-b)}\right)
    = \frac p{s_0} (1-\lambda).
  \]
  This nonnegative integral also verifies $\lambda\leq1$;
  $\lambda\geq0$ follows from $a\leq p$.
  Normalizing gives exactly the reverse density and branch probability.
  Induction and composition therefore prove \eqref{eq:adjoint} for
  every branch, including its recursive choices. Integrating over
  $x$ shows that a uniform-input forward call succeeds with probability
  $p/\sigma_{e,H}$ and has uniform output conditional on success.

  \medskip\noindent\textbf{Survivors and coordinates.}
  In the base cases, every coordinate and survivor is unchanged. When
  $f$'s spanning time is unchanged, $f$ keeps its coordinate and its
  survival status. In the main case, in the first branch, $f$ survives in both experiments and keeps its
  coordinate. In the other branches it does not survive, and $y_f\leq x_f$.
  For the affine branch this follows from
  \[
    y_f-x_f=\frac{(a-b)(x_f-p)}{p-a}\leq0.
  \]
  In each branch, the recursive backgrounds record exactly when $f$ is
  present before and after its coordinate update. Every smaller call
  preserves its survivor set, fixes survivors' coordinates, and
  decreases other coordinates forward. Their composition gives
  properties (ii) and (iii). Reverse calls read
  these same pairs in the opposite direction.

  \medskip\noindent\textbf{Number of recursive calls.}
  On a uniform forward input, the remaining coordinates are uniform conditional on $x_f$ and the chosen branch.
  By the induction hypothesis, the output of a smaller call conditioned on success is uniform.
  Inductively, every recursive call also receives uniform input.

  In the main case, let $P_i$ be the probability of making exactly
  $i$ direct recursive calls.
  Only the third branch can make two; making no
  child call means an immediate failure in the surviving range.
  Consequently,
  \begin{align*}
    P_0 & ~=~ \underbrace{\vphantom{\bigg|} \frac{a-b}{t}}_{x_f \in (b, a]} + \underbrace{\vphantom{\bigg|} \frac{b}{t}}_{x_f \in [0, b]} \Bigl( \underbrace{\vphantom{\bigg|} 1-\frac{s_1}{s_0}}_{\substack{\text{prob.\ of}\\ \text{failure}}} \Bigr) ~=~ \frac{a-(s_1/s_0)b}{t}, \\[1ex]
    P_2 & ~=~ \frac{1}{t} \int_p^t ~ \biggl( \underbrace{\vphantom{\bigg|} 1-\frac{h(u)}{s_0}}_{\substack{\text{prob.\ of}\\ \text{branch 3}}} \biggr) \underbrace{\frac{p}{h(u)}}_{\substack{\text{success prob.}\\ \text{of 1st call}}} du
    ~=~ \frac{1}{t} \int_{s_1}^{s_0}\frac{p(v-p)}{v^2}\,dv
    \leq \frac{a-b}{t} \leq P_0.
  \end{align*}
  Thus the expected number of children is $1-P_0+P_2\leq1$. Each recursive call has one fewer element in its candidate set, so induction bounds the expected total number of recursive forward calls by $1+|O|$.
  The unchanged-spanning-time case has one child and satisfies the same bound; the base cases have none.

  The branchwise balance identities also match successful forward
  and reverse trees of recursive calls, with the same number of
  nodes. Hence a reverse call on uniform input has the
  recursive-call count of a successful forward call on uniform input.
  Since that success probability is
  $p/\sigma_{e,H}\geq p=1/\ee$, its expected number of recursive reverse calls is at most
  $\ee(|O|+1)$.

  Finally, the spanning times used by a call are found by checking
  rank at the finitely many fixed background times. Each call
  therefore needs only polynomially many independence queries and
  arithmetic operations outside its recursive calls. The only
  repeated scalar sampling takes at most two trials in expectation.
  All elements involved belong to the original call's candidates, background, or added element $e$.
  This proves property (iv) and completes the lemma.
\end{proof}

\section{Extensions and Consequences}\label{sec:extensions}

We first use the characterization to compare probability and
ordinal competitiveness on special classes and examine finite-size
guarantees. We then extend the snapshot algorithm to intersections
of matroids and derive a submodular consequence of the optimal
single-matroid algorithm. In these last two applications,
competitiveness refers to the expected objective value.

\subsection{Examples and finite-size evidence}
\label{sec:lp-examples}

The characterization in \Cref{sec:lp-objective}
supports both algorithm construction and
impossibility proofs. Uniform matroids separate the two competitive
objectives, while linear matroids admit stronger finite-size guarantees.

\paragraph{Uniform matroids: separating the two objectives.}
Let $U_{k,n}$ be the matroid on $n$ elements whose independent
sets have size at most $k$.

\begin{theorem}[Uniform matroids]\label{cor:uniform-separation}
The optimal per-element selection fraction among ordinal algorithms
satisfies
$\frac1{\rho(U_{k,n})}
=1-\Theta\!\big(\sqrt{\frac{\log k}{k}}\big)$
as $k\to\infty$, uniformly over $n\geq2k$.
In contrast, an ordinal algorithm can achieve expected-value
fraction $1-O(k^{-1/2})$.
\end{theorem}
\begin{proof}
The probability statement combines
\Cref{prop:uniform-lp,prop:uniform-lp-lower}.
The expected-value guarantee follows from
Kleinberg's algorithm~\cite{Kleinberg-SODA05,SotoTV-SODA18}.
\end{proof}

Thus the extra $\sqrt{\log k}$ loss in the per-element guarantee
is necessary, answering the question of Soto, Turkieltaub, and
Verdugo~\cite{SotoTV-SODA18}. The constructive bound uses
relative-rank thresholds and independent thinning; its parameters
and analysis are given in \Cref{prop:uniform-lp}.

The matching lower bound already holds for $n=2k$ and uses a
single midpoint snapshot. If each optimum element is selected
with probability at least $1-\varepsilon$, capacity permits at
most $k\varepsilon$ expected selections outside the optimum,
while the $k$-th largest element can be missed with probability
at most $\varepsilon$. Let $X$ count the top $k$ elements among
the first $k$ arrivals, so
$X\sim\operatorname{Hypergeometric}(2k,k,k)$.
Combining these two requirements and eliminating the algorithm's
midpoint acceptance marginals gives
\[
2k\varepsilon\geq
\sum_{r=1}^k\min\{\Pr[X<r],\,r\Pr[X=r]\}.
\]
For $\Omega(\sqrt{k\log k})$ values of $r$, the summand is
bounded below by an absolute positive constant. This forces
$\varepsilon=\Omega(\sqrt{\log k/k})$.
\Cref{prop:uniform-lp-lower} proves the certificate and extends
the bound to every $n\geq2k$. No threshold or monotonicity
assumption on the algorithm is needed.

\paragraph{Linear matroids: the finite-size rank-one guarantee.}
For $n\geq2$ and $1\leq s<n$, write
$c_n(s)=\frac sn\sum_{i=s}^{n-1}\frac1i$
for the success probability of the classical secretary rule
that rejects the first $s$ arrivals.

\begin{theorem}[Linear matroids]\label{thm:linear}
For every linear matroid on $n\geq2$ elements, known in advance,
and every $1\leq s<n$, there is an ordinal algorithm accepting
each element of $\OPT(V)$ with probability exactly $c_n(s)$,
for every nonnegative value assignment.
\end{theorem}

Choosing the best cutoff gives
$\rho(\calM)\leq\rho(U_{1,n})<\ee$ whenever $\calM$ has positive
rank. Thus the guarantee improves on $1/\ee$ for every finite
$n$. This covers representable classes such as graphic,
transversal, and laminar matroids~\cite{FifeOxley2019}, but
only when the matroid is known in advance.

The construction bounds the expected dimension of the intersection
of the accepted span with every subspace. The one-dimensional
constraints ensure sufficient mass on sets that do not span
an arriving candidate; the higher-dimensional constraints allow
these availability guarantees to be maintained together.
\Cref{sec:linear} gives the feasible family and proof.
The construction is finite, with no polynomial running-time
bound asserted.

\paragraph{Computational evidence beyond linear matroids.}
Let $\rho_{\mathrm{ord}}(\calM)$ denote the optimal expected-value
competitive ratio among ordinal algorithms.
Numerically optimizing the LP of \Cref{cor:lp-ordinal} suggested
the stronger finite-size comparison
\[
\rho_{\mathrm{ord}}(\calM)
\leq \rho_{\mathrm{ord}}(U_{1,n}).
\]
Every tested instance satisfied this inequality up to numerical
precision. The computations covered every positive-rank matroid
on seven elements and $1655$ of the $1723$ positive-rank matroids
on eight elements, including $43$ of the $44$ nonrepresentable
ones. These are floating-point observations for the expected-value
objective, rather than proofs of the stronger per-element
guarantee. \Cref{sec:lp-computations} gives the computational
reductions, coverage, and numerical details.

\subsection{Intersection of matroids}\label{sec:intersection}

Fix matroids $\calM_j=(V,\calI_j)$ for $j\in[k]$, with $k\geq1$, and fixed nonnegative values $v_e$. Let $I^\star$ be a maximum-value common independent set. 
For an auxiliary value vector $a=(a_e)_{e\in V}$, write
$a(A)=\sum_{e\in A}a_e$ and let $\OPT_j(T;a)$ be greedy in
$\calM_j$ on $T$, using values $a$ and the fixed tie-breaking order.
The individual optima need not have a valuable intersection, even when $I^\star$ has large value. The preprocessing of Feldman, Svensson, and Zenklusen~\cite{FSZ-SICOMP22} supplies auxiliary values for which their overlap is valuable in expectation.

\begin{theorem}\label{thm:intersection}
There is an $8k^2(k+1)$-competitive secretary algorithm under the intersection of $k$ matroids. It needs only $n=|V|$, comparisons of values, and independence queries on arrived elements in each matroid, and uses $O(kn^2)$ independence queries.
\end{theorem}
\begin{proof}
\noindent\textbf{Preprocessing.}
Let $p_0=1-1/(2k)$, draw $L\sim\Bin(n,p_0)$, and reject the first $L$ arrivals, forming a sample $P$. Each element belongs to $P$ independently with probability $p_0$. Let $\OPT_\cap(T)$ be decreasing-value greedy on $T$ subject to all $k$ constraints: it adds each element whenever the resulting set is independent in every matroid. Define
\[
a_e=\begin{cases}
v_e,&e\notin P\text{ and }e\in\OPT_\cap(P+e),\\
0,&\text{otherwise}.
\end{cases}
\]
Greedy scans every element with positive auxiliary value before
the zero-valued elements. Thus the overlap guarantee of
Feldman, Svensson, and Zenklusen~\cite[Theorem~3.3]{FSZ-SICOMP22}
gives
\begin{equation}\label{eq:overlap}
\E_P[a(B^+(a))]\geq\frac{v(I^\star)}{4k^2},
\qquad
B^+(a)=\{e:a_e>0\}\cap\bigcap_{j=1}^k\OPT_j(V;a).
\end{equation}
In particular, $B^+(a)\subseteq V\setminus P$. The value $a_e$ is computable at the arrival of $e$ using only $P+e$.
It therefore suffices to select each element of $B^+(a)$ with
conditional probability at least $1/(2(k+1))$, given $P$.
The snapshot argument below supplies this probability by requiring
the element to precede at most one exchange partner per matroid.

\medskip\noindent\textbf{A common reference set.}
Conditional on $P$, let $U=V\setminus P$ and $m=|U|$.
If $m=0$, return the empty set. Otherwise, run
\Cref{alg:matroidSec} on all $m$ remaining elements with values
$a$ and fresh randomness. Its initial reference set is a new
sample from $U$, not the preprocessing sample $P$.
Require the preservation test in every matroid:
\[
\OPT_j(C';a)\cap F=\OPT_j(C;a)\cap F\qquad(j\in[k]).
\]
If the test passes, accept the current element exactly when it belongs to every proposed greedy set, and perform the reference-set update whether or not it is accepted. After each processing step, add its element to $F$. As in \Cref{sec:snapshot}, sample steps cannot be accepted because they are deleted from the proposed set.

We retain all elements of $U$ in the stream, including those assigned auxiliary value zero. Thus the second algorithm knows its input size $m$ without knowing how many future elements will pass the preprocessing test. Its invariant
$\ALG\subseteq\bigcap_{j=1}^k\OPT_j(C;a)$
proves feasibility in all $k$ matroids.

\medskip\noindent\textbf{Snapshot and acceptance probability.}
Conditional on $P$, the auxiliary values $a$ are fixed and the
remaining arrival order is uniform. The conjunction of the
$k$ preservation tests has the same answer at $C$ and $C\oplus e$.
Thus each update can be undone by the same rule, and
\Cref{lem:snapshot,cor:snapshot-independent} apply unchanged.
For each fixed target $e$, the configuration $C_e$ immediately
before its processing step is therefore uniform on $2^U$,
independently of the uniformly random processing order.

Fix $e\in B^+(a)$ and condition on $C_e=T$ with $e\notin T$. This is a real arrival, since no earlier step changes the membership of $e$. By greedy monotonicity (\Cref{lem:greedy}(i)), $e\in\OPT_j(T+e;a)$ for every $j$. Inserting $e$ displaces at most one element in each matroid. Let $D$ be the set of distinct displaced elements. Once $P,T,e$ are fixed, $D$ is fixed and $|D|\leq k$.

The test passes exactly when no element of $D$ has been processed, or equivalently when $e$ precedes them all in the processing order. The snapshot property leaves that order uniform, giving
\[
\Pr[e\in\ALG\mid P,C_e=T]=\frac{1}{|D|+1}\geq\frac{1}{k+1}.
\]
Since $\Pr[e\notin C_e\mid P]=1/2$, each element of $B^+(a)$ is accepted with conditional probability at least $1/(2(k+1))$. Therefore,
\[
\E[v(\ALG)]\geq\E[a(\ALG)]
\geq\frac{\E_P[a(B^+(a))]}{2(k+1)}
\geq\frac{v(I^\star)}{8k^2(k+1)}.
\]
Finally, preprocessing uses at most $n$ common-greedy scans, each requiring $O(kn)$ independence queries. The second stage uses at most two greedy scans per matroid at each of its $m$ processing steps. All queries involve only $P+e$ or the current and proposed reference sets, hence only arrived elements. The total is $O(kn^2)$.
\end{proof}

\subsection{Submodular objectives}

A set function $f:2^V\to\mathbb R_{\geq0}$ is \emph{submodular} if $f(A+e)-f(A)\geq f(B+e)-f(B)$ whenever $A\subseteq B$ and $e\notin B$; it need not be monotone. In the submodular matroid secretary problem, the goal is to maximize $\E[f(\ALG)]$, compared with $\max_{I\in\calI}f(I)$. The algorithm knows $n$ and may query $f$ and independence on sets of arrived elements. The reduction below additionally requires an upper bound on every
element's selection probability. Our survivor construction provides
this because only elements of the auxiliary early sample can survive.

\begin{theorem}[{Feldman and Zenklusen~\cite{FZ-SICOMP18}}]
\label{thm:fz}
Let $\alpha\geq1$. Suppose an additive matroid secretary algorithm
is $\alpha$-competitive and selects every element with probability
at most $1/\alpha$. Then there is a $96\alpha$-competitive algorithm
for nonnegative submodular objectives in the same information model.
The reduction preserves expected polynomial arithmetic and oracle
complexity.
\end{theorem}

\begin{corollary}\label{cor:submodular}
For every matroid and every nonnegative submodular objective, there
is a $96\ee$-competitive secretary algorithm with expected polynomial
arithmetic and oracle complexity. It needs only the number of
elements and independence and value-oracle queries on subsets of
arrived elements.
\end{corollary}
\begin{proof}
The algorithm of \Cref{thm:survival-efficient} is
$\ee$-competitive and has terminal distribution $\Law_{V,1}$.
In the survivor experiment defining this distribution, every survivor belongs to the
auxiliary early sample $E$. Hence, for every $e\in V$,
\[
\Pr[e\in\ALG]\leq\Pr[e\in E]=p=1/\ee.
\]
Apply \Cref{thm:fz} with $\alpha=\ee$.
\end{proof}

\section*{Acknowledgments and AI Methodology Statement}
\addcontentsline{toc}{section}{Acknowledgments and AI Methodology Statement}

\noindent\textbf{Relation to the predecessor manuscripts.}
This paper merges and extends the preprints
\cite{Singla-arXiv26,ABHM-arXiv26,Huang-arXiv26}
into a unified treatment.
The combinatorial snapshot algorithm was presented by
Singla~\cite{Singla-arXiv26}; the optimal single-sample prophet
inequality by Abdi, Banihashem, Hajiaghayi, and
Mittal~\cite{ABHM-arXiv26}; and the optimal general-matroid
secretary algorithm and its efficient implementation by
Huang~\cite{Huang-arXiv26}. The paper \cite{ABHM-arXiv26} also gave, in the
known-matroid model, an optimal secretary algorithm for linear
matroids and a black-box reduction yielding a $64$-competitive
algorithm for arbitrary matroids. 
This treatment organizes the results around snapshot invariants,
acceptance distributions, and survival simulation.

\medskip\noindent\textbf{AI usage in research.}
The results came from combining the authors' direction with the
technical capabilities of AI models. The authors made judgment calls
about which approaches to pursue, used linear-program computations
to guide the search, and directed the exploration toward simulating
an offline distribution of accepted elements. The linear program in
\Cref{sec:lp} and the computations with it were devised and carried
out without AI models, starting in February 2026.\footnote{The
implementation and its outputs are available at
\url{https://github.com/TlatoaniHJ/MatroidSecretary}.}
Interactions with models available through ChatGPT Pro, especially
GPT Astra, were used in obtaining the main algorithmic guarantees,
including \Cref{thm:four-secretary,thm:ssmpi-tight,thm:linear,thm:tight-secretary};
GPT Sol~5.6 was also used in developing the single-sample prophet
and linear-matroid results. Along the way, the authors explored
single-sample algorithms~\cite{Li-24,DKLRS-FOCS24}, pairings of
sample and real elements~\cite{RS-STOC26}, the minimax
principle~\cite{KMPS-FOCS26}, and reductions to free-order secretary
through tangential sequences~\cite{FSW-arXiv26}. Recognizing that
the survival-comparison and online-realization arguments are
algorithmic led from partial, computationally inefficient guarantees
to the efficient implementation. The authors checked the resulting
proofs.

\medskip\noindent\textbf{AI usage in writing.}
This manuscript was prepared with the assistance of ChatGPT and Claude, which were used to draft and revise the exposition, including the unified presentation of results that first appeared in separate papers. The authors have checked all mathematical arguments and take full responsibility for the contents.

\medskip\noindent\textbf{Acknowledgments.}
Singla is grateful to Anupam Gupta and S.~Matthew Weinberg for discussions on the matroid secretary problem over the past decade, and thanks Thomas Kesselheim, Robert Kleinberg, Wenzheng Li, Thiago Oliveira, Kalen Patton, Rebecca Reiffenh\"auser, Aviad Rubinstein, Siddharth M.~Sundaram, and Ola Svensson for related discussions, and Anupam Gupta, S.~Matthew Weinberg, and Robert Kleinberg for comments on early versions of~\cite{Singla-arXiv26}. Banihashem thanks Neel Patel for pointing out the existence of the reduction from matroid secretary to constant-sample prophet inequalities. Banihashem, Hajiaghayi, and Mittal thank Dariusz Kowalski, Piotr Krysta, and Jan Olkowski for early discussions and joint work on the matroid secretary problem, which led to~\cite{BHKKMO-ESA25} on graphic matroids. 
Huang thanks Dongchen Li, Zahra Parsaeian, and Zixuan Zhu
  for early discussions and joint work on the matroid
  secretary problem.

{\small\bibliographystyle{alpha}\bibliography{references}}

\appendix

\crefalias{section}{appendix}
\crefalias{subsection}{subappendix}

\clearpage

\section{Why Natural Greedy Rules Fail}
\label{sec:lookingBack}

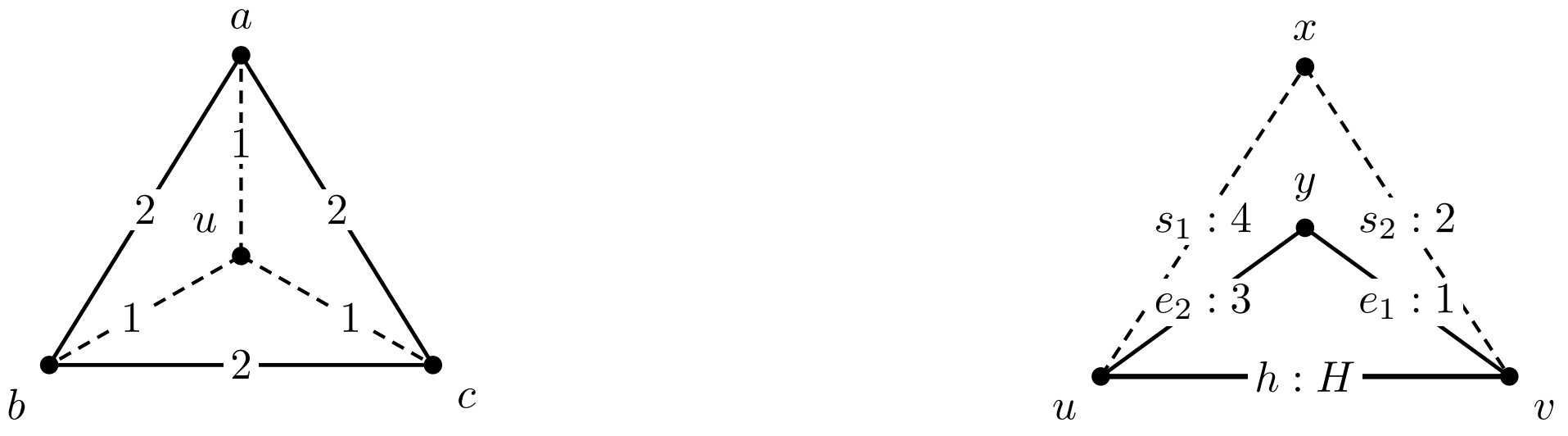
\begin{figure}[htbp]
\centering
\begin{minipage}[t]{0.47\linewidth}
\centering
\begin{tikzpicture}[
    x=1cm,y=1cm,
    vertex/.style={circle,fill=black,inner sep=1.5pt},
    sample/.style={densely dashed,line width=0.8pt},
    real/.style={line width=0.9pt},
    wt/.style={fill=white,inner sep=1.5pt,font=\small},
    every node/.style={font=\small}
]
\coordinate (a) at (0,1.50);
\coordinate (b) at (-1.55,-1.00);
\coordinate (c) at (1.55,-1.00);
\coordinate (u) at (0,-0.12);
\draw[real] (a)--node[wt,pos=0.5] {$2$} (b);
\draw[real] (b)--node[wt,pos=0.5] {$2$} (c);
\draw[real] (c)--node[wt,pos=0.5] {$2$} (a);
\draw[sample] (u)--node[wt,pos=0.56] {$1$} (a);
\draw[sample] (u)--node[wt,pos=0.57] {$1$} (b);
\draw[sample] (u)--node[wt,pos=0.57] {$1$} (c);
\node[vertex,label=above:$a$] at (a) {};
\node[vertex,label=below left:$b$] at (b) {};
\node[vertex,label=below right:$c$] at (c) {};
\node[vertex,label=above left:$u$] at (u) {};
\end{tikzpicture}

\smallskip
{\small (a) The na\"ive rule accepts a cycle.}
\end{minipage}
\hfill
\begin{minipage}[t]{0.49\linewidth}
\centering
\begin{tikzpicture}[
    x=1cm,y=1cm,
    vertex/.style={circle,fill=black,inner sep=1.5pt},
    sample/.style={densely dashed,line width=0.8pt},
    real/.style={line width=0.9pt},
    wt/.style={fill=white,inner sep=1.5pt,font=\small},
    every node/.style={font=\small}
]
\coordinate (u) at (-1.65,-1.00);
\coordinate (v) at (1.65,-1.00);
\coordinate (x) at (0,1.50);
\coordinate (y) at (0,0.20);
\draw[sample] (u)--node[wt,pos=0.5] {$s_1:4$} (x);
\draw[sample] (x)--node[wt,pos=0.5] {$s_2:2$} (v);
\draw[real] (v)--node[wt,pos=0.5] {$e_1:1$} (y);
\draw[real] (y)--node[wt,pos=0.5] {$e_2:3$} (u);
\draw[real,line width=1.2pt] (u)--node[wt] {$h:H$} (v);
\node[vertex,label=below left:$u$] at (u) {};
\node[vertex,label=below right:$v$] at (v) {};
\node[vertex,label=above:$x$] at (x) {};
\node[vertex,label=above:$y$] at (y) {};
\end{tikzpicture}

\smallskip
{\small (b) A hat; real arrival order $e_1,e_2,h$.}
\end{minipage}
\caption{Dashed edges belong to the sample; solid edges are real.
In (a), every real edge individually belongs to the optimum when
added to the sample. In (b), contracting $e_1$ allows greedy to
accept $e_2$, whereas ordinary greedy would drop $e_1$.}
\label{fig:looking-back}
\end{figure}

Many sample-based algorithms for matroid secretary exploit greedy monotonicity (\Cref{lem:greedy}), i.e., an element
$e\in\OPT(V)$ of the global optimum is also part of the
optimal solution on any subset containing $e$. This suggests
a na\"ive algorithm that discards the first roughly $n/2$
elements to use as a sample $S$ and then accepts an element $e$ if
and only if $e\in\OPT(S\cup\{e\})$. In the secretary setting, this
algorithm has good total value since 
it accepts every $e\in\OPT(V)\setminus S$, but its
accepted set need not be independent.
For example, consider the graphic matroid in
\Cref{fig:looking-back}(a): the sample consists of edges
$(u,a),(u,b),(u,c)$ of weight $1$, and the real elements are
$(a,b),(b,c),(c,a)$ of weight $2$. The na\"ive algorithm accepts
all three real edges, which form a cycle.

\medskip
\noindent\textbf{Greedy algorithms.}
A natural approach to go beyond the na\"ive algorithm maintains
an independent set $I_t$ at all times $t$, initialized to the
optimum solution on the sample, and containing all accepted
elements. It accepts the next real element $e$ if and only if the
optimum solution on $I_t\cup\{e\}$ \emph{after contracting the accepted
elements} contains $e$, and updates $I_t$ to this new solution
(with the contracted elements included).

The authors of \cite{BBSW-WINE21} considered this and related
``greedy'' algorithms. 
More formally, their algorithms maintain an independent set
$I_t$ that spans the previously observed elements and satisfies
two properties: (i)~$I_t$ contains all
accepted elements, and (ii)~$I_t$ contains no real element rejected
after the sampling stage. The choice of $I_t$ is flexible, but the
contracted-greedy acceptance rule above is mandatory. They show
that this class cannot be constant-competitive on graphic matroids, using hat-shaped instances.

A small hat illustrates the failure mechanism; see
\Cref{fig:looking-back}(b). The sample contains a $u$--$v$ path
with edge weights $4,2$, while the real elements form another
$u$--$v$ path with weights $1,3$, followed by a base edge of
arbitrarily large weight $H$. The greedy algorithm accepts the
weight-$1$ edge since it preserves independence. It then accepts
the weight-$3$ edge: after contracting the first acceptance, the
optimum keeps weights $4,3,1$ and drops the weight-$2$ sample.
The two accepted edges now connect $u$ to $v$, making the heavy
base edge infeasible.

The combinatorial secretary algorithm (\Cref{alg:matroidSec})
builds on two ideas.

\medskip
\noindent\textbf{Idea 1: Preserve processed greedy decisions (Step~\ref{alg:direct-test} of \Cref{alg:matroidSec}).}
Rather than making the greedy decision from a single independent
set $I_t$ after contracting the accepted elements, maintain a
reference set $C$
and only consider $e$ if the proposed update preserves greedy
membership on the processed elements. In particular, we \emph{do not
contract} the already accepted elements; instead, Step~\ref{alg:direct-test} ensures
that these greedy memberships do not change. This
immediately guarantees feasibility, since the accepted elements
always belong to the optimum of the current reference set.

On the hat above, after sampling $s_1,s_2$, consider the
processing order $e_1,e_2,h,s_1,s_2$, so both sample elements
are processed after all real arrivals.
The weight-$1$ edge $e_1$
is accepted, but inserting the weight-$3$ edge $e_2$ would change
the ordinary optimum from weights $4,2,1$ to $4,3,2$, removing
$e_1$. Step~\ref{alg:direct-test} therefore rejects this update, and the heavy base
can subsequently be accepted. Thus, unlike contracted greedy,
our algorithm can reject an element even when it improves the
optimum subject to retaining earlier acceptances.

Although this idea avoids the harmful exchange above, if we
update the reference set only when we accept an element,
the algorithm can get stuck on simple matroids. For example,
consider a uniform matroid of rank $r$ on $n=4r$ elements with
values in $[1,1+\varepsilon]$, with distinct perturbations to
avoid ties. In the secretary algorithm, this modification keeps
$C=S\cup\ALG$: sample elements are never deleted. Once the
smallest element of $\OPT(C)$ is an accepted element, every
improving insertion would remove it and is forbidden. The algorithm can then remain permanently stuck with many unused slots. A heuristic calculation suggests that it accepts only $O(\sqrt r)$ elements in expectation, rather than $\Omega(r)$.\footnote{Roughly, ignoring the extra restrictions from processed sample elements, after $\sqrt r$ acceptances the smallest accepted
element has only $O(\sqrt r)$ elements below it in the current
greedy solution, on average. Every further acceptance must
replace a sample element below this accepted element, and there
are only $O(\sqrt r)$ such elements left.}

\medskip
\noindent\textbf{Idea 2: Update reversibly (Step~\ref{alg:direct-test}(b) of \Cref{alg:matroidSec}).}
To overcome this rigidity, the algorithm updates
the reference set whenever the element passes Step~\ref{alg:direct-test},
even when it is not accepted. In the secretary algorithm, this
allows consistent deletions of sample elements. 
Such a deletion
can bring a smaller, not-yet-processed sample element
 into the optimum, allowing a
later real arrival to replace that sample without removing an
earlier acceptance. This also explains why we retain the whole
reference set rather than only its current optimum: sampled
elements outside that optimum can become relevant after later
updates.

Because Step~\ref{alg:direct-test} compares the same two optima in either direction, every permitted update is reversible; this is the basis of the snapshot lemma (\Cref{lem:snapshot}).

\section{Feasible Families for Special Classes}
\label{sec:special-class-families}

This appendix proves the special-class results in
\Cref{sec:lp-examples}. For uniform matroids, a threshold-and-thinning
construction and a matching midpoint lower bound establish the
probability guarantee in \Cref{cor:uniform-separation}.
For linear matroids, a recursive feasible family based on a
subspace invariant proves \Cref{thm:linear}.

\subsection{Uniform matroids}
\label{sec:uniform-lp-proof}

We prove the two probability bounds used in
\Cref{cor:uniform-separation}. Recall that $U_{k,n}$ has all
subsets of size at most $k$ as its independent sets. We first
recover the known $1-O(\sqrt{\log k/k})$ per-element probability
guarantee~\cite{SotoTV-SODA18}, then prove the matching lower bound.

\begin{proposition}\label{prop:uniform-lp}
For $n\geq k\geq2$, the probability LP for the uniform matroid
$U_{k,n}$ has value at least
$1-O(\sqrt{\log k/k})$, with an absolute constant independent
of $n$.
\end{proposition}
\begin{proof}
It suffices to consider $k\geq3$, since bounded values of $k$
can be absorbed into the constant. Set
\[
b=\sqrt{k\log k},\qquad
q=\frac{k-b}{k+6b},\qquad
r_i=\min\left\{i,\left\lfloor
          \frac{ki}{n}+3b\sqrt{\frac{i}{n}}
          \right\rfloor\right\}.
\]
Call an arrival at position $i$ eligible if it is among the
first $r_i$ elements of the observed value order.

Starting from $\mu_\emptyset(\emptyset)=1$, define the accepting
masses by
\[
y_{O,e}(A)
=
q\,\mu_{O-e}(A)\,
\ind\{e\text{ is eligible},\ |A|<k\},
\]
and define each $\mu_O$ recursively by consistency
\eqref{eq:lp-flow}. Capacity holds directly, and induction gives
nonnegative, normalized distributions supported on sets of size at most
$k$. The realizing algorithm accepts each eligible arrival with
probability $q$, independently, until it has accepted $k$ elements.

Consider this algorithm without the limit of $k$ acceptances,
and let $Z$ be its total number of acceptances. 
The successive relative ranks, with rank $1$ denoting the largest
element observed so far, are independent, with the rank at
position $i$ uniform on $[i]$. Thus $Z$ is a sum of independent Bernoulli variables,
and
\[
\E Z
=q\sum_{i=1}^n\frac{r_i}{i}
\leq q\left(k+\frac{3b}{\sqrt n}
                     \sum_{i=1}^n i^{-1/2}\right)
\leq q(k+6b)=k-b.
\]
A Chernoff bound therefore gives
\[
\Pr[Z\geq k]
\leq
\exp\left(-\frac{(k-\E Z)^2}{2k}\right)
\leq k^{-1/2}.
\]

Fix $e\in\OPT(V)$ and condition on its arrival at position $i$.
Its relative rank is $1+H$, where $H$ counts the larger elements
among the preceding $i-1$ arrivals. Thus $H$ is hypergeometric,
with mean at most $a=ki/n$. Suppose $a\geq\log k$.
If $r_i=i$, eligibility is automatic. Otherwise, failure of
the rank test implies
\[
H\geq r_i
\geq a+3\sqrt{a\log k}-1
\geq a+2\sqrt{a\log k}.
\]
The hypergeometric Chernoff bound gives
\[
\Pr[e\text{ is ineligible}\mid e\text{ arrives at }i]
\leq
\exp\left(
-\frac{4a\log k}
       {2(a+\frac23\sqrt{a\log k})}
\right)
\leq k^{-1}.
\]
The positions with $ki/n<\log k$ have total probability at most
$\log k/k$. Hence $e$ passes the rank test and its independent acceptance coin succeeds with probability at least
$q(1-\log k/k-k^{-1})$.
Outside the event $Z\geq k$, all such acceptances fit within
capacity. Consequently,
\[
\Pr[e\in\ALG]
\geq
q\left(1-\frac{\log k}{k}-\frac1k\right)-\frac1{\sqrt k}
=
1-O\left(\sqrt{\frac{\log k}{k}}\right).
\]
This holds for every full value order and every
$e\in\OPT(V)$, proving all terminal inclusion constraints.
\end{proof}

\begin{proposition}[Matching probability lower bound]
\label{prop:uniform-lp-lower}
For $n\geq2k$ and $k\geq128$, the probability LP for $U_{k,n}$
has value at most
$1-\frac1{64}\sqrt{\frac{\log k}{k}}$.
Consequently, together with \Cref{prop:uniform-lp}, its value is
$1-\Theta(\sqrt{\log k/k})$, uniformly over $n\geq2k$ as
$k\to\infty$.
\end{proposition}
\begin{proof}
First let $n=2k$. By \Cref{thm:polytope-realization}, a feasible
point of value $1-\varepsilon$ yields an ordinal algorithm
selecting each of the $k$ largest elements with probability at
least $1-\varepsilon$. Average the algorithm over permutations
of the element labels; this preserves its guarantee.
Number the elements $1,\ldots,2k$ by decreasing value, for analysis
only, and let $S$ be the first $k$ arrivals.

The relative arrival order within $S$ is independent of its set
of global ranks. By symmetry and ordinality, there are numbers
$a_r\in[0,1]$ such that, conditional on any possible $S$, its
$r$-th largest element has been accepted by this midpoint with
probability $a_r$.
Put
\[
X=|S\cap[k]|,\qquad
 g_r=\Pr[X=r]=\frac{\binom{k}{r}^2}{\binom{2k}{k}},\qquad
 b_r=\Pr[X<r].
\]
Let $W$ count the accepted elements of $S$ outside $[k]$, and let
$D$ be the event that element $k$ belongs to $S$ but is rejected.
The terminal inclusion constraints and the capacity bound give
$\E W\leq k\varepsilon$, while $\Pr[D]\leq\varepsilon$.
The $r$-th largest element of $S$ is outside $[k]$ exactly when
$X<r$. Conditional on $X=r$, element $k$ belongs to $S$ with
probability $r/k$ and is then its $r$-th largest element. Hence
\[
\E W=\sum_{r=1}^k b_ra_r,
\qquad
\Pr[D]=\sum_{r=1}^k\frac rk g_r(1-a_r).
\]
Combining these two inequalities eliminates the algorithm:
\begin{equation}\label{eq:uniform-midpoint-dual}
2k\varepsilon
\geq\E W+k\Pr[D]
=\sum_{r=1}^k\bigl[b_ra_r+rg_r(1-a_r)\bigr]
\geq\sum_{r=1}^k\min\{b_r,rg_r\}.
\end{equation}

Set $m=\lceil k/2\rceil$ and
$L=\lfloor\sqrt{k\log k}/8\rfloor$.
For $r=m+1,\ldots,m+L$, symmetry gives $b_r\geq1/2$.
We claim that $g_r\geq1/k$ throughout this interval.
Indeed, $g_m$ is a maximum point mass; since
$\operatorname{Var}(X)\leq k/4$, Chebyshev's inequality gives
$g_m\geq1/(4\sqrt{k})$.
For $1\leq s\leq L\leq m/2$, the binomial-coefficient ratio and
$\log((1-x)/(1+x))\geq-3x$ for $0\leq x\leq1/2$ give
\[
\frac{g_{m+s}}{g_m}
=\prod_{j=1}^s\left(\frac{k-m-j+1}{m+j}\right)^2
\geq\exp\left(-\frac{6s(s+1)}k\right)
\geq k^{-3/16}.
\]
Thus $g_{m+s}\geq\frac14 k^{-11/16}\geq1/k$ for $k\geq128$.
On these $L$ indices, both $b_r$ and $rg_r$ are at least $1/2$.
Equation~\eqref{eq:uniform-midpoint-dual} therefore yields
\[
\varepsilon\geq\frac{L}{4k}
\geq\frac1{64}\sqrt{\frac{\log k}{k}}.
\]

Finally, an ordinal algorithm on $n>2k$ elements can be simulated
on $2k$ real elements by interleaving $n-2k$ fictitious elements,
all smaller than the real ones, at uniformly random positions
and in a uniformly random relative order. Fictitious selections
are discarded but still consume capacity. The combined order is
uniform, and its $k$ largest elements are real, so the simulation
preserves the per-element guarantee. The bound for $2k$ elements
therefore applies to every $n\geq2k$.
\end{proof}

\begin{remark}[A midpoint certificate]
Equation~\eqref{eq:uniform-midpoint-dual} is a dual certificate
for a relaxation of the probability LP: retain only the
symmetrized midpoint marginals $a_r$, the terminal guarantee for
element $k$, and the sum of the terminal guarantees for all
$k$ optimum elements. It gives the finite bound
\[
\frac1{\rho(U_{k,2k})}
\leq 1-\frac1{2k}\sum_{r=1}^k
\min\left\{\Pr[X<r],\,r\Pr[X=r]\right\},
\qquad X\sim\operatorname{Hypergeometric}(2k,k,k).
\]
No threshold or monotonicity assumption on the algorithm is used.
\end{remark}

\subsection{Linear matroids}
\label{sec:linear}\label{sec:linear-proofs}

We prove \Cref{thm:linear} by recursively constructing a family
in $\mathcal P(\calM)$, using a subspace invariant in place of the
survivor distributions of \Cref{sec:survival}. Throughout, $\calM$ is a
linear matroid on $n\geq2$ elements, known in advance, and
$1\leq s<n$ is a fixed cutoff.
Recall that $c_n(s)$, defined in \Cref{sec:lp-examples}, is the
success probability of the classical cutoff rule. Its maximum
over $s$ is the optimal success probability for the classical
$n$-element secretary problem~\cite{Lindley1961,Dynkin1963},
and hence equals $1/\rho(U_{1,n})$.
The construction below is finite; no polynomial running-time
bound is asserted.

\medskip\noindent\textbf{The subspace invariant.}
By a theorem of Rado~\cite{Rado1957}, every linear matroid has a representation over a finite field. Fix such a representation $(u_e)_{e\in V}$ in $\mathbb F^d$, for a finite field $\mathbb F$; the algorithm can find one by exhaustive search before the arrivals. For $A\subseteq V$, let $\langle A\rangle$ be the span of $\{u_e:e\in A\}$, and for a vector $u$ let $\langle u\rangle$ be its span. For a subspace $U\leq\mathbb F^d$, the \emph{load} of $A$ at $U$ is $\ell_U(A)=\dim(\langle A\rangle\cap U)$. A distribution $\mu$ on independent sets is \emph{$\beta$-bounded} if
\begin{equation}\label{eq:lin-bounded}
\E_{A\sim\mu}[\ell_U(A)]\leq\beta\dim U\qquad\text{for every subspace }U\leq\mathbb F^d.
\end{equation}
There are finitely many subspaces, so this is a finite system of linear inequalities.

To motivate the invariant, recall the cutoff rule in rank one. After $t\geq s$ arrivals it has accepted an element with probability exactly $\beta_t=1-s/t$, and it accepts a record arriving at time $t>s$ with probability $a_t=s/(t-1)$. These numbers satisfy
\begin{equation}\label{eq:lin-schedule}
\beta_{t-1}+a_t=1,\qquad\beta_{t-1}+\frac{a_t}t=\beta_t\qquad(t>s).
\end{equation}
For a nonloop $e$ of a linear matroid, the constraint \eqref{eq:lin-bounded} at the line spanned by $u_e$ states that $u_e\in\langle A\rangle$ with probability at most $\beta$. With $\beta=\beta_{t-1}$, at least $a_t$ of the probability mass can therefore accept $e$ at time $t$. Lines bound the mass available for an acceptance but not its cost: bounding the increase of the load at $U$ requires the constraint at $U+\langle u_e\rangle$. To maintain many constraints at once, the proof replaces families of subspaces by chains of intersections and sums, and intersections of subspaces spanned by elements need not be spanned by elements. The invariant is therefore imposed on every subspace.

\begin{definition}[The family for linear matroids]\label{def:linear-family}
Fix $1\leq s<n$. For $|O|\leq s$, let $\mu_O$ be the point mass at $\emptyset$ and $y_{O,\cdot}=0$. For $|O|=t>s$, assume that $\mu_{O-e}$ has been defined for every $e\in O$, and let $y_{O,\cdot}$ be a canonical (for instance, lexicographically smallest) solution of the following finite linear system:
\begin{enumerate}[label=(\alph*)]
\item $0\leq y_{O,e}(A)\leq\mu_{O-e}(A)$ for every $e\in O$ and independent $A\subseteq O-e$;
\item $y_{O,e}(A)=0$ unless $e\in\OPT(O)$ and $u_e\notin\langle A\rangle$;
\item $\sum_Ay_{O,e}(A)=a_t$ for every $e\in\OPT(O)$;
\item the distribution $\mu_O$ defined by \eqref{eq:lp-flow} is $\beta_t$-bounded.
\end{enumerate}
\end{definition}

The family depends only on the ordered sets and the fixed representation. If the systems are feasible, normalization follows inductively from \eqref{eq:lp-flow}; together with (a) and (b), this places the family in $\mathcal P(\calM)$. By \Cref{thm:polytope-realization}, it is realized by \eqref{eq:lp-rule}. The only issue is feasibility of the system for $t>s$.

\begin{proposition}\label{prop:linear-feasible}
For every ordered set $O$ with $|O|>s$, the system in \Cref{def:linear-family} is feasible. For $|O|\geq s$, the distribution $\mu_O$ is a $\beta_{|O|}$-bounded probability distribution on independent subsets of $O$.
\end{proposition}

We prove the proposition below. Accepting $e$ at a state $A$ raises the load at $U$ exactly when $u_e\in\langle A\rangle+U$. Along a chain of subspaces these sets of states are nested, so a single order of filling the acceptance mass preserves every constraint of the chain. Any weighted family of subspaces can be replaced by a chain without decreasing any load or changing the total dimension, and a minimax argument then gives one solution for all subspaces. The bound holds only on average over the identity of the last arrival. This is where the uniformly random order is used: the candidates $\OPT(O)$ form an independent set, which has at most $\dim U$ vectors in any subspace $U$.

\medskip\noindent\textbf{Admitting an element along a chain.}
The following two identities hold for subspaces $Y,U$ and a vector $u$. Both follow from $\dim(X\cap Y)=\dim X+\dim Y-\dim(X+Y)$:
\begin{alignat}{2}
\dim((Y+\langle u\rangle)\cap U)-\dim(Y\cap U)&=\ind\{u\in Y+U\}\qquad&&\text{if }u\notin Y,\label{eq:lin-id1}\\
\dim(Y\cap(U+\langle u\rangle))-\dim(Y\cap U)&=\ind\{u\in Y+U\}\qquad&&\text{if }u\notin U.\label{eq:lin-id2}
\end{alignat}
By \eqref{eq:lin-id1}, accepting $e$ at a state $A$ with $u_e\notin\langle A\rangle$ raises the load at $U$ by one exactly when $u_e\in\langle A\rangle+U$. We call such states \emph{expensive} for $U$. By \eqref{eq:lin-id2}, if $u_e\notin U$, the probability that $u_e\in\langle A\rangle+U$, that is, of an expensive or infeasible state, is the difference between the expected loads at $U+\langle u_e\rangle$ and at $U$.

An \emph{addition rule} for $e$ on a distribution $\mu$ assigns masses $0\leq z(A)\leq\mu(A)$, with $z(A)=0$ whenever $u_e\in\langle A\rangle$. It moves mass $z(A)$ from $A$ to $A+e$; its total mass is $\sum_Az(A)$.

\begin{lemma}[Chain admission]\label{lem:chain}
Let $\mu$ be $\beta$-bounded, let $u_e\neq0$, and let $0\leq a\leq1-\beta$. For every chain of subspaces $C_1\subseteq\cdots\subseteq C_m$, there is an addition rule of total mass $a$ whose updated distribution $\mu'$ satisfies $\E_{\mu'}[\ell_U]\leq\beta\dim U+a\ind\{u_e\in U\}$ for every $U$ in the chain.
\end{lemma}
\begin{proof}
For a chain member $U$, let $D_U=\{A:u_e\in\langle A\rangle+U\}$ be its set of expensive or infeasible states. These sets are nested along the chain. Give each feasible state the index $i(A)=\min\{i:A\in D_{C_i}\}$, or $m+1$ if there is no such $i$. The feasible mass is $1-\Pr_\mu[u_e\in\langle A\rangle]\geq1-\beta\geq a$, by \eqref{eq:lin-bounded} at $\langle u_e\rangle$. Place mass $a$ on feasible states in decreasing order of index, splitting the last atom if necessary.

Fix a chain member $U$ and let $p_U=\mu(D_U)$. Every state outside $D_U$ is feasible and is filled before any state of $D_U$. Hence the mass placed in $D_U$ is $(a-1+p_U)_+$, and by \eqref{eq:lin-id1} this is the increase of the expected load at $U$. If $u_e\in U$, then $D_U$ contains every state and the increase is $a$. Otherwise, \eqref{eq:lin-id2} gives $\E_\mu[\ell_U]+p_U=\E_\mu[\ell_{U+\langle u_e\rangle}]\leq\beta(\dim U+1)$. Using $\beta+a\leq1$, the updated load is
\[
\max\bigl\{\E_\mu[\ell_U],\ \E_\mu[\ell_U]+p_U+a-1\bigr\}\leq\beta\dim U.\qedhere
\]
\end{proof}

\Cref{lem:chain} needs a chain; the next lemma reduces any weighted family of subspaces to one.

\begin{lemma}[Uncrossing]\label{lem:uncross}
For nonnegative rational weights $(\lambda_U)_{U\leq\mathbb F^d}$, there are a chain $C_1\subseteq\cdots\subseteq C_m$ and nonnegative weights $\kappa_1,\ldots,\kappa_m$ such that, for every subspace $Y$,
\[
\sum_U\lambda_U\dim(Y\cap U)\leq\sum_i\kappa_i\dim(Y\cap C_i),
\qquad
\sum_U\lambda_U\dim U=\sum_i\kappa_i\dim C_i.
\]
\end{lemma}
\begin{proof}
Clear denominators and regard the weights as a multiset of subspaces; at the end, $\kappa_i$ will be the multiplicity of $C_i$ divided by the common denominator. While two members $U,T$ are incomparable, replace them by $U\cap T$ and $U+T$. This preserves the total dimension. It does not decrease the left side for any $Y$: the subspaces $Y\cap U$ and $Y\cap T$ have intersection $Y\cap U\cap T$ and sum contained in $Y\cap(U+T)$. With $d_0=\dim(U\cap T)$, the sum of squared dimensions increases by $2(\dim U-d_0)(\dim T-d_0)>0$. This integer is bounded, so the process terminates, at a chain.
\end{proof}

\begin{lemma}[Averaged extension]\label{lem:averaged}
Let $|O|=t\geq1$, let $B\subseteq O$ be independent, and for each $e\in O$ let $\mu_e$ be a $\beta$-bounded distribution on independent subsets of $O-e$. If $0\leq a\leq1-\beta$, there are addition rules $z_e$ for $\mu_e$, of total mass $a$ for $e\in B$ and zero for $e\notin B$, such that the average over $e\in O$ of the updated distributions is $(\beta+a/t)$-bounded.
\end{lemma}
\begin{proof}
Let $P$ be the set of families $z=(z_e)_{e\in O}$ of addition rules with the prescribed total masses. It is a nonempty polytope: the elements of the independent set $B$ are nonloops, so by the availability bound in the proof of \Cref{lem:chain}, each $\mu_e$ with $e\in B$ has feasible mass at least $1-\beta\geq a$. For $z\in P$ and a subspace $U$, let $g_U(z)$ be the expected load at $U$ of the averaged updated distribution; it is affine in $z$. We need $z\in P$ with $g_U(z)\leq(\beta+a/t)\dim U$ for all $U$.

Fix nonnegative rational weights $\lambda$, and obtain a chain from \Cref{lem:uncross}. Apply \Cref{lem:chain} to this chain in every row $e\in B$, and add nothing in the other rows. For each chain member $C_i$, averaging the row bounds gives
\[
g_{C_i}(z)\leq\beta\dim C_i+\frac at\,\bigl|\{e\in B:u_e\in C_i\}\bigr|\leq\Bigl(\beta+\frac at\Bigr)\dim C_i,
\]
since the independent set $B$ has at most $\dim C_i$ vectors in $C_i$. Taking expectations in the pointwise domination of \Cref{lem:uncross} gives $\sum_U\lambda_U g_U(z)\leq(\beta+a/t)\sum_U\lambda_U\dim U$.

Thus $\min_{z\in P}\sum_U\lambda_U\bigl(g_U(z)-(\beta+a/t)\dim U\bigr)\leq0$ for every rational nonnegative $\lambda$, and hence for every $\lambda$ in the probability simplex: the minimum over the compact set $P$ of a jointly continuous function is continuous in $\lambda$. The function is linear in $\lambda$ and affine in $z$, on a product of two polytopes. By the minimax theorem~\cite{vonNeumann1928}, some $z\in P$ makes the maximum over the simplex nonpositive, so every constraint holds at that $z$.
\end{proof}

The lemma controls only the average of the updated distributions, not each row separately; the independence of $B$ bounds the average increase.

\begin{proof}[Proof of \Cref{prop:linear-feasible}]
We show by induction on $t=|O|\geq s$ that the system in \Cref{def:linear-family} is feasible when $t>s$ and that $\mu_O$ is a $\beta_t$-bounded probability distribution on independent sets. For $t=s$, the point mass at $\emptyset$ is $0$-bounded and $\beta_s=0$. For $t>s$, apply \Cref{lem:averaged} with $B=\OPT(O)$, $\mu_e=\mu_{O-e}$, $\beta=\beta_{t-1}$, and $a=a_t=1-\beta_{t-1}$. Elements of the independent set $B$ are nonloops, so $u_e\neq0$. The resulting rules satisfy (a)--(c), and (d) holds because $\beta_{t-1}+a_t/t=\beta_t$ by \eqref{eq:lin-schedule}. The distribution defined by \eqref{eq:lp-flow} is the averaged updated distribution, a probability distribution on independent sets.
\end{proof}

\begin{proof}[Proof of \Cref{thm:linear}]
By \Cref{prop:linear-feasible}, the family in \Cref{def:linear-family} is well defined and belongs to $\mathcal P(\calM)$. Apply \Cref{thm:polytope-realization}.

Fix $e\in\OPT(V)$. By \Cref{lem:greedy}(i), $e\in\OPT(O)$ for every observed ordered set $O$ containing $e$. Conditional on $e$ arriving at position $t$ and on the observed order $O$, its acceptance probability is therefore $\sum_Ay_{O,e}(A)=a_t$ when $t>s$, and zero when $t\leq s$. Since $e$ arrives at each position with probability $1/n$,
\[
\Pr[e\in\ALG]=\frac1n\sum_{t=s+1}^n\frac s{t-1}=c_n(s).
\]
Choosing $s$ optimally gives $\rho(\calM)\leq\rho(U_{1,n})$. It remains to show $\max_sc_n(s)>1/\ee$. For $n=2$, $c_2(1)=1/2$. For $n\geq3$, let $\zeta=n/\ee$, $j=\lfloor\zeta\rfloor$, and $\vartheta=\zeta-j$; then $1\leq j<j+1\leq n-1$. Using $\ln x\leq x-1$, first for $x=(j+1)/\zeta$ and then for $x=(i+1)/i$, with strict inequality in the second case,
\[
\frac1\ee=\frac\zeta n\ln\frac n\zeta<\frac{j+1-\zeta}n+\frac\zeta n\sum_{i=j+1}^{n-1}\frac1i=(1-\vartheta)c_n(j)+\vartheta c_n(j+1).
\]
Hence one of the two cutoffs achieves more than $1/\ee$.
\end{proof}

In a loopless matroid of rank one, the system in \Cref{def:linear-family} has a unique solution and the algorithm is exactly the classical cutoff rule. The subspace invariant replaces the single number $\beta_t$ by a bound on every subspace.

\medskip\noindent\textbf{Open problem.}
Does every matroid $\calM$ on $n\geq2$ elements with a nonloop, known in advance, satisfy $\rho(\calM)\leq\rho(U_{1,n})$? Equivalently, is there always an ordinal algorithm that selects every element of $\OPT(V)$ with probability at least $\max_s c_n(s)$? \Cref{thm:linear} answers this affirmatively for linear matroids. The computations of \Cref{sec:lp-computations} suggest the weaker comparison for the ordinal expected-value objective.

\section{Computational Insights}
\label{sec:lp-computations}

This appendix gives the numerical findings and computational
reductions summarized in \Cref{sec:lp-examples}.
Write $\rho_{\mathrm{ord}}(\calM)$ for the optimal expected-value
competitive ratio among ordinal algorithms on the known matroid
$\calM$. By \Cref{cor:lp-ordinal}, its reciprocal is the value of
the program with prefix constraints. The ratio $\rho(\calM)$ in
\Cref{sec:lp-objective} instead requires a guarantee for every
element of $\OPT(V)$. The computations reported here concern
$\rho_{\mathrm{ord}}$.

\subsection{Numerical findings and coverage}

Using the catalogue of matroids on at most nine elements of
Mayhew and Royle~\cite{MR-JCTB08}, we solved the program with
Gurobi in floating-point arithmetic. The coverage is summarized
below.\footnote{The implementation and its outputs are available
at \url{https://github.com/TlatoaniHJ/MatroidSecretary}.}

\begin{center}
\small
\begin{tabular}{@{}cll@{}}
\toprule
Ground-set size & Positive-rank matroids solved
                & Nonrepresentable matroids solved\\
\midrule
$7$ & All & None exist\\
$8$ & $1655$ of $1723$ & $43$ of $44$\\
\bottomrule
\end{tabular}
\end{center}

The $68$ unsolved eight-element matroids all have rank four,
and all but one have no nontrivial automorphism. For these $67$
matroids, the symmetry reduction described below does not shrink
the programs. Every matroid on at most seven elements is representable
over some field~\cite{Oxley2011}; on eight elements, there are
$44$ matroids not representable over any field~\cite{MR-JCTB08}.

Every computed ratio was below $\ee$. More strongly, every tested
instance satisfied
\[
\rho_{\mathrm{ord}}(\calM)
\leq \rho_{\mathrm{ord}}(U_{1,n}),
\]
with equality, up to numerical precision, only for $U_{1,n}$
itself. For $n=8$, the computed values of
$1/\rho_{\mathrm{ord}}(\calM)$ exceed
$1/\rho_{\mathrm{ord}}(U_{1,8})$ by at least $0.004$ except for
$U_{1,8}$, far above the solver tolerance.
These are floating-point observations, not exact certificates.
For $n\leq7$, representability and \Cref{thm:linear} also prove
$\rho_{\mathrm{ord}}(\calM)\leq\rho_{\mathrm{ord}}(U_{1,n})<\ee$
for every positive-rank matroid. The solved eight-element
instances include $43$ nonrepresentable matroids, beyond the
scope of that theorem.

These computations predate the proofs in this paper. At the
time, they were evidence for the strong conjecture beyond the
classes for which it was known, and they guided the search for
the single-sample prophet inequality of \Cref{sec:ssmpi-tight}
and the construction of \Cref{sec:linear}.

\subsection{LP representation and reductions}

\paragraph{Acceptance marginals.}
For a full value order $\omega$ and an element $e$, define
\begin{equation}\label{eq:computational-marginal}
p_{\omega,e}(y)=
\sum_{\substack{S\subseteq V\\e\in S}}
\frac{1}{n\binom{n-1}{|S|-1}}
\sum_{A\subseteq S-e}y_{\omega|_S,e}(A).
\end{equation}
The element $e$ arrives at each position with probability $1/n$;
conditional on position $|S|$, its observed set is uniform among
the $\binom{n-1}{|S|-1}$ sets of that size containing $e$.
Thus \eqref{eq:computational-marginal} is its inclusion probability,
and equals $\sum_{A\ni e}\mu_\omega(A)$ for a feasible family.
The ordinal objective can therefore be imposed through
\[
\sum_{e\in H}p_{\omega,e}(y)\geq\gamma\rank(H)
\qquad\text{for every full order $\omega$ and prefix $H$}.
\]
This expression uses the total accepting mass for each element,
which is preserved by the state aggregation below.

\paragraph{Computational reductions.}
We use the unconditional state and accepting-transition probabilities
\[
\frac{\mu_O(A)}{\binom n{|O|}}
\qquad\text{and}\qquad
\frac{y_{O,e}(A)}{n\binom{n-1}{|O|-1}},
\]
respectively. This rescaling preserves the optimum.
Three further reductions make the program tractable
for small matroids. First, only prefixes ending with an element
of $\OPT(\omega)$ need constraints: between successive such
prefixes, the rank is unchanged and the left side only increases.
Second, averaging a feasible family over the automorphism group
of $\calM$ preserves its objective, so an optimal family can be
restricted to orbit representatives.

Third, call independent sets $A,A'\subseteq O$ \emph{equivalent at
$O$} if $A\cup J$ and $A'\cup J$ have the same independence status
for every $J\subseteq V\setminus O$. They admit the same feasible
extensions. A class $Q$ at $O-e$ lies in one class at $O$ after
rejecting $e$. Moreover, either every $A\in Q$ can accept $e$ or
none can; in the former case all sets $A+e$, for $A\in Q$, lie in
one class at $O$. Both facts follow directly from the definition,
using extension sets $J$ and $J+e$, respectively.

The aggregated program has a probability variable for each class
at each ordered set and an accepting-mass variable for each
predecessor class. It imposes the initial point mass at
$\emptyset$, class capacities, independent support, and
\eqref{eq:lp-flow} summed over each destination class. In
\eqref{eq:computational-marginal}, replace the sum over states by
the sum over predecessor classes. Summing a feasible family over
classes gives an aggregated feasible solution with exactly the
same accepting totals and objective. Conversely, at a state in
a predecessor class $Q$, accept with probability
$y_{O,e}(Q)/\mu_{O-e}(Q)$, with $0/0=0$. The class determines both
possible destination classes, so induction in \eqref{eq:lp-flow}
shows that this algorithm realizes the prescribed class distributions
and accepting masses. Formula \eqref{eq:computational-marginal}
then gives the same per-element marginals and hence the same
objective. The aggregation preserves the optimum; it need not
preserve each individual terminal set probability.

\subsection{The truncation hypothesis}

Before the general result was known, the computations also suggested structural statements that would imply the strong conjecture. We describe one of them. For $1\leq k\leq\rank(\calM)$, the \emph{truncation} $\calM^{(k)}$ has the independent sets of $\calM$ of size at most $k$. The truncation $\calM^{(1)}$ is the classical secretary problem on the nonloops of $\calM$, so $\rho_{\mathrm{ord}}(\calM^{(1)})\leq\ee$. We say that $\calM$ satisfies \emph{truncation monotonicity} if $\rho_{\mathrm{ord}}(\calM^{(k)})$ is nonincreasing in $k$; then $\rho_{\mathrm{ord}}(\calM)\leq\rho_{\mathrm{ord}}(\calM^{(1)})\leq\ee$, which would have implied the strong conjecture. The computed values satisfy truncation monotonicity, up to solver tolerance. \Cref{sec:truncation} shows that it holds, strictly, for uniform matroids, but fails for the graphic matroid of a large complete bipartite graph $K_{2,N}$. The failure of this route is part of why the construction of \Cref{sec:linear} controls the span of the accepted set directly instead of inducting on the rank.

Truncation monotonicity would also imply the comparison with rank one observed above. Indeed, $\calM^{(1)}$ is the classical problem on the $n'\leq n$ nonloops of $\calM$, and inserting $n-n'$ fictitious elements, as in the proof of \Cref{prop:trunc-three}, shows that $\rho_{\mathrm{ord}}(U_{1,n'})\leq\rho_{\mathrm{ord}}(U_{1,n})$. The counterexample does not contradict this weaker statement: there, $\rho_{\mathrm{ord}}(\calM)$ is only shown to exceed $100/53$, while $\rho_{\mathrm{ord}}(U_{1,n})$ is close to $\ee$. \Cref{thm:linear} proves it for every linear matroid.

\section{Ordinal Ratios under Truncation}
\label{sec:truncation}

This appendix proves the two statements about truncation mentioned in \Cref{sec:lp-computations}. Recall that $\rho_{\mathrm{ord}}(\calM)$ is the optimal ordinal competitive ratio of a known matroid $\calM$, and that $\calM^{(k)}$ is its truncation to rank $k$. Truncation monotonicity asserts that $\rho_{\mathrm{ord}}(\calM^{(k)})\leq\rho_{\mathrm{ord}}(\calM^{(k-1)})$ for $2\leq k\leq\rank(\calM)$. \Cref{thm:trunc-uniform} proves it, with strict inequality, for uniform matroids. \Cref{thm:trunc-graphic} refutes it for a graphic matroid; there, the bound on $\calM$ holds even for algorithms that observe numerical values.

\subsection{Uniform matroids}

For $1\leq k\leq n$, let $r(n,k)$ be the largest expected fraction of the $k$ largest elements that a possibly randomized algorithm can select, with capacity $k$, when $n$ elements arrive in uniformly random order and the algorithm observes only the relative order of the arrived elements. The maximum exists, and is attained by a deterministic rule, because the problem is a finite Markov decision problem.

Let $R_t\in\{1,\ldots,t\}$ be the \emph{relative rank} of the $t$-th arrival among the first $t$, with $1$ the largest. The relative ranks are independent, $R_t$ is uniform on $\{1,\ldots,t\}$, and conditional on $R_1,\ldots,R_t$ the set of global ranks of the first $t$ arrivals is a uniform $t$-subset of $\{1,\ldots,n\}$. The $t$-th arrival is the $R_t$-th largest of this subset. Hence, conditional on the entire ordinal history, it is among the $k$ largest elements with probability
\begin{equation}\label{eq:trunc-posterior}
p_k(t,R_t),\qquad p_k(t,s)=\Pr[\Xi_{k,t}\geq s],\quad\Xi_{k,t}\sim\operatorname{Hypergeometric}(n,k,t),
\end{equation}
the number of successes among $t$ draws without replacement from $n$ items, $k$ of which are successes.

\begin{lemma}\label{lem:trunc-topk}
For $1\leq k\leq n$, $\rho_{\mathrm{ord}}(U_{k,n})=1/r(n,k)$.
\end{lemma}
\begin{proof}
Let an ordinal algorithm be $c$-competitive on $U_{k,n}$. Run it on a uniformly random relabeling of the elements; every permutation of the ground set is an automorphism of $U_{k,n}$, so this preserves feasibility. For each value order, its behavior is an average of its behavior on other value orders, so it remains $c$-competitive, and the sequence of labels it observes is now independent of $(R_1,\ldots,R_n)$. The labels can therefore be generated by internal randomness, so the relabeled algorithm observes only relative order, as in the definition of $r(n,k)$. \Cref{cor:lp-ordinal}, applied to the prefix of the $k$ largest elements, shows that it selects at least $k/c$ of them in expectation. Hence $r(n,k)\geq1/c$.

Conversely, since the relative ranks are independent, the optimal continuation value depends only on the time and the number of free slots. By backward induction, there is an optimal deterministic rule for $r(n,k)$ that accepts the $t$-th arrival, when $b\geq1$ slots remain, exactly when $p_k(t,R_t)$ is at least a threshold depending only on $t$ and $b$. Since $p_k(t,s)$ is nonincreasing in $s$, a smaller relative rank is accepted whenever a larger one is. We claim that under this rule, the probability that the element of global rank $j$ is selected is nonincreasing in $j$. Pair each arrival order with the order exchanging the positions of the elements of ranks $j$ and $j+1$. Every other element compares identically with these two elements. Hence the two runs see the same relative ranks, except at the later of the two positions, where the relative rank is smaller by one when the later element has rank $j$. The decisions coincide before that position, and at that position the run with the smaller relative rank accepts whenever the other does. Summing over the pair shows that rank $j$ is selected at least as often as rank $j+1$, and averaging over all pairs proves the claim.

Consequently, for a prefix $H$ of size $m\leq k$, the average selection probability over $H$ is at least that over the $k$ largest elements, which is $r(n,k)$. For $m>k$, the expected number selected from $H$ is at least $kr(n,k)$. In both cases it is at least $r(n,k)\rank(H)$, and \Cref{cor:lp-ordinal} gives $\rho_{\mathrm{ord}}(U_{k,n})\leq1/r(n,k)$.
\end{proof}

We reformulate $r(n,k)$ as an optimal stopping problem. For independent nonnegative random variables $Z=(Z_1,\ldots,Z_n)$ with finite support, let $\Phi_b(Z)$ be the largest expected total of at most $b$ selections made online, where $Z_t$ is observed at time $t$. Let $Z^k_t=p_k(t,R_t)$. Then
\begin{equation}\label{eq:trunc-stopping}
k\,r(n,k)=\Phi_k(Z^k).
\end{equation}
Indeed, by \eqref{eq:trunc-posterior}, the expected number of the $k$ largest elements selected by a rank-observing algorithm is the expected sum of $Z^k_t$ over the selected times. Every rule based on $Z^k$ is rank-observing. Conversely, by independence, the Bellman decision of a rank-observing rule at time $t$ depends on $R_t$ only through $Z^k_t$, so observing the relative ranks does not increase the optimal value.

\begin{lemma}[Comparison]\label{lem:trunc-compare}
Let $Z,Z'$ have independent coordinates with $\E[(Z_t-c)_+]\leq\E[(Z'_t-c)_+]$ for every $t$ and every $c\geq0$. Then $\Phi_b(Z)\leq \Phi_b(Z')$ for every $b$.
\end{lemma}
\begin{proof}
Let $v_t(b)$ be the optimal value from time $t$ onward with $b$ slots, with $v_{n+1}=0$ and $v_t(0)=0$. Then $v_t(b)\geq v_t(b-1)$ and
\[
v_t(b)=v_{t+1}(b)+\E\bigl[(Z_t-[v_{t+1}(b)-v_{t+1}(b-1)])_+\bigr].
\]
The map $(x,x')\mapsto x+\E[(Z_t-(x-x'))_+]$ is nondecreasing in both arguments, and the threshold $x-x'$ is nonnegative. Backward induction on $t$ gives $v_t(b)\leq v'_t(b)$ for all $b$.
\end{proof}

\begin{lemma}[Thinning and capacity]\label{lem:trunc-thin}
Let $1\leq m<K$, and let $\xi_1,\ldots,\xi_n$ be independent Bernoulli$(m/K)$ variables, independent of $Z$. Then
\[
\frac Km\,\Phi_m\bigl((\xi_tZ_t)_t\bigr)\leq(1-\delta)\Phi_K(Z)+\delta \Phi_m(Z),\qquad\delta=K^{1-n}.
\]
\end{lemma}
\begin{proof}
Give each time an independent uniform color $C_t\in\{1,\ldots,K\}$. For each set $J$ of $m$ colors, the variables $Z_t\ind\{C_t\in J\}$ form an $(m/K)$-thinning of $Z$. Run on each of them the same optimal deterministic $m$-selection rule, chosen never to select a zero reward, and let $I^J_t$ indicate its selections. Let $a_t=\binom{K-1}{m-1}^{-1}\sum_JI^J_t$. Only the $\binom{K-1}{m-1}$ sets containing $C_t$ can select at time $t$, so $0\leq a_t\leq1$. Each copy selects at most $m$ times, so $\sum_ta_t\leq K$. The expected value $\E[\sum_ta_tZ_t]$ equals $\binom Km\binom{K-1}{m-1}^{-1}\Phi_m((\xi_tZ_t)_t)=\frac Km\Phi_m((\xi_tZ_t)_t)$.

Round these fractional decisions online with one independent uniform $U\in[0,1)$. With $Q_t=\sum_{i\leq t}a_i$, select at time $t$ when $\lfloor Q_t+U\rfloor>\lfloor Q_{t-1}+U\rfloor$. This selects at most $K$ times, at time $t$ with probability $a_t$ conditional on $Z$ and the colors. For every fixed coloring, it is an online rule for $Z$. With probability $\delta$, all colors coincide; then all active copies see $Z$ itself and make identical decisions, so at most $m$ selections are made. Bounding the value by $\Phi_m(Z)$ on this event and by $\Phi_K(Z)$ otherwise proves the lemma.
\end{proof}

\begin{lemma}[Adjacent posteriors]\label{lem:trunc-adjacent}
For $2\leq k\leq n$, every $t$, and every convex $\varphi$ with $\varphi(0)=0$,
\[
\E[\varphi(Z^{k-1}_t)]\leq\frac{k-1}k\,\E[\varphi(Z^k_t)].
\]
\end{lemma}
\begin{proof}
Couple the two hypergeometric counts by turning one uniformly random success of the population into a failure. Given that the sample contains $s$ successes, the removed success was sampled with probability $s/k$. Hence $\Pr[\Xi_{k-1,t}\geq s]=\Pr[\Xi_{k,t}\geq s+1]+(1-s/k)\Pr[\Xi_{k,t}=s]$, that is,
\[
p_{k-1}(t,s)=\Bigl(1-\frac sk\Bigr)p_k(t,s)+\frac skp_k(t,s+1)\qquad(1\leq s\leq k-1),
\]
while $p_{k-1}(t,s)=0$ for $s\geq k$. Jensen's inequality and summation over $s$ give
\[
\sum_{s=1}^{k-1}\varphi(p_{k-1}(t,s))\leq\sum_{s=1}^{k-1}\Bigl[\Bigl(1-\frac sk\Bigr)\varphi(p_k(t,s))+\frac sk\varphi(p_k(t,s+1))\Bigr]=\frac{k-1}k\sum_{s=1}^k\varphi(p_k(t,s)).
\]
Since $R_t$ is uniform on $\{1,\ldots,t\}$, and $p_k(t,s)=0$ whenever $s>\min\{t,k\}$, we have $t\,\E[\varphi(Z^k_t)]=\sum_{s=1}^k\varphi(p_k(t,s))$ and $t\,\E[\varphi(Z^{k-1}_t)]=\sum_{s=1}^{k-1}\varphi(p_{k-1}(t,s))$. Dividing by $t$ proves the claim.
\end{proof}

\begin{theorem}\label{thm:trunc-uniform}
For $2\leq k\leq n$, $r(n,k)>r(n,k-1)$. Equivalently, $\rho_{\mathrm{ord}}(U_{k,n})<\rho_{\mathrm{ord}}(U_{k-1,n})$.
\end{theorem}
\begin{proof}
Let $m=k-1$, and let $\xi_t$ be independent Bernoulli$(m/k)$ variables. For $c\geq0$, the function $(x-c)_+$ is convex and vanishes at zero. By \Cref{lem:trunc-adjacent},
\[
\E[(Z^m_t-c)_+]\leq\frac mk\E[(Z^k_t-c)_+]=\E[(\xi_tZ^k_t-c)_+].
\]
By \eqref{eq:trunc-stopping} and \Cref{lem:trunc-compare,lem:trunc-thin},
\[
m\,r(n,m)=\Phi_m(Z^m)\leq \Phi_m\bigl((\xi_tZ^k_t)_t\bigr)\leq\frac mk\Bigl[\Phi_k(Z^k)-\delta\bigl(\Phi_k(Z^k)-\Phi_m(Z^k)\bigr)\Bigr].
\]
It remains to show $\Phi_k(Z^k)>\Phi_m(Z^k)$. For $t\leq k$ and every $s\leq t$, we have $Z^k_t\geq p_k(t,t)=\binom{n-t}{k-t}/\binom nk\geq\binom nk^{-1}$. Follow an optimal $m$-selection rule, and use one additional slot at its first rejection among the first $k$ times; such a rejection exists because at most $m=k-1$ of these times are selected. This gains at least $\binom nk^{-1}$ on every realization. Hence $m\,r(n,m)<\frac mk\,k\,r(n,k)$.
\end{proof}

Together with \Cref{lem:trunc-topk}, the theorem gives $\rho_{\mathrm{ord}}(U_{k,n})\leq\rho_{\mathrm{ord}}(U_{1,n})\leq\ee$. This recovers, as a guarantee on value, the known ratio $\ee$ for uniform matroids; the ordinal algorithm of \cite{BIKK-APPROX07} even selects each of the $k$ largest elements with the classical cutoff probability.

\subsection{A graphic counterexample}

\begin{theorem}\label{thm:trunc-graphic}
Let $N=10^{42}$, and let $\calM$ be the graphic matroid of the complete bipartite graph $K_{2,N}$. Then
\[
\rho_{\mathrm{ord}}(\calM^{(3)})<\frac{100}{53}<\rho_{\mathrm{ord}}(\calM).
\]
Since $\rank(\calM)=N+1$, there is $4\leq k\leq N+1$ with $\rho_{\mathrm{ord}}(\calM^{(k)})>\rho_{\mathrm{ord}}(\calM^{(k-1)})$. The upper bound on $1/\rho_{\mathrm{ord}}(\calM)$ holds even for algorithms that observe numerical values and know the input distribution used in the proof.
\end{theorem}

The graph $K_{2,N}$ has no cycle of length at most three, so every set of at most three edges is independent and $\calM^{(3)}$ is the uniform matroid $U_{3,2N}$. The theorem follows from the next two propositions and \Cref{lem:trunc-topk}.

\begin{proposition}\label{prop:trunc-three}
For every $m\geq3$, $r(m,3)\geq L^\star$, where $L^\star=0.5326\ldots>53/100$ is given by \eqref{eq:trunc-Lstar}.
\end{proposition}
\begin{proof}
Consider $h\geq6$ elements and the following rule: reject the first $\lfloor h/6\rfloor$ arrivals; until time $\lfloor h/2\rfloor$, accept records; afterwards, accept arrivals among the three largest so far; stop after three acceptances. Call the $t$-th arrival \emph{eligible} if the rule would accept it with a free slot: $R_t=1$ in the first active phase, and $R_t\leq3$ in the second. Eligibility events are independent, with probabilities $1/t$ and $3/t$, respectively, and a slot is free at time $t$ exactly when fewer than three earlier arrivals were eligible. This event depends only on $R_1,\ldots,R_{t-1}$, so it is independent of $R_t$.

Apply \eqref{eq:trunc-posterior} with $h$ in place of $n$. Given a free slot, time $t$ contributes in expectation $p_3(t,1)/t$ selected top-three elements in the first phase, and $\frac1t\sum_{s\leq3}p_3(t,s)=\E[\Xi_{3,t}]/t=3/h$ in the second. Write $x=t/h$. As $h\to\infty$, $p_3(t,1)\to1-(1-x)^3$ uniformly, and the expected number of eligible arrivals before $t$ converges uniformly to $\lambda(x)=\ln(6x)$ for $x\leq1/2$ and to $\lambda(x)=\ln3+3\ln(2x)$ for $x\geq1/2$. By Le Cam's inequality~\cite{LeCam1960}, the number of eligible arrivals before $t$ differs in total variation from a Poisson variable with the same mean by at most the sum of the squared eligibility probabilities, which is $O(1/h)$. Hence the expected fraction of the top three selected converges to
\begin{equation}\label{eq:trunc-Lstar}
L^\star=\int_{1/6}^{1/2}\Bigl(1+\lambda+\frac{\lambda^2}2\Bigr)\frac{3-3x+x^2}{18x}\,dx+\int_{1/2}^1\Bigl(1+\lambda+\frac{\lambda^2}2\Bigr)\frac{dx}{24x^3},
\end{equation}
where $\lambda=\lambda(x)$. With $\alpha_2=\ln2$ and $\alpha_3=\ln3$, integration gives
\[
L^\star=-\frac{3\alpha_2^2}{32}-\frac{\alpha_2\alpha_3}{16}-\frac{5\alpha_2}{32}+\frac{\alpha_3^3}{36}+\frac{11\alpha_3^2}{144}+\frac{47\alpha_3}{144}+\frac{425}{1728}=0.53261\ldots
\]
Enclosures $0.693<\alpha_2<0.694$ and $1.098<\alpha_3<1.099$, used with the unfavorable endpoint in each term, give $L^\star>0.531$.

To apply this to $m$ elements, choose $h\geq\max\{m,6\}$ and insert $h-m$ fictitious elements, smaller than all real ones, at uniformly random positions and in a uniformly random relative order; both are sampled before the arrivals. Comparisons involving fictitious elements are known to the algorithm. Fictitious selections are discarded but count against the capacity. The combined order is uniform and its three largest elements are real, so $r(m,3)$ is at least the value of the rule on $h$ elements. Letting $h\to\infty$ proves the proposition.
\end{proof}

\begin{proposition}\label{prop:trunc-K2N}
With $N=10^{42}$, every online algorithm for the graphic matroid of $K_{2,N}$, even one observing numerical values, has expected value less than $\frac{53}{100}$ times the expected optimum on some random instance. Hence $\rho_{\mathrm{ord}}(\calM)>100/53$.
\end{proposition}
\begin{proof}
We use the construction of~\cite{BHKKMO-ESA25}. First consider two elements. Fix integers $Q>1$ and $L\geq1$. Choose $i\in\{0,\ldots,L-1\}$ with probability $Q^{-(i+1)}/\Gamma$, where $\Gamma=\sum_{i=0}^{L-1}Q^{-(i+1)}$, and present the values $Q^i$ and $Q^{i+1}$ in uniformly random order. The expected maximum is $\nu=L/\Gamma$. Suppose an algorithm accepts a first value $Q^j$ with probability $p_j$. Granting it the second value whenever it rejects the first, its expected reward at scale $i$ is at most
\[
\frac12\bigl[p_iQ^i+(1-p_i)Q^{i+1}\bigr]+\frac12\bigl[p_{i+1}Q^{i+1}+(1-p_{i+1})Q^i\bigr]
=Q^{i+1}\Bigl[\frac{1+1/Q}2+\frac{(1-1/Q)(p_{i+1}-p_i)}2\Bigr].
\]
Each scale has probability times $Q^{i+1}$ equal to $1/\Gamma$, so the differences $p_{i+1}-p_i$ telescope. Using $p_L-p_0\leq1$, the expected reward is at most $\alpha\nu$, where $\alpha=(1+1/Q)/2+(1-1/Q)/(2L)$.

Now let $o_1,o_2$ be the two vertices of degree $N$ in $K_{2,N}$, and let $w_1,\ldots,w_N$ be the others. The edges $o_1w_i,o_2w_i$ form the $i$-th pair. Draw an independent two-element instance for each pair, assign its values to the two edges in random order, and present all edges in uniformly random order. For each pair, consider the first selected edge of that pair. Condition on the values of the other pairs, the arrival order of the edges, and the algorithm's randomness; these are independent of the pair's scale and of the assignment of its values, so the first arriving edge of the pair carries either value with probability $1/2$, as in the two-element instance. The decision on that edge depends only on its value. The bound above therefore applies, and the first selections have total expected value at most $N\alpha\nu$. A forest contains both edges of at most one pair, since two complete pairs form a cycle of length four. The remaining selections are worth at most $Q^L$. On the other hand, the heavier edges of the pairs form a forest, so the expected optimum is at least $N\nu$. The ratio is at most $\alpha+Q^L/(N\nu)$.

For $Q=100$, $L=21$, and $N=Q^L$, we have $\Gamma<1/99$, so $\nu>2079$ and $\alpha=37/70$. The ratio is at most $37/70+1/2079<53/100$. A $c$-competitive algorithm has expected value at least $1/c$ times the optimum on every fixed instance, and hence on average. Therefore every such $c$ satisfies $c\geq(37/70+1/2079)^{-1}>100/53$, and so does the infimum $\rho_{\mathrm{ord}}(\calM)$.
\end{proof}

\section{From Single-Sample Prophet Inequalities to Secretary}\label{sec:reduction}

This appendix shows how to turn a single-sample matroid prophet
algorithm into a matroid secretary algorithm. We give a
self-contained presentation of the construction attributed to
Wenzheng Li by \cite{FLTWWZ-EC24}. We work in the secretary model with the matroid $\calM=(V,\calI)$ and $n=|V|$ known in advance; the values $v\in\mathbb R_{\geq0}^V$ are fixed and arrive in uniformly random order. As in \Cref{sec:snapshot}, $\OPT(T)$ is decreasing-value greedy on $T$ with ties broken by labels, and $f\succ e$ means that $f$ precedes $e$ in this order. For a nonnegative vector $w$ indexed by $T\subseteq V$, write $\operatorname{opt}(w)=\max\{w(I):I\in\calI,\ I\subseteq T\}$ for the optimum value of the restriction to $T$; thus $\operatorname{opt}(v|_T)=v(\OPT(T))$.

A single-sample prophet algorithm is \emph{$c$-competitive in random order} if, for every matroid and independent nonnegative distributions, given one independent sample from each distribution, it selects an independent set of expected value at least $1/c$ times the expected optimum, when the arrival order is uniformly random and independent of the samples and values and is not known to the algorithm. By averaging over orders, \Cref{thm:ssmpi-tight} gives such an algorithm with $c=2$.

\begin{theorem}[Secretary from single-sample prophet inequalities]\label{thm:reduction}
A single-sample prophet algorithm that is $c$-competitive in random order yields a $16c^2$-competitive secretary algorithm for every matroid known in advance. The reduction preserves polynomial running time. With \Cref{alg:ssmpi-tight}, it gives an ordinal $64$-competitive secretary algorithm.
\end{theorem}

The natural idea is to reject a prefix of the arrivals, use their values as samples, and run the prophet algorithm on the remaining elements. The difficulty is that the prophet algorithm may select an element whose value was used as a sample, and such an element has already been rejected. The reduction makes these conflicts rare: each element is presented to the prophet algorithm with a value that is nonzero only with a small probability $q$, independently as a sample and as a value. The useful reward is then linear in $q$, while a conflict requires both activations and costs only order $q^2$. A preliminary sample ensures that this quadratic loss is charged against the optimum rather than against the total value of all elements.

\medskip\noindent\textbf{A preliminary sample.}
Draw $K_0\sim\Bin(n,1/2)$ and reject the first $K_0$ arrivals, forming a set $P$; it contains each element independently with probability $1/2$. An element $e\notin P$ is \emph{eligible} if $e\notin\cl(\{f\in P:f\succ e\})$. Let $T$ be the set of eligible elements. Eligibility of $e$ can be decided when $e$ arrives, since all values in $P$ have been observed.

\begin{lemma}\label{lem:reduction-sample}
$\E[v(T)]=\E[v(\OPT(P))]\leq v(\OPT(V))$ and $\E[\operatorname{opt}(v|_T)]\geq v(\OPT(V))/2$.
\end{lemma}
\begin{proof}
Whether $e\in\cl(\{f\in P:f\succ e\})$ is independent of whether $e\in P$. If it fails, then $e\in\OPT(P)$ when $e\in P$, and $e\in T$ when $e\notin P$; both happen with probability $1/2$. Summing over $e$ gives the equality, and $\OPT(P)$ is independent, which gives the inequality. Each $e\in\OPT(V)$ is not spanned by the elements preceding it, so $\OPT(V)\setminus P$ is an independent subset of $T$, of expected value $v(\OPT(V))/2$.
\end{proof}

\medskip\noindent\textbf{Rare activations.}
Condition on $P$, and let $V'=V\setminus P$ and $N=|V'|$. The remaining arrival order is uniform on $V'$. For $e\in V'$ let $\bar v_e=v_e\ind\{e\in T\}$, which can be computed when $e$ arrives. Fix $q\in(0,1)$. We run the prophet algorithm on the restriction of $\calM$ to $V'$ with the independent distributions $\Dist_e=(1-q)\delta_0+q\delta_{\bar v_e}$. Its sample and value of $e$ are $X_e\bar v_e$ and $Y_e\bar v_e$, where all $X_e,Y_e$ are independent Bernoulli($q$) bits.

\begin{enumerate}[leftmargin=*]
\item Draw independent Bernoulli($q$) bits $b_1,\ldots,b_N$, and let $K=\sum_ib_i$. Reject the next $K$  arrivals, forming a set $R$, and set $X_e=\ind\{e\in R\}$. Give the prophet algorithm the sample $\bar v_e$ for $e\in R$ and $0$ for $e\in V'\setminus R$, and draw the bits $Y_e$ independently.
\item For $i=1,\ldots,N$: if $b_i=1$, take the next element $e$ of $R$ in its arrival order; if $b_i=0$, let $e$ be the next  arrival and compute $\bar v_e$. Present $e$ to the prophet algorithm with value $Y_e\bar v_e$. If the prophet algorithm selects $e$ and $b_i=0$, accept $e$.
\end{enumerate}

All selections are recorded in the internal state of the prophet algorithm, including selections of elements of $R$, which are discarded. Hence the accepted set is a subset of the prophet algorithm's independent output. Every decision on an arrival is made immediately.

\begin{lemma}\label{lem:reduction-simulation}
Conditional on $P$, the set $R$ contains each element of $V'$ independently with probability $q$, and the order $\Sigma$ in which elements are presented to the prophet algorithm is uniform and independent of $R$ and of the bits $Y$.
\end{lemma}
\begin{proof}
Fix $D\subseteq V'$ and an order $\sigma$ of $V'$. The outcome $(R,\Sigma)=(D,\sigma)$ occurs for exactly one arrival order and one bit string: the arrival order lists $D$ in the order $\sigma$ followed by $V'\setminus D$ in the order $\sigma$, and the bits mark the positions of $D$ in $\sigma$. Hence
\[
\Pr[R=D,\ \Sigma=\sigma\mid P]=\frac{q^{|D|}(1-q)^{N-|D|}}{N!}.
\]
The bits $Y$ are drawn independently of both.
\end{proof}

By \Cref{lem:reduction-simulation}, the prophet algorithm sees independent samples and values from the distributions $\Dist_e$, in a uniformly random order independent of them, exactly as its guarantee requires.

\begin{proof}[Proof of \Cref{thm:reduction}]
A selected element of $R$ contributes to the prophet algorithm's value only if $X_e=Y_e=1$, so the discarded value is at most $\sum_{e\in V'}X_eY_e\bar v_e$, whose conditional expectation is $q^2v(T)$. Every accepted element has value $v_e\geq Y_e\bar v_e$. The prophet guarantee therefore gives
\[
\E[v(\ALG)\mid P]\geq\frac1c\,\E\bigl[\operatorname{opt}((Y_e\bar v_e)_{e\in V'})\bigm|P\bigr]-q^2v(T)\geq\frac qc\,\operatorname{opt}(v|_T)-q^2v(T),
\]
where the second inequality retains the activated elements of $\OPT(T)$. By \Cref{lem:reduction-sample}, $\E[v(\ALG)]\geq(q/(2c)-q^2)\,v(\OPT(V))$. Choosing $q=1/(4c)$ gives $v(\OPT(V))/(16c^2)$.

Eligibility requires one independence query per arrival against a prefix of $\OPT(P)$, and the remaining bookkeeping is polynomial, so the reduction preserves polynomial running time. For \Cref{alg:ssmpi-tight}, every value presented to the prophet algorithm is either $0$ or an observed value of an eligible element, and the algorithm only compares values. Treat every eligible element as larger than every artificial zero. For positive distinct values, the comparisons and the eligibility tests then depend only on the relative order of the observed values, so the secretary algorithm is ordinal; its guarantee extends to all nonnegative values, with ties resolved by labels, by taking limits of positive distinct value vectors with the same order. With $c=2$, the ratio is $64$.
\end{proof}

\end{document}